\documentclass{article}
\usepackage{PRIMEarxiv}

\usepackage[utf8]{inputenc} 
\usepackage[T1]{fontenc}    
\usepackage{hyperref}       
\usepackage{url}            
\usepackage{booktabs}       
\usepackage{amsmath,amssymb,amsfonts}%
\usepackage{amsthm}%
\usepackage{mathrsfs}%
\usepackage{nicefrac}       
\usepackage{microtype}      
\usepackage{lipsum}
\usepackage{fancyhdr}       
\usepackage{graphicx}       
\usepackage{algorithm}%
\usepackage{algorithmicx}%
\graphicspath{{media/}}     

\title{Meta-Processing: A robust framework for multi-tasks seismic processing
}

\author{
  Shijun Cheng, Randy Harsuko, Tariq Alkhalifah\\
  King Abdullah University of Science and Technology \\
  Thuwal 23955-6900\\
  \texttt{{sjcheng.academic@gmail.com}, {mochammad.randycaesario@kaust.edu.sa}, tariq.alkhalifah@kaust.edu.sa} \\
}

\begin{document}
\maketitle

\begin{abstract}
Machine learning-based seismic processing models are typically trained separately to perform specific seismic processing tasks (SPTs), and as a result, require plenty of training data. However, preparing training data sets is not trivial, especially for supervised learning (SL). Nevertheless, seismic data of different types and from different regions share generally common features, such as their sinusoidal nature and geometric texture. To learn the shared features, and thus, quickly adapt to various SPTs, we develop a unified paradigm for neural network-based seismic processing, called Meta-Processing, that uses limited training data for meta learning a common network initialization, which offers universal adaptability features. The proposed Meta-Processing framework consists of two stages: meta-training and meta-testing. In the meta-training stage, each SPT is treated as a separate task and the training dataset is divided into support and query sets. Unlike conventional SL methods, here, the neural network (NN) parameters are updated by a bilevel gradient descent from the support set to the query set, iterating through all tasks. In the meta-testing stage, we also utilize limited data to fine-tune the optimized NN parameters in an SL fashion to conduct various SPTs, such as denoising, interpolation, ground-roll attenuation, image enhancement, and velocity estimation, aiming to converge quickly to ideal performance. Comprehensive numerical examples are performed to evaluate the performance of Meta-Processing on both synthetic and field data. The results demonstrate that our method significantly improves the convergence speed and prediction accuracy of the NN.
\end{abstract}


\section{Introduction}\label{sec1}
 Seismic processing is an important step in the seismic survey value chain, dedicated to improving the quality of the seismic data. It involves a range of operations that include filters and corrections, such as seismic denoising, deconvoluton, interpolation, and velocity estimation. These techniques are critical for extracting useful information from recorded data that can be used to understand the geological structure and characteristics of the Earth's subsurface \cite{gan2022ewr, sun2023multichannel, yang2023fwigan}. 

 Commonly, seismic processing techniques can be classified into two broad categories: theory-driven and data-driven approaches, which are based on the fundamental principles that guide each technique and the underlying mathematical models that drive the analysis. Theory-driven approaches leverage the nature of the physical processes that govern the propagation of seismic waves in the subsurface. For example, the $f\raisebox{0mm}{-}k$ filtering method, which relies on the difference in frequency distribution between noise and signal, uses Fourier transform to convert seismic data into the frequency-wavenumber domain in which noise is potentially separated from signal \cite{askari2008ground, liu2019seismic}. Radon transform-based interpolation is used to estimate the missing data by fitting linear events in the seismic data \cite{trad2002accurate, trad2003latest, yu2007wavelet, wang2010seismic, shao2022seismic}. Seismic normal moveout corrections utilize the relationship between the travel time of seismic waves and the recording distance away from the source to correct the time delay of seismic reflection data \cite{de1988normal, shatilo2000constant}. Considerable attention has been devoted to the development of theory-driven approaches for seismic processing \cite{spitz1991seismic, wang2002seismic, herrmann2008curvelet, herrmann2008non, naghizadeh2009f, fomel2010seislet, liu2010oc, chen2014random, chen2015random, chen2014iterative, geng2020relative, zhang20223}, as they have the potential to provide optimal solutions based on mathematical and physical principles. 
 
 However, theory-driven approaches are usually developed based on some assumptions, which means that the accuracy of the processed data can be significantly impacted if the assumptions are incorrect. Moreover, the complexity of the processing involved in this approach can be a significant obstacle, requiring a high level of expertise and knowledge. Consequently, processing can be time-consuming and costly. Furthermore, the computational challenges associated with theory-driven seismic processing cannot be overlooked either. The calculations involved in some of the processing tasks, like velocity estimation, are frequently complicated and require significant computing power and resources. 

 On the other hand, data-driven seismic processing techniques have been gaining popularity in recent years due to their ability to process and interpret large amounts of seismic data quickly and efficiently. Unlike theory-driven approaches that rely on a priori assumptions about the underlying physical mechanism, data-driven methods typically leverage machine learning algorithms to identify patterns and extract meaningful information from the seismic data. One of the primary strengths of data-driven methods is their flexibility and adaptability to different types of seismic processing tasks (SPTs). Hence, it has been used to perform a wide range of SPTs, such as noise attenuation \cite{yu2019deep, dong2019desert, dong2022seismic, yuan2020ground, birnie2021potential, liu2022accelerating, saad2020deep, liu2022coherent, saad2023unsupervised}, interpolation \cite{jia2017can, wang2019deep, wang2020seismic, tang2023simultaneous, zhang2020can}, deconvolution \cite{alaudah2018learning, gao2021deep}, imaging \cite{zhang2021consistent, zhang2021least, zhang2022improving, yu2023enhancing, cheng2023seismic, cheng2023elastic}, inversion \cite{yang2019deep, du2022deep, li2022target, yang2023well}, and interpretation \cite{zheng2019applications, waldeland2018convolutional, wu2019faultseg3d, shi2019saltseg}, and are often not limited by specific assumptions of any given processing approach. Meanwhile, the increasing amounts of seismic data being acquired has made it increasingly necessary to find faster and more efficient ways of processing and interpreting such data. Data-driven approaches have emerged as a viable solution to this problem, as they can automate many of the routine and repetitive tasks involved in theory-driven seismic processing workflows, reducing processing time and improving the accuracy of results. 

Although machine learning-based data-driven methods have shown successful applications in many SPTs, most of the developments involved developing networks for specific tasks or training them on specific datasets. This leads to many drawbacks. Firstly, different neural network (NN) models need to be trained and optimized separately for each SPT, which can be time-consuming and computationally expensive. Secondly, they may not fully capture the complexity and interrelationships among different SPTs. As a result, their generalization ability to other tasks or datasets is often limited. Moreover, training NN for various SPTs may require a large amount of labeled data for each specific task, which can be challenging and costly to obtain. This limitation may restrict the applicability of these methods, particularly in cases where data are scarce or expensive to acquire \cite{birnie2021potential, liu2022coherent, saad2023unsupervised}, like field data. Furthermore, training networks for SPTs independently may lead to suboptimal performance and inefficiencies. By considering the interrelationships among different tasks, it may be possible to improve the overall processing efficiency and accuracy. Therefore, it is important to develop more integrated approaches that can effectively capture the complex interrelationships among different SPTs. Multi-task learning (usually involving two tasks) have demonstrated improvements in performance compared to the common case in which the networks are trained for the two tasks seperately \cite{birnie2022transfer}. 

Obviously, multi-task learning is a challenging. However, it is feasible as seismic data may share common feature information, such as their sinusoidal nature and geometric behavior. This feature was demonstrated by Harsuko and Alkhalifah \cite{harsuko2022storseismic}, as they utilized a neural network, borrowed from natural language processing, to perform many SPTs. Specifically, they pretrain an NN for seismic reconstruction in a self-supervised fashion to extract seismic features. The pretrained NN is then fine-tuned via supervised learning for various SPTs, including denoising, velocity estimation, first arrival picking, and normal moveout. This work involves reusing knowledge from the source domain to improve performance on multiple target domains, synonymous to transfer learning. The hope is that the pretrained model will provide a good starting point for learning the various tasks, reducing the training time required to achieve high performance. Nonetheless, the utilization of transfer learning comes with certain limitations and challenges that impede its full potential. Firstly, transfer learning assumes that the source and target domains are similar, and that the knowledge learned in the source domain is relevant to the target domain, but this is not always the case. If the target domain is significantly different from the source domain, the performance of the transfer learning model may suffer. Even when the source and target domains are similar, the tasks in seismic processing can be different. For example, a model trained for event detection may not perform well for seismic inversion, which requires estimating the subsurface properties. Secondly, transfer learning models may suffer from overfitting to the source domain, leading to poor performance on the target domain. Furthermore, conventional transfer learning requires a large amount of labeled data in the source domain to train the model, which may not be readily available due to the high cost and time required to collect and label seismic data. Finally, it is difficult to determine which features or representations are transferable between different tasks and domains in seismic processing due to the complex nature of the data. 

In contrast, meta-learning (MetaL) \cite{schmidhuber1987evolutionary, hospedales2021meta} could be a promising alternative in seismic processing \cite{mousavi2022deep}, as it focuses on learning how to learn and adapt to new tasks and domains efficiently. MetaL models can be highly flexible and robust, allowing them to quickly adapt to new tasks and generalize across domains. Also, it can leverage prior knowledge and experience to learn new tasks more efficiently, without requiring a large amount of labeled data. Moreover, MetaL models can be trained on multiple tasks simultaneously, making them more scalable than transfer learning models. This flexibility of MetaL models is particularly useful in seismic processing, where the subsurface properties can vary greatly across different geological settings, and the tasks may differ substantially. The robustness of MetaL models helps them to be less susceptible to overfitting and task-specific biases, which is critical in seismic processing due to the limited availability of labeled data. Furthermore, the efficiency of MetaL in leveraging prior knowledge and experience can be beneficial for reducing the time and cost required to develop effective seismic processing models. Besides, the scalability of MetaL models can facilitate the handling of the numerous tasks required in seismic processing, allowing for more efficient and effective processing of seismic data. However, limited attention has been devoted to the application of MetaL in the field of seismology. Yuan et al. \cite{yuan2020adaptive} addressed the challenge of adaptation for first arrival picking among different seismic datasets using MetaL algorithms and demonstrated more accurate picking results than transfer learning methods. Sun and Alkhalifah \cite{sun2020ML} employ the concept of MetaL to develop an optimization algorithm that accelerates the convergence of full waveform inversion. 

Hence, in this paper, inspired by the many features that MetaL posses, we develop a unified paradigm for various SPTs, referred to as Meta-Processing, which uses limited training data and provides a common network initialization for universal adaptation. We, specifically, use the UNet \cite{ronneberger2015u}, which is the most popular and classic deep convolutional NN in the field of seismic processing \cite{wu2019faultseg3d, shi2019saltseg, yang2019deep, liu2022coherent, yu2023enhancing}, as our basic network architecture. Within the framework of Meta-Processing, this network will undergo a two-stage training: meta-training and meta-testing. In meta-training, each seismic processing is regarded as an independent task, whose limited training data is separated as two sets: support and query data sets. Different from conventional training approaches that optimize an NN model for a specific data set, a bilevel gradient updating from the support set to query sets is utilized to train a meta-learner model using various SPTs, including seismic denoising, interpolation, ground roll attenuation, imaging enhancement, and velocity estimation. In meta-testing, the meta-learner is used to quickly adapt the model to SPTs by fine-tuning its parameters on a few training data. Following that, the fine-tuned model is evaluated on a test set of examples from the various SPTs. Hence, the key idea of our method is to train an initialized set of parameters for a designated model across various SPTs. By doing so, we aim to achieve maximal performance on the corresponding task's test set with only a small number of gradient updates, which are computed using a small amount of training data specific to each task. 

The rest of the paper is structured as follows. We begin with reviewing the fundamental workflow of neural network-based seismic processing. Then, we detail the idea of the Meta-Processing algorithm, also, illustrate the data set establishment, network architecture, and the loss functions used in this study. Subsequently, we present the results from implementing our method on both synthetic and field data. Finally, we conclude by summarizing our work. \\ 
\section{Methodologies}\label{sec2}
In this section we will highlight the concept of NN-based processing then describe the components of the Meta-processing including the algorithm, the data set, the network architecture, and loss function. 

\subsection{Neural network-based seismic processing}
\begin{figure}[htp]
\centering
\includegraphics[width=0.4\textwidth]{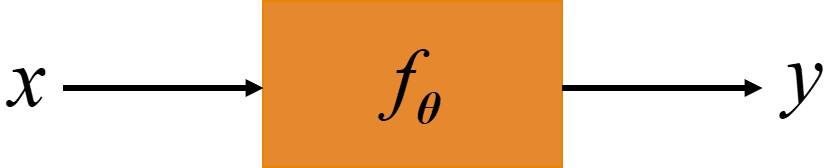}
\caption{The basic idea behind NN-based seismic processing, where $x$ and $y$ represent the input seismic data and the ideal processing results, respectively, and ${f_{\boldsymbol{\theta}}}$ denotes a parameterized function to represent the mapping relationship from $x$ to $y$.}
\label{fig1}
\end{figure}

Seismic processing is a complex and data-intensive task that aims to extract valuable information about the subsurface structure of the Earth by analyzing the signals embedded in recorded seismic waves. Machine learning algorithms, particularly NN, have emerged as a promising tool for addressing the challenges posed by seismic processing. The basic idea behind NN-based seismic processing, which is shown in Figure \ref{fig1}, is to leverage the power of machine learning algorithms to learn the underlying mapping relationships between input seismic data $x$ and the ideal processing results $y$ as follows:
\begin{equation}
\setlength{\abovedisplayskip}{3pt}
\setlength{\belowdisplayskip}{3pt}
y=\mathrm{NN}(x;\boldsymbol{\theta}),
\end{equation}
where the mapping relationships are represented by a parameterized function ${f_{\boldsymbol{\theta}}}$ with the learned parameters $\boldsymbol{\theta}$ of the NN. 

\begin{figure}[htp]
\centering
\includegraphics[width=0.75\textwidth]{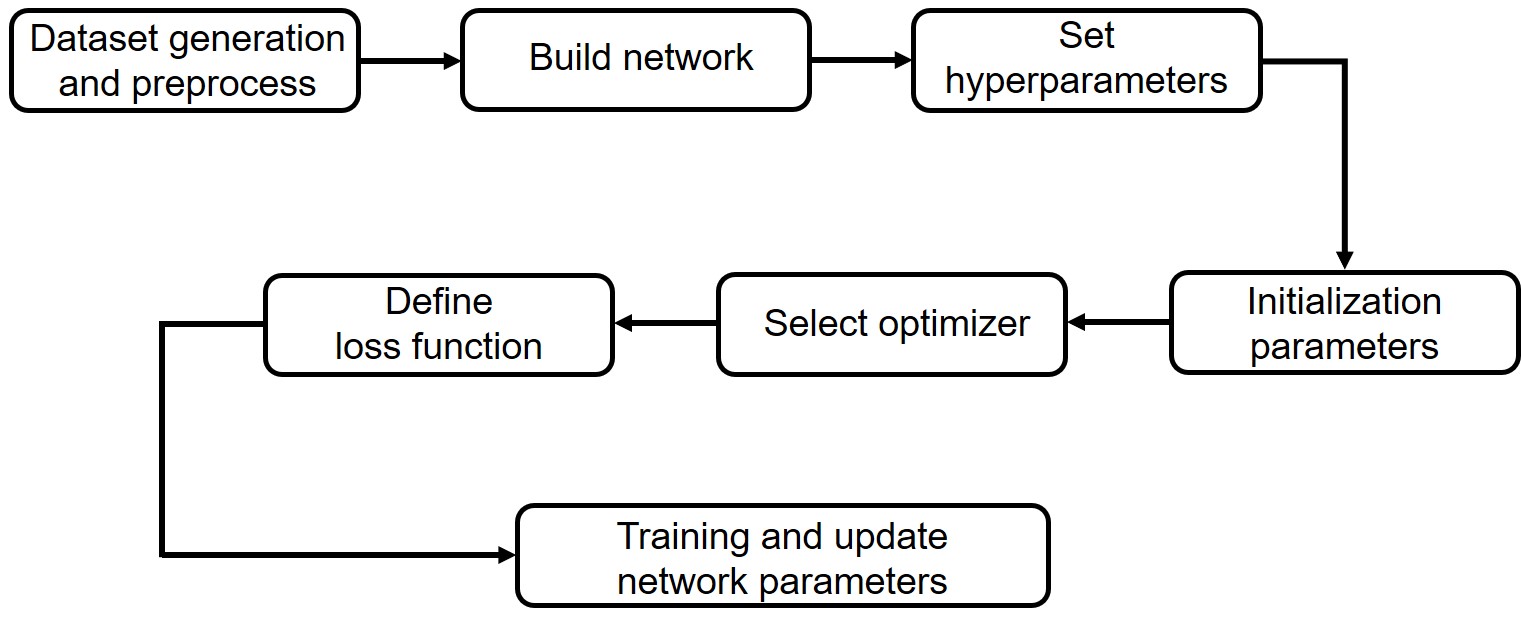}
\caption{The workflow for training NN.}
\label{fig2}
\end{figure}

To obtain the parameterized function ${f_{\boldsymbol{\theta}}}$ for a specific SPT, we usually need to train the NN from scratch, whose workflow is reviewed in Figure \ref{fig2}. In brief, a large dataset is prepared and preprocessed, for example by normalizing it to $[-1,1]$. Next, we design an appropriate network architecture appropriate for the specific task at hand. Afterwards, careful consideration must be taken to set the hyperparameters (e.g., learning rate), NN initialization, the selection of optimizer, and the definition of a loss function. Initialization is a crucial component that affects the network's convergence speed and final optimization results, and therefore, it is essential to provide an appropriate initialization for the network. In practice, however, many initialization techniques rely on random or default values. Once these steps have been completed, the network can be trained using gradient descent to update the network's parameters iteratively. 

While the NN can effectively process input seismic data to predict desired output variables of interest after being trained, its efficacy is limited to the specific dataset and the task it was trained on. Furthermore, even for the same task, performance can be severely impacted when applied the network on seismic data from other regions. This occurs because the NN is only able to capture a limited number of seismic features from the training dataset, resulting in restricted performance and generalizability. Therefore, re-optimizing the NN is necessary to achieve advanced performance, although this process can be time-consuming. 

\begin{figure}[htp]
\centering
\includegraphics[width=0.5\textwidth]{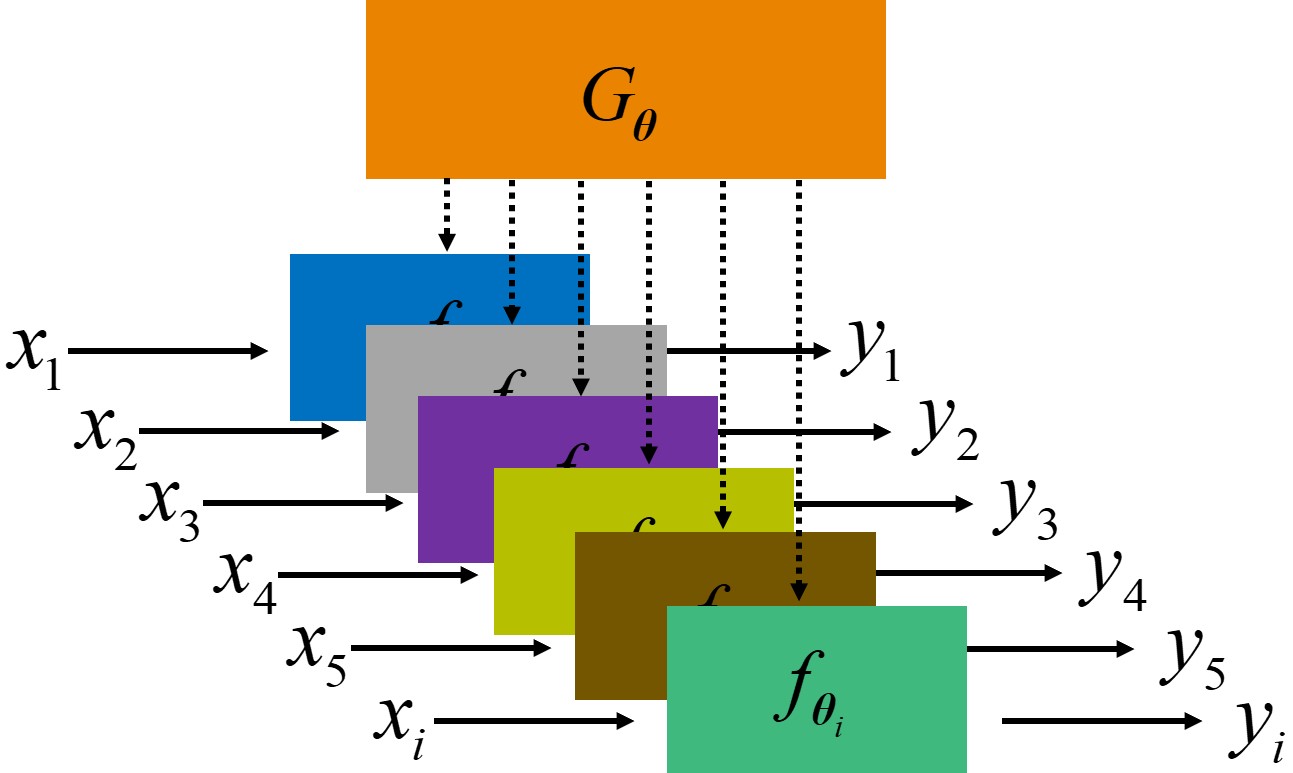}
\caption{A schematic of our motivation, where $x_i$ and $y_i$ are the input seismic data and the ideal processing results, respectively, for various SPTs, and ${G_{\boldsymbol{\theta}}}$ denotes the designated task mapping relationship, which provides robust initialization to the parameterized function ${f_{{\boldsymbol{\theta}}_i}}$ of each target SPT.}
\label{fig3}
\end{figure}

In light of these challenges, we are compelled to inquire whether it is possible to present a multi-task processing network with exceptional generalization abilities, capable of achieving superior performance via minimal gradient updates. In other words, our objective transcends the confines of a solitary SPT and instead seeks a task-level mapping relationship by standing on the shoulders of various SPTs. We illustrate this motivation in Figure \ref{fig3}. As seen, we aim to train a task-level parameterized function ${G_{\boldsymbol{\theta}}}$ that captures the designated task mapping relationship. Such a function can serve as a robust initialization for various target SPTs, enabling them to converge to the corresponding optimal solution ${f_{{\boldsymbol{\theta}}_i}}$ using limited training dataset and a small number of gradient updates. In the subsequent section, we will illustrate the application of MetaL algorithms to realize this goal.

\subsection{Meta-Processing algorithm}
MetaL provides a different framework in machine learning by enhancing the learning algorithm itself through multiple learning episodes across a distribution of related tasks \cite{finn2017model}. Inspired by this concept, we propose a Meta-Processing algorithm (see Algorithm 1) to provide a good initial NN model for various SPTs. In the following, we briefly introduce the algorithm. 

First, we need to split the collected training data into a support data set and a query data set for each task. The support and query data sets are similar to training and test sets in the context of conventional supervised learning (SL), but there are some important differences. In conventional SL, the training data set is used to optimize the model for a specific objective, while the test data set is used to evaluate the performance of the trained model on new, unseen data. In contrast, here, the support data set is used to train the model on the task-specific objective during the inner loop optimization, while the query data set is used to evaluate the performance of the adapted model after the inner loop optimization, i.e., in the outer loop optimization, which includes all the tasks. 

Following that, we can start the meta-training stage of the Meta-Processing algorithm. Let's use a parameterized function ${G_{\boldsymbol{\theta}}}$ to represent the NN with learned parameters ${\boldsymbol{\theta}}$. First, we randomly initialize the NN parameters ${\boldsymbol{\theta}}$. Then, we sample an SPT  ${\mathcal{T}_i}$ (like denoising, interpolation, and image enhancement), which includes limited support and query data sets, for a given task set ${p \left( \mathcal{T} \right)}$. For each sample task, we first perform a few iterations of gradient descent, namely the inner loop, to optimize the Meta-Processing model parameters. That is, we evaluate the performance of the NN on the support data set from the sampled task, and use gradient descent over single or multiple iterations to obtain the updated network parameters ${\boldsymbol{\theta}}^{'}$ as follows:
\begin{equation}\label{eq2}
\setlength{\abovedisplayskip}{3pt}
\setlength{\belowdisplayskip}{3pt}
\boldsymbol{\theta}_i^{'} = {\boldsymbol{\theta}} - lr_{inner} \cdot \nabla_{\boldsymbol{\theta}} \mathcal{L}_{\mathcal{T}_i} \left( G_{\boldsymbol{\theta}} \right),
\end{equation}
where ${lr_{inner}}$ denotes the learning rate for inner iterations as a hyperparameter, and ${\mathcal{L}}$ is the loss function. It should be emphasized that the inner loop is performed for each task separately, using a copy of the Meta-Processing model that is initialized with the current parameters of the Meta-Processing model. In essence, we are only updating the parameters of a copy of the meta-model, and not the meta-model itself. Specifically, we calculate the gradients based on the loss, and employ these gradients to update the model parameters using Equation (\ref{eq2}). Here, we do not use backpropagation to update the meta-model. 

Subsequently, the updated network parameters ${{\boldsymbol{\theta}}_i^{'}}$ are assessed on the query data set to calculate the corresponding loss value $\mathcal{L}_{\mathcal{T}_i}( G_{{\boldsymbol{\theta}}_i^{'}})$, i.e., the outer loop. This marks the completion of the training process for one task. As we repeat this process for all other tasks, we accumulate the losses evaluated on the query data sets of all tasks. Here, the accumulated loss 
\begin{equation}\label{eq3}
\setlength{\abovedisplayskip}{3pt}
\setlength{\belowdisplayskip}{3pt}
{\boldsymbol{\theta}} \leftarrow {\boldsymbol{\theta}} - lr_{meta} \cdot \nabla_{\boldsymbol{\theta}} \sum_{\mathcal{T}_{i} \sim p (\mathcal{T})}^{} \mathcal{L}_{\mathcal{T}_i} ( G_{{\boldsymbol{\theta}}_i^{'}}),
\end{equation}
is finally used to update the desired Meta-Processing model parameters ${\boldsymbol{\theta}}$, where ${lr_{meta}}$ denotes the meta (outer iterations) learning rate. The aforementioned steps constitute the complete process of one epoch of training in the meta-training stage. As seen, the key difference between meta-training and conventional SL is that in meta-training, the model is trained to learn a good initialization that can be quickly adapted to new tasks, whereas the idea of conventional SL is to ensure the trained model can provide the accurate predictions on new, unseen data. 

After completing meta-training, we will perform the meta-testing stage to fine-tune the meta-based initialization model on each task, which is also trained with new limited data, and evaluate the model's convergence speed for each task, as well as its prediction accuracy on the corresponding test sets. The meta-testing stage follows a similar procedure as in conventional SL, with the only difference being that we provide the NN with a better and more robust initialization that can adapt to various SPTs. 

\begin{algorithm}
\caption{Meta-Processing}\label{alg:Framwork}
\textbf{Input:} ${p(\mathcal{T})}$: Different seismic processing tasks with the corresponding support and query datasets. \\
\textbf{Input:} ${lr_{inner}, lr_{meta}}$: Learning rate for inner and outer loops, respectively. \\
\textbf{Input:} ${iter}$: The number of iterations in the support dataset for every task. \\
\textbf{--------------------------------------- Meta-training stage ------------------------------------} \\
\textbf{Output:} Meta-based initialization of the NN model 
\begin{algorithmic}
\State 1: Randomly initialize network parameters ${\boldsymbol{\theta}}$
\State 2: \textbf{while} all tasks ${p(\mathcal{T})}$ \textbf{do}
\State 3: \quad Sample batch of tasks ${\mathcal{T}_i \sim p ( \mathcal{T})}$
\State 4: \quad \textbf{for} every $\mathcal{T}_i$ \textbf{do}
\State 5: \quad \quad \textbf{for} ${i}$ \textbf{in} ${iter}$ \textbf{do}
\State 6: \quad \quad \quad Evaluate $\nabla_{\boldsymbol{\theta}} \mathcal{L}_{\mathcal{T}_i} \left( G_{\boldsymbol{\theta}} \right)$ with respect to the support dataset for the sample task $\mathcal{T}_i$
\State 7: \quad \quad \quad Compute adapted parameters with gradient descent: \\
\quad \quad \quad \quad \quad \quad \quad \quad \quad ${\boldsymbol{\theta}}_i^{'} = {\boldsymbol{\theta}} - lr_{inner} \cdot \nabla_{\boldsymbol{\theta}} \mathcal{L}_{\mathcal{T}_i} \left( G_{\boldsymbol{{\boldsymbol{\theta}}}} \right)$
\State 8: \quad \quad \textbf{end for}
\State 9: \quad \quad Evaluate $ \mathcal{L}_{\mathcal{T}_i}( G_{{\boldsymbol{\theta}}_i^{'}})$ with respect to the query dataset from the sample task $\mathcal{T}_i$
\State 10: \quad \textbf{end for}
\State 11: \quad Sum the loss of all tasks on the query dataset: $\mathcal{L}_{sum} = \sum_{\mathcal{T}_i \sim p \left( \mathcal{T} \right)} \mathcal{L}_{\mathcal{T}_i} ( G_{{\boldsymbol{\theta}}_i^{'}})$
\State 12: \quad Update the Meta-Processing ${\boldsymbol{\theta}} \leftarrow {\boldsymbol{\theta}} - lr_{meta} \cdot \nabla_{\boldsymbol{\theta}} \mathcal{L}_{sum}$
\State 13: \textbf{end while}
\State 14: \textbf{Return:} Meta-Processing parameters ${\boldsymbol{\theta}}$
\end{algorithmic}
\textbf{--------------------------------------- Meta-testing stage ------------------------------------}\\
\textbf{Output:} Task-specific NN model 
\begin{algorithmic}
\State 15: Fine-tune the Meta-Processing parameters ${\boldsymbol{\theta}}$ on each specific task
\State 16: Testing the updated model to obtain the seismic processing results
\end{algorithmic}
\end{algorithm}

\subsection{Data set establishment}
As previously stated, unlike conventional SL, our algorithm requires a support data set and a query data set to be provided during the meta-training stage, and a training data set and a test data set to be utilized during the meta-testing stage. Therefore, for each of the five tasks specified in this study, we generate 200 pairs of input-label data for both the support and query data sets. Likewise, for the meta-testing stage, we also collect 200 pairs of input-label data for each task. It is worth emphasizing that all the training data used in our experiments are synthetic, and the size of the dataset is deliberately limited. This is due to the significant challenge of acquiring labeled data in real-world scenarios.  We will evaluate the effectiveness of our algorithm on both synthetic and field data to assess its performance in practical settings. 

\subsection{Network architecture}
\begin{figure}[htp]
\centering
\includegraphics[width=1\textwidth]{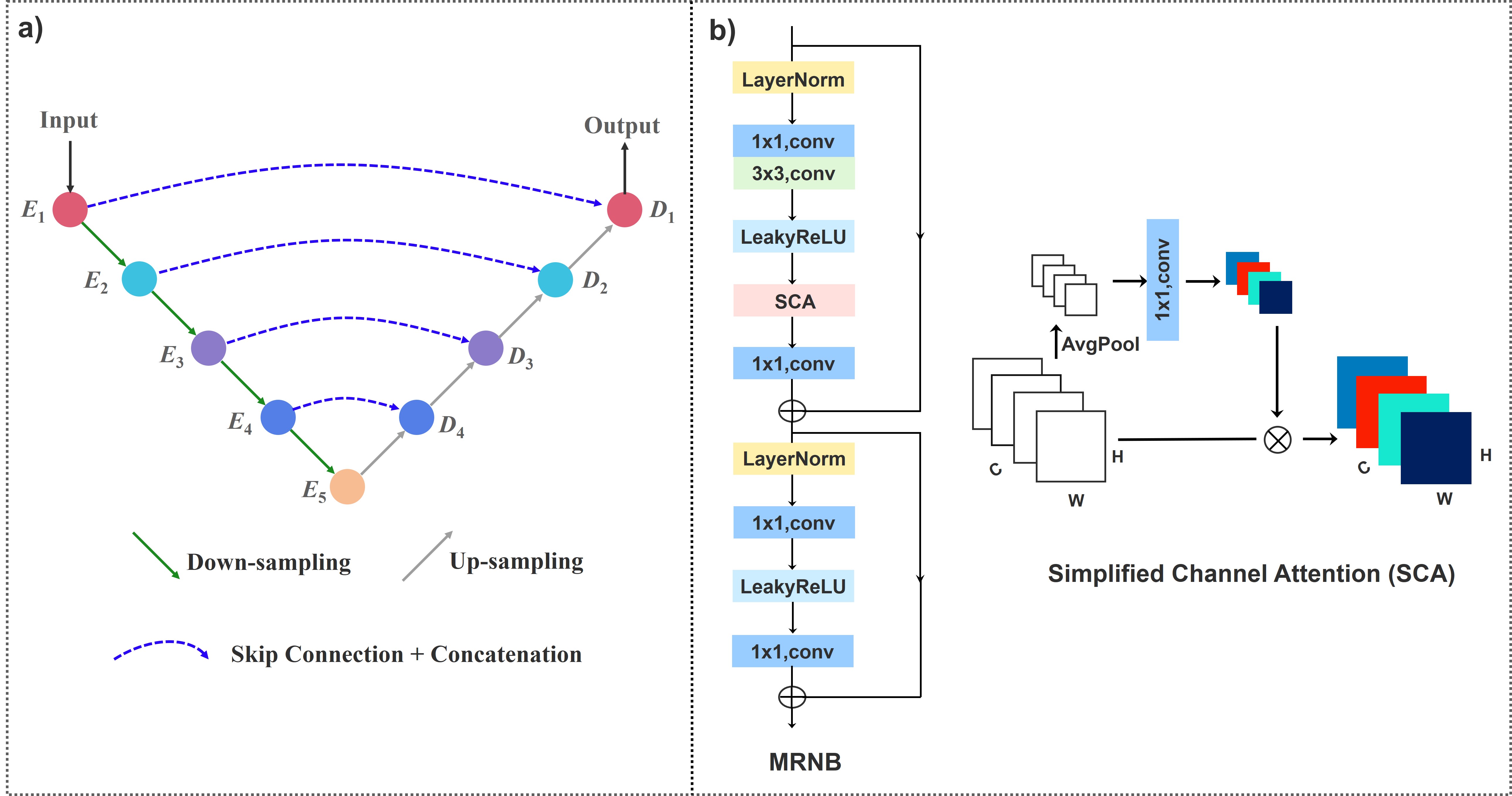}
\caption{The neural network architecture used in our study. (a) The UNet architecture. (b) The illustration of the modified residual network baseline (MRNB). }
\label{fig4}
\end{figure}

The UNet is a type of convolutional neural network architecture commonly applied in the field of seismology, and has demonstrated excellent performance in numerous SPTs. Here, we also adopt the UNet network architecture, as shown in Figure \ref{fig4}a. The UNet architecture consists of a contracting path and an expanding path. The contracting path is a series of encoders ($E_1$, $E_2$, $E_3$, $E_4$, and $E_5$) and pooling layers that extract high-level features from the input seismic data, while reducing its resolution. The expanding path is a series of decoders ($D_1$, $D_2$, $D_3$, and $D_4$) and upsampling layers that reconstruct the original resolution. Also, the architecture includes skip connections to connect corresponding layers in the contracting and expanding paths, which allow the network to reconstruct detailed structures that might be lost in the down-sampling process.  

To further improve UNet's performance, we utilize a modified residual network baseline (MRNB) to replace the conventional convolutional layers in the encoder and decoder. We present the structure of MRNB in Figure \ref{fig4}b. As we can see, MRNB consists of two residual blocks that combine Layer Normalization (LayerNorm), convolutional layers (1x1 and 3x3 conv), a simplified channel attention (SCA) module, and nonlinear activation functions LeakyReLU. In the first residual block, the input data are first normalized using Layer Normalization and then processed through a 1x1 convolutional layer to double the number of channels. Next, a 3x3 convolutional layer is applied to extract features from the input data, followed by a nonlinear activation function LeakyReLU, which introduces nonlinearity into the network. An SCA module is then applied to the output of the LeakyReLU layer to perform channel-wise feature recalibration. Finally, another 1x1 convolutional layer is applied to restore the number of channels to the original input. Each channel of the resulting output is multiplied by a corresponding coefficient, which is updated during the network's training process, and then added to the input data as the input to the second residual block. The second residual block is similar to the first, but without the 3x3 convolutional layer and SCA module. This is done to reduce the number of trainable network parameters while maintaining the depth in the network. Here, the encoders $E_1$, $E_2$, $E_3$, $E_4$, and $E_5$ include 2, 2, 4, 8, and 12 MRNBs, respectively, with the corresponding number of feature maps are 64, 128, 256, 512, and 1024, respectively. The decoders $D_1$, $D_2$, $D_3$, and $D_4$ all utilize 2 MRNBs corresponding to 64, 128, 256, and 512 feature maps, respectively. 

\subsection{Loss functions}
Loss functions are a crucial component of NN training as they provide a measure of how well the network is performing on a particular task. Selecting the appropriate loss function can help improve the model's accuracy, convergence speed, and generalization ability. Hence, for both the meta-training and meta-testing stages, we combine the mean square error (MSE) and multiscale structure similarity index measure (MS-SSIM) to optimized the NN training. In which, the MSE loss is a common metric for evaluating the regression models' performance and can be expressed as: 
\begin{equation}\label{eq4}
\setlength{\abovedisplayskip}{3pt}
\setlength{\belowdisplayskip}{3pt}
\begin{gathered}
\mathcal{L}_{\rm{MSE}}\left (L,O \right)=\frac{1}{N}\displaystyle \sum^{N}_{i=1}{\left|L_{i}-O_{i} \right|^2},
\end{gathered}
\end{equation}
where $L$ and $O$ represent the label and the output of the network, respectively, and $N$ is the total number of samples. 

MS-SSIM is a sophisticated metric that assesses the degree of structural similarity between the prediction and label by considering its multiscale nature from a visual perspective \cite{du2022deep, geng2022deep}. The calculation process of MS-SSIM involves multiple steps, beginning with the division of the input data into non-overlapping patches at various scales using a Gaussian filter. Subsequently, the local luminance $l(\cdot)$, contrast $c(\cdot)$, and structural information $s(\cdot)$ at different scales are computed using the following equation
\begin{equation}\label{eq5}
\setlength{\abovedisplayskip}{5pt}
\setlength{\belowdisplayskip}{5pt}
\left\{\begin{array}{c}
l(L, O)=\frac{2 \mu_L \mu_O+C_1}{\mu_L^2+\mu_O^2+C_1} 
\\[8pt]
c(L, O)=\frac{2 \sigma_L \sigma_O+C_2}{\sigma_L^2+\sigma_O^2+C_2} 
\\[8pt]
\setlength{\belowdisplayskip}{5pt}
s(L, O)=\frac{\sigma_{L O}+C_3}{\sigma_L \sigma_O+C_3}
\end{array}\right.,
\end{equation}
where $\mu_L$ and $\mu_O$ denote the mean values of the local patch corresponding to the label and output, respectively, $\sigma_L$ and $\sigma_O$ are the variance of the local patch, and $\sigma_{L O}$ represents the covariance of local patch. $C_1$, $C_2$, and $C_3$ are small constants to stabilize the division, where $C_2=C_3$. The MS-SSIM loss is then calculated by comparing the local luminance, contrast, and structural information at different scales, employing a weighted average exponents $\alpha_M$, $\beta_j$, and $\gamma_j$ across scales as follows
\begin{equation}\label{eq6}
\setlength{\abovedisplayskip}{5pt}
\setlength{\belowdisplayskip}{5pt}
\begin{split}
\mathcal{L}_{\mathrm{MS\mbox{-}SSIM}}\left (L,O \right) = 1 -
\left [l_{M} \left (L,O \right) \right]^{\alpha_{M}} \cdot \displaystyle \prod^{M}_{j=1} \left [c_{M} \left (L,O \right) \right]^{\beta_{j}} \left [s_{M} \left (L,O \right) \right]^{\gamma_{j}}.
\end{split}
\end{equation}
Here, referring to Wang et al. \cite{wang2003multiscale}, we set five scales, and the exponents are $\beta_1=\gamma_1=0.0448$, $\beta_2=\gamma_2=0.2856$, $\beta_3=\gamma_3=0.3001$, $\beta_4=\gamma_4=0.2363$, and $\alpha_5=\beta_5=\gamma_5=0.1333$. 

Utilizing MSE and MS-SSIM losses, the total loss function is defined as
\begin{equation}\label{eq7}
\setlength{\abovedisplayskip}{3pt}
\setlength{\belowdisplayskip}{3pt}
\begin{gathered}
\mathcal{L}=c \cdot (\epsilon_1 \cdot \mathcal{L}_{\rm{MSE}}+\epsilon_2 \cdot \mathcal{L}_{\mathrm{MS\mbox{-}SSIM}}).
\end{gathered}
\end{equation}
where the hyperparameters $\epsilon_1$ and $\epsilon_2$ are used to balance the two losses. Here, for simplification, both $\epsilon_1$ and $\epsilon_2$ are set to 1. $c$ represents a scaling factor that is employed to adjust the magnitude of the loss value. During the meta-training stage, large loss values can trigger instability issues in the training process, so we, in this paper, set $c=0.1$ as a measure to prevent optimization failures. However, in the meta-testing stage, the network initialization provided by the meta-training stage is already sufficiently robust, and hence $c$ is set to 1.  \\ 
\section{Numerical examples}\label{sec3}
In our work, the meta-training stage is executed for 40000 epochs utilizing the AdamW optimizer, where the initial values of the meta-learning rate and inner-loop learning rate are set to 1e-3 and 5e-3, respectively. The meta-learning rate is gradually reduced by a factor of 0.8 every 2000 epochs, while the inner-loop learning rate undergoes a similar reduction for the first 20000 epochs, remaining constant thereafter. During the meta-testing stage, the optimized network parameters from the meta-training stage are exposed to individual fine-tuning for each SPT, with a total of 300 epochs executed and an initial learning rate of 1e-4. Every fine-tuning process consists of a total of 300 epochs with an initial learning rate of 1e-4. We need to emphasize, however, that such time-intensive training is definitely unnecessary to the task initialized from Meta-Processing; rather, we do it for the sole purpose of facilitating better comparisons with randomly initialized network. 

We now present the results from our approach on both synthetic and field data. To validate the effectiveness of the Meta-Processing algorithm, we compare its prediction results with those of randomly initialized networks. Our assessment begins with a thorough evaluation of the performance of our approach to synthetic data. Subsequently, we present the results of further testing conducted on field data.

\subsection{Synthetic data}

\subsubsection{Denoising}
\begin{figure}[htp]
\centering
\includegraphics[width=0.35\textwidth]{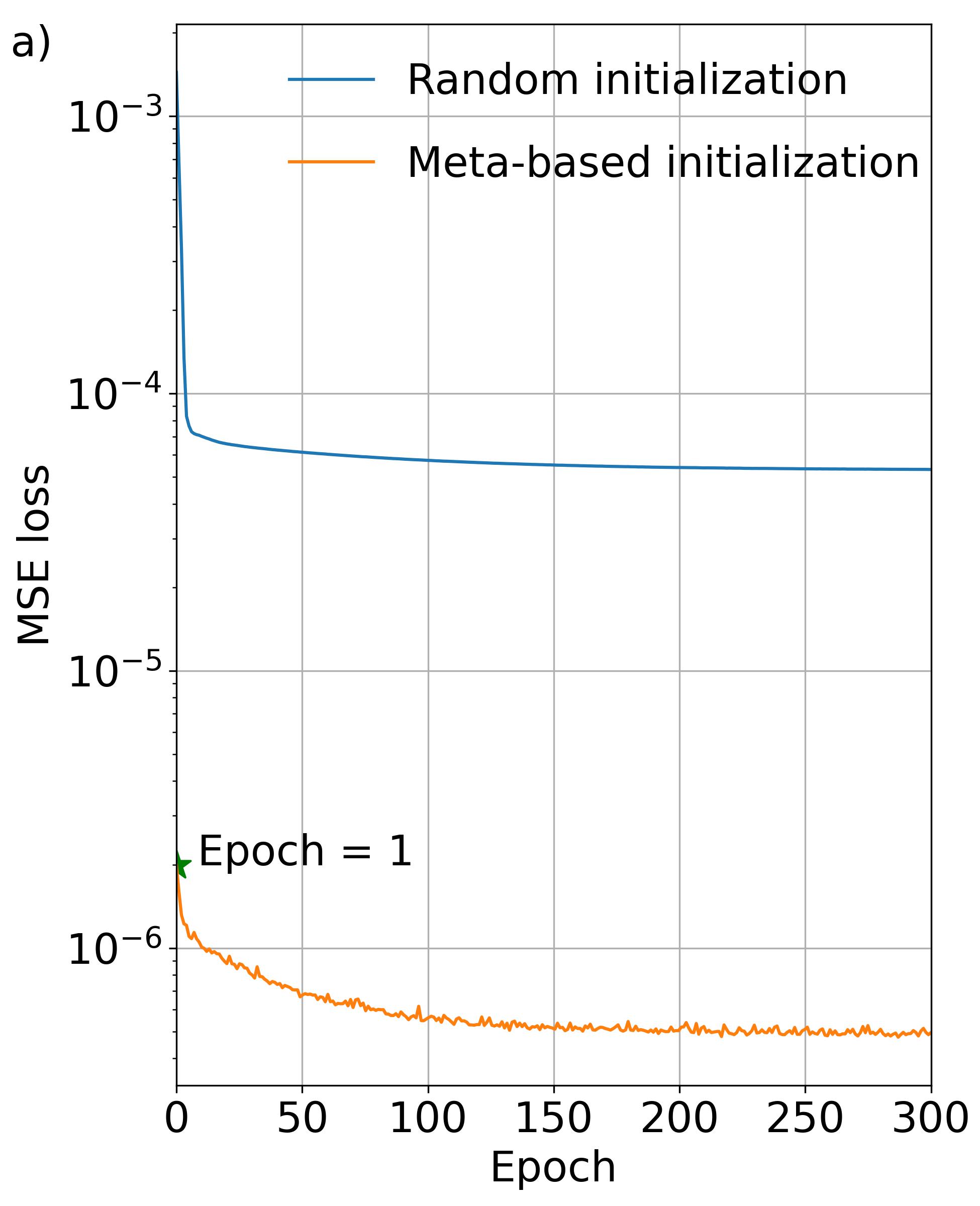}
\hspace{1cm}
\includegraphics[width=0.35\textwidth]{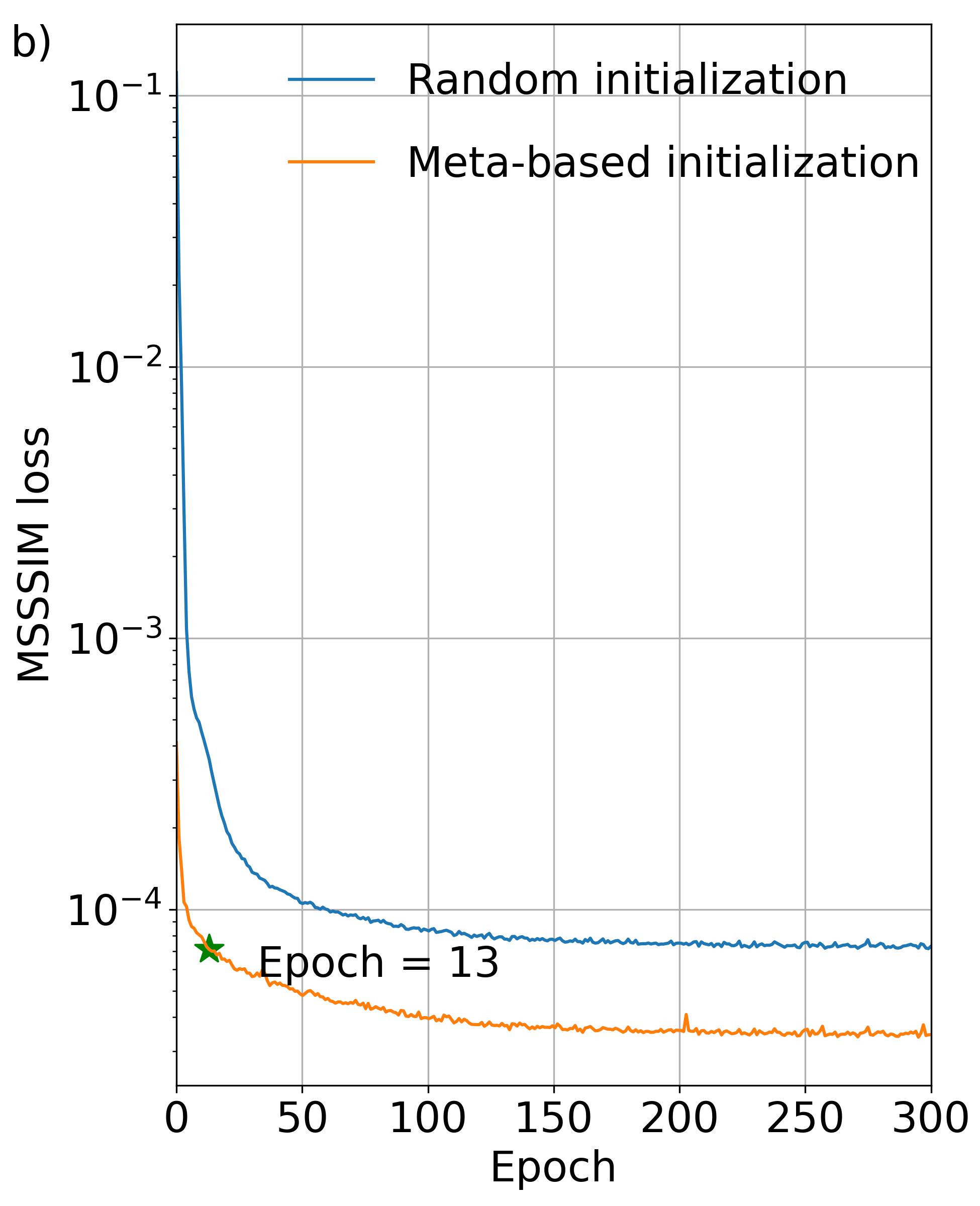}
\caption{The MSE (a) and MSSSIM (b) loss function curves of the neural networks training with meta-learning initialization and random initialization of seismic denoising task. }
\label{fig5}
\end{figure}

In the first example, we focus on removing random noise, which is the most common type of noise. To compare the convergence speed and accuracy of the meta-learning initialization-based network (MLIN) and the randomly initialized network (RIN) during the fine-tuning stage, MSE and MSSSIM losses are utilized as evaluation metrics. The metrics are plotted in Figures \ref{fig5}a and \ref{fig5}b, where the epochs marked with a star indicate the number of epochs of training required for the MLIN to achieve the same metric as the RIN. Same notation will be used later. We can see that, the MLIN achieves significantly smaller MSE loss than the RIN after only one epoch of gradient descent updates, which is far less than the 300 epochs required by the RIN. Moreover, from the perspective of the MSSSIM loss, MLIN surpasses the performance of the RIN after only 13 epochs of optimization. These demonstrate that the MLIN outperformed the RIN in terms of both convergence speed and accuracy. 

\begin{figure}[htp]
\centering
\includegraphics[width=0.3\textwidth]{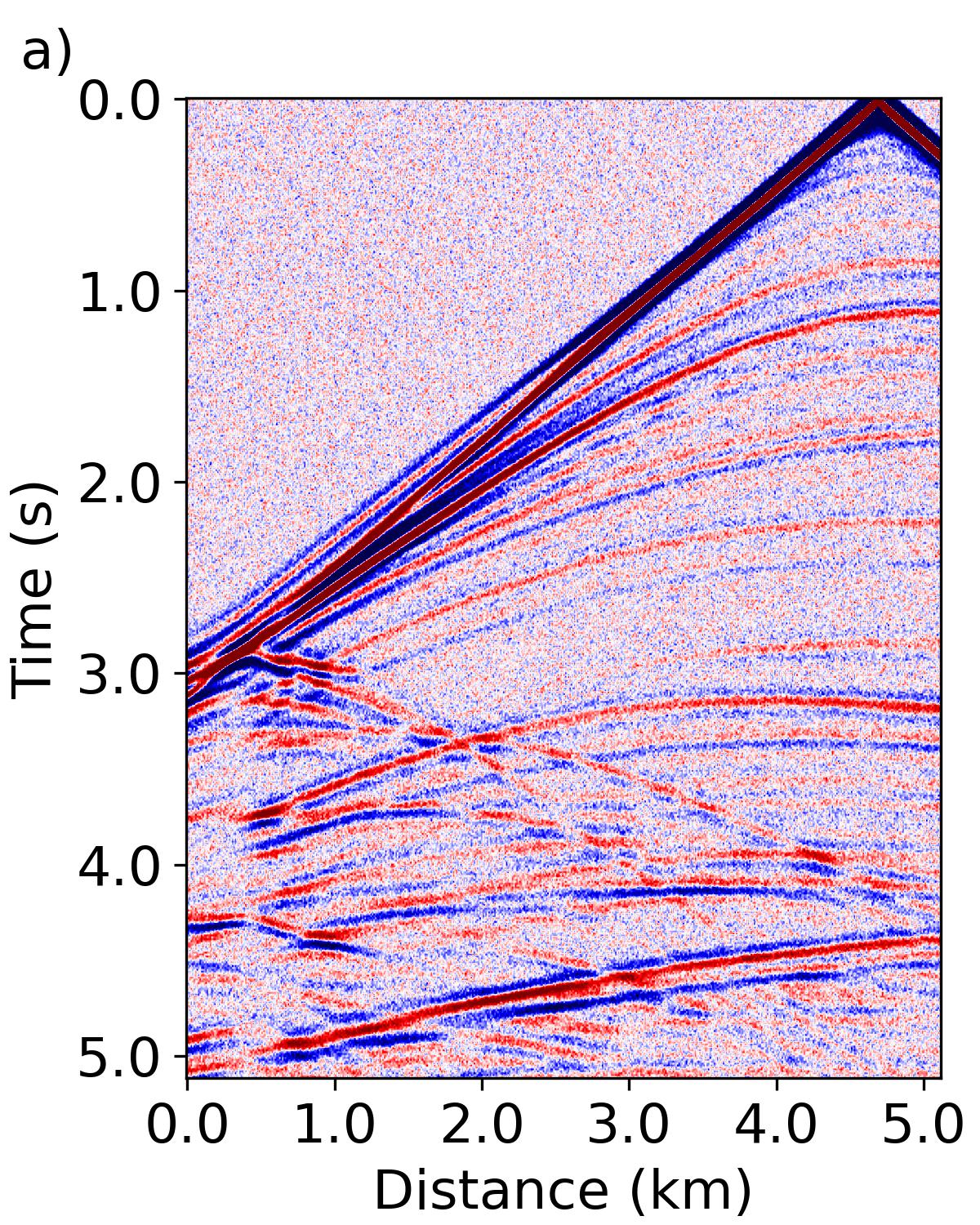}
\hspace{1cm}
\includegraphics[width=0.3\textwidth]{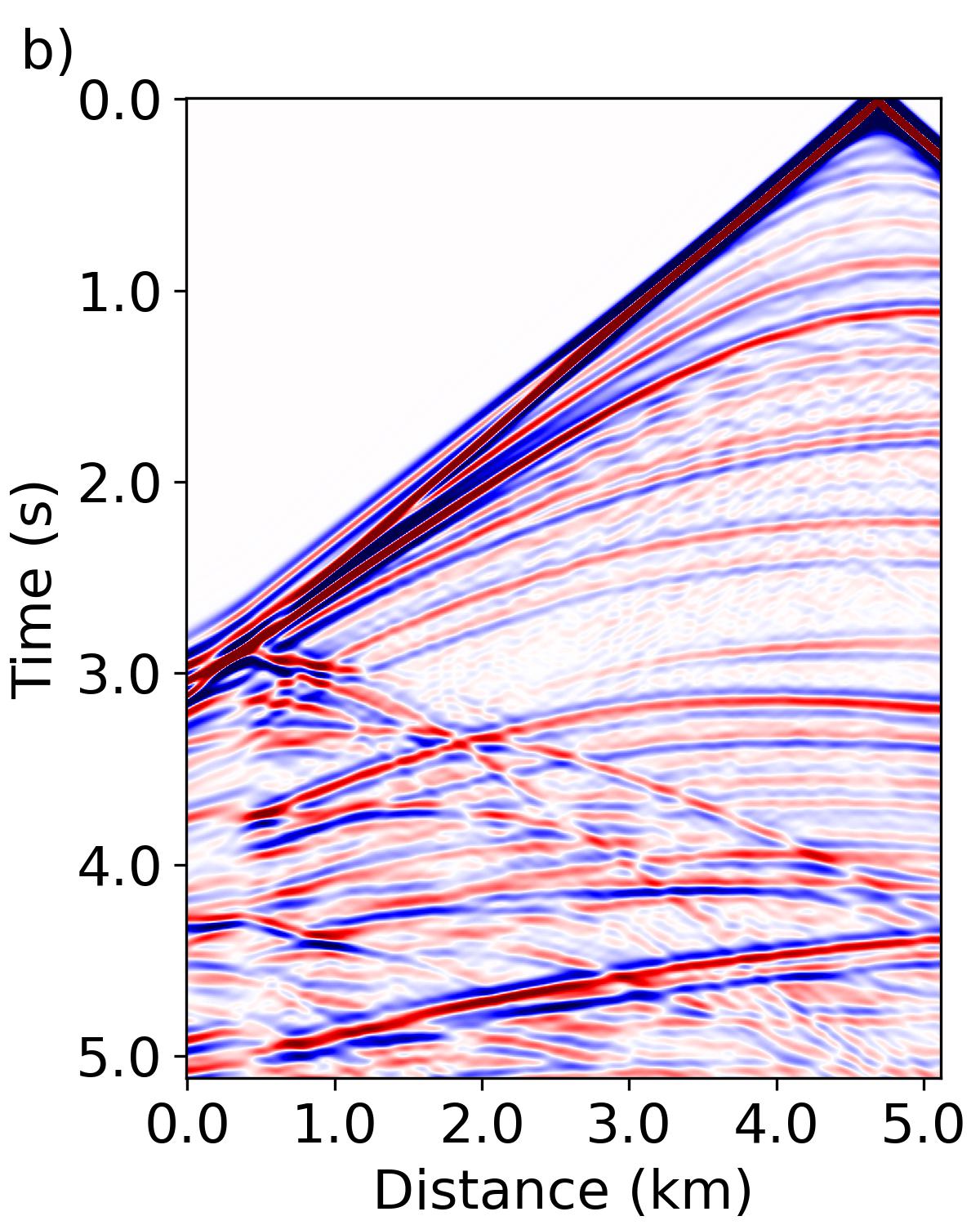}
\caption{The noisy (a) and clean (b) data of the synthetic test dataset. }
\label{fig6}
\end{figure}

\begin{figure}[htp]
\centering
\includegraphics[width=0.25\textwidth]{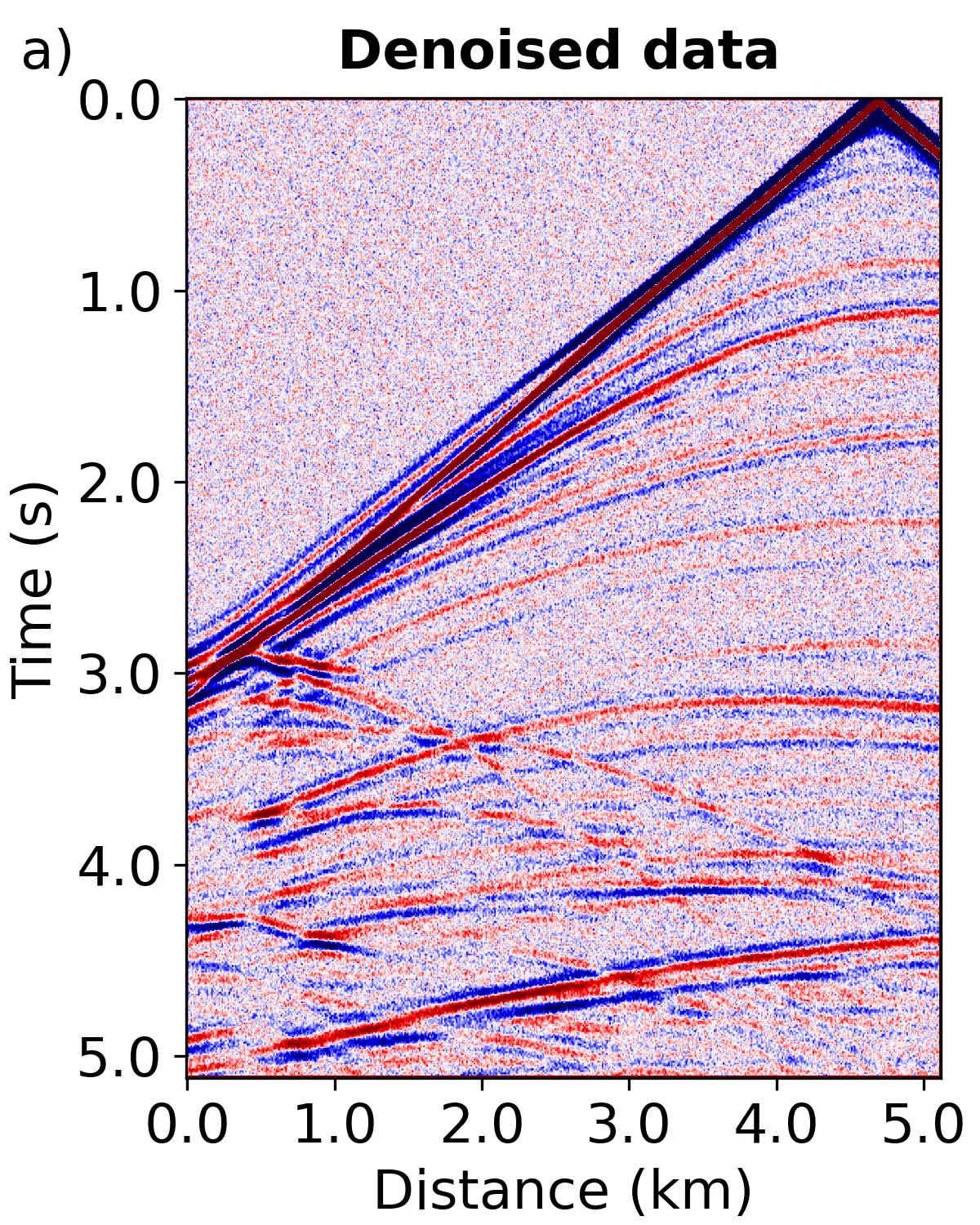}
\hspace{0.3cm}
\includegraphics[width=0.25\textwidth]{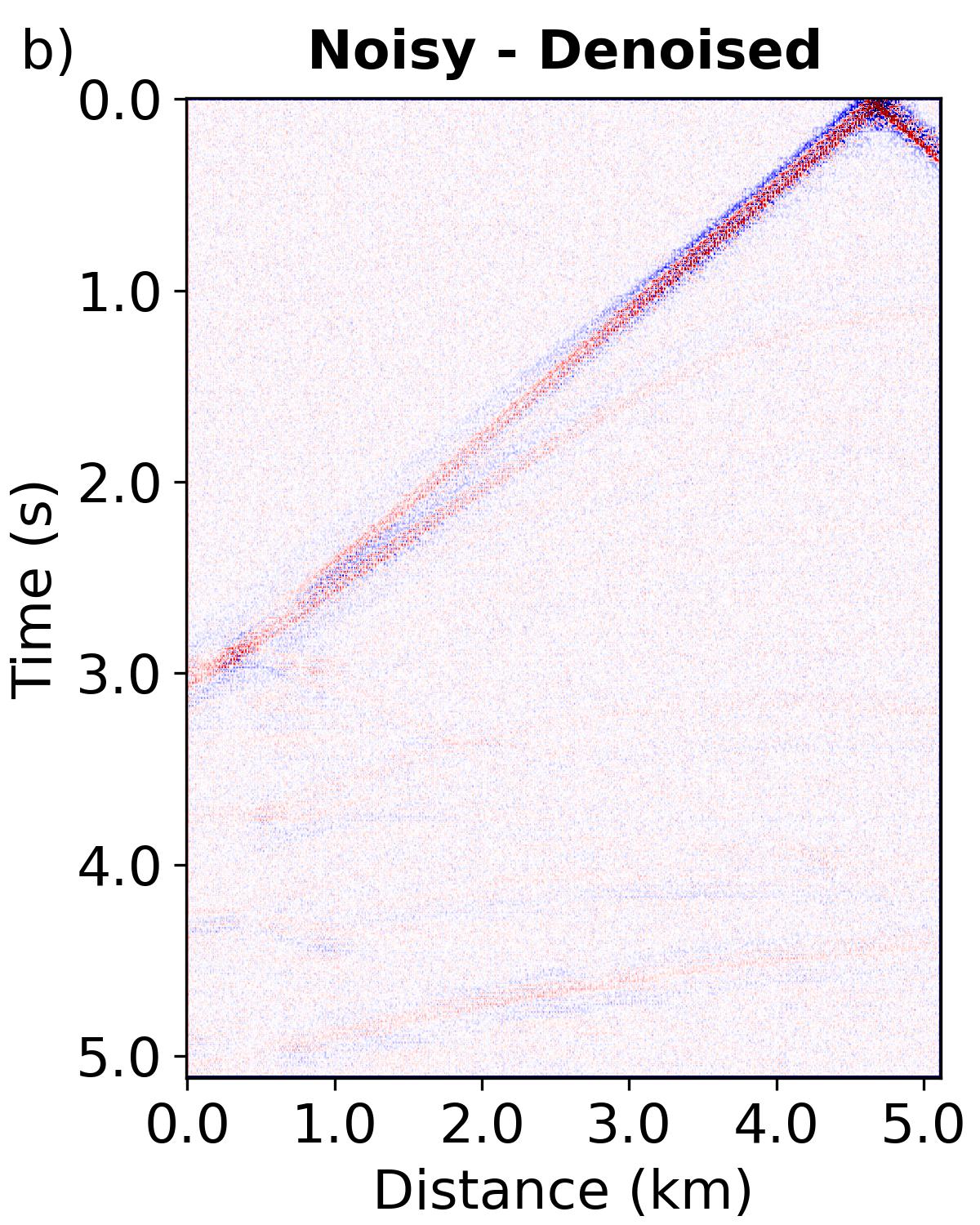}
\hspace{0.3cm}
\includegraphics[width=0.25\textwidth]{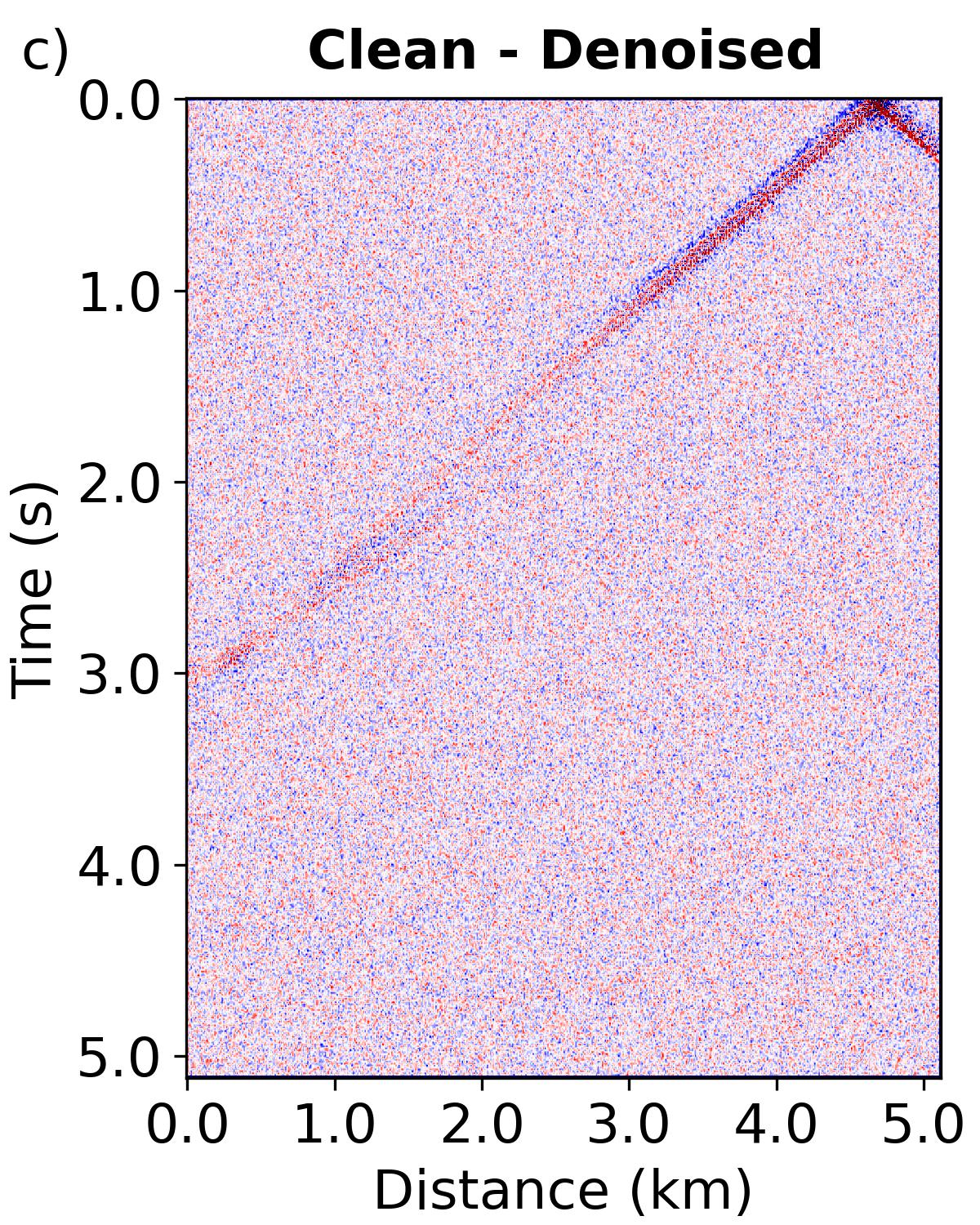}
\includegraphics[width=0.25\textwidth]{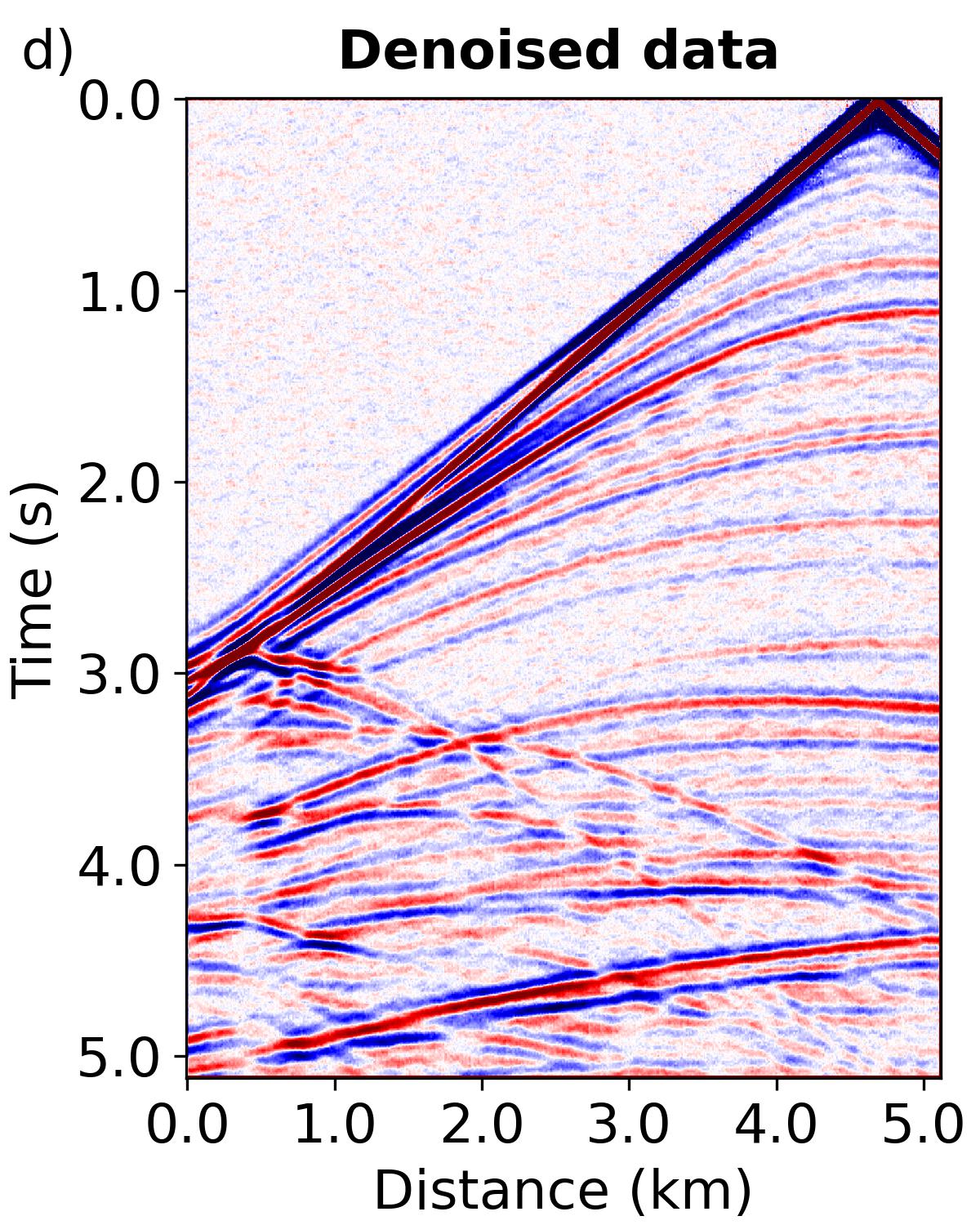}
\hspace{0.3cm}
\includegraphics[width=0.25\textwidth]{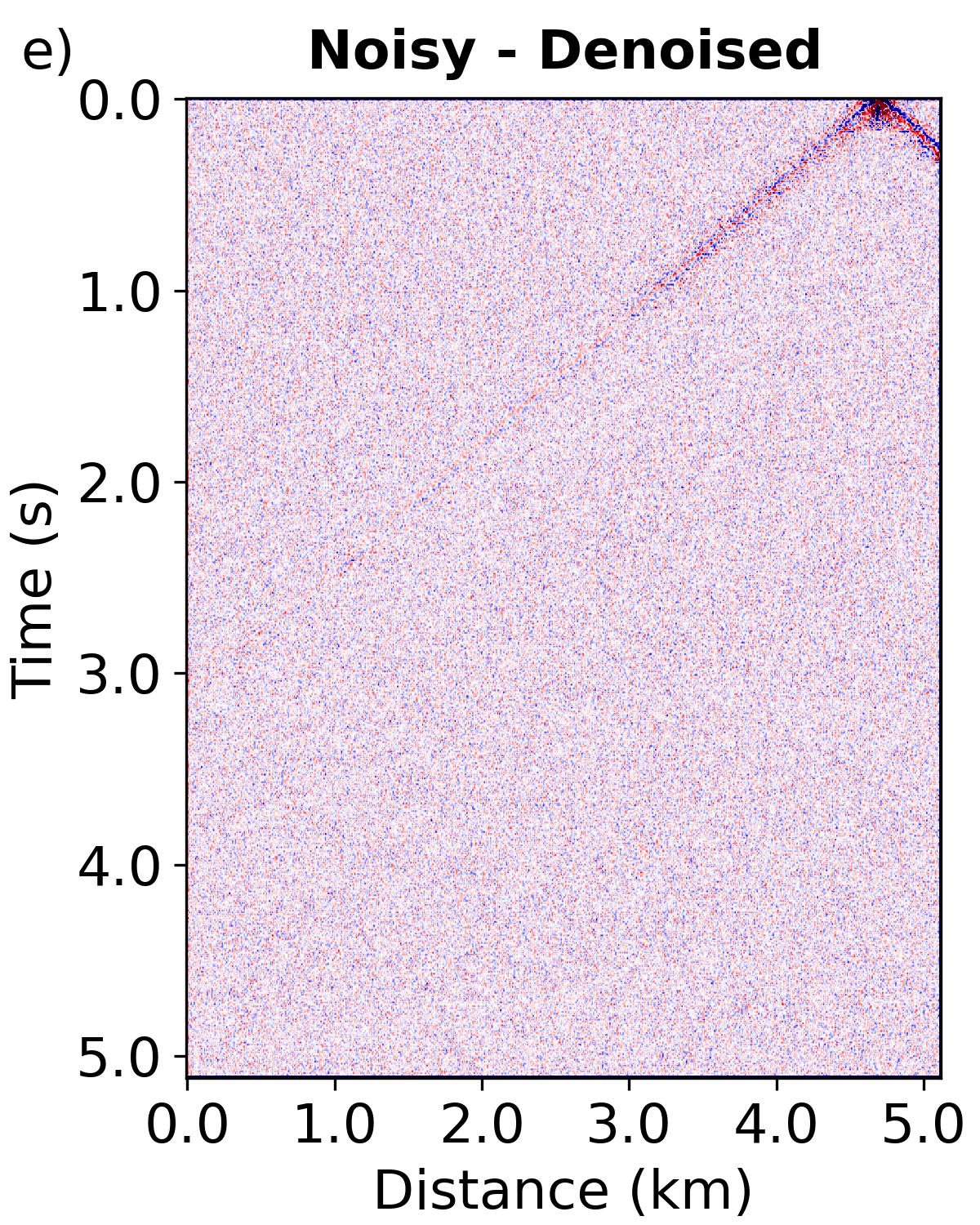}
\hspace{0.3cm}
\includegraphics[width=0.25\textwidth]{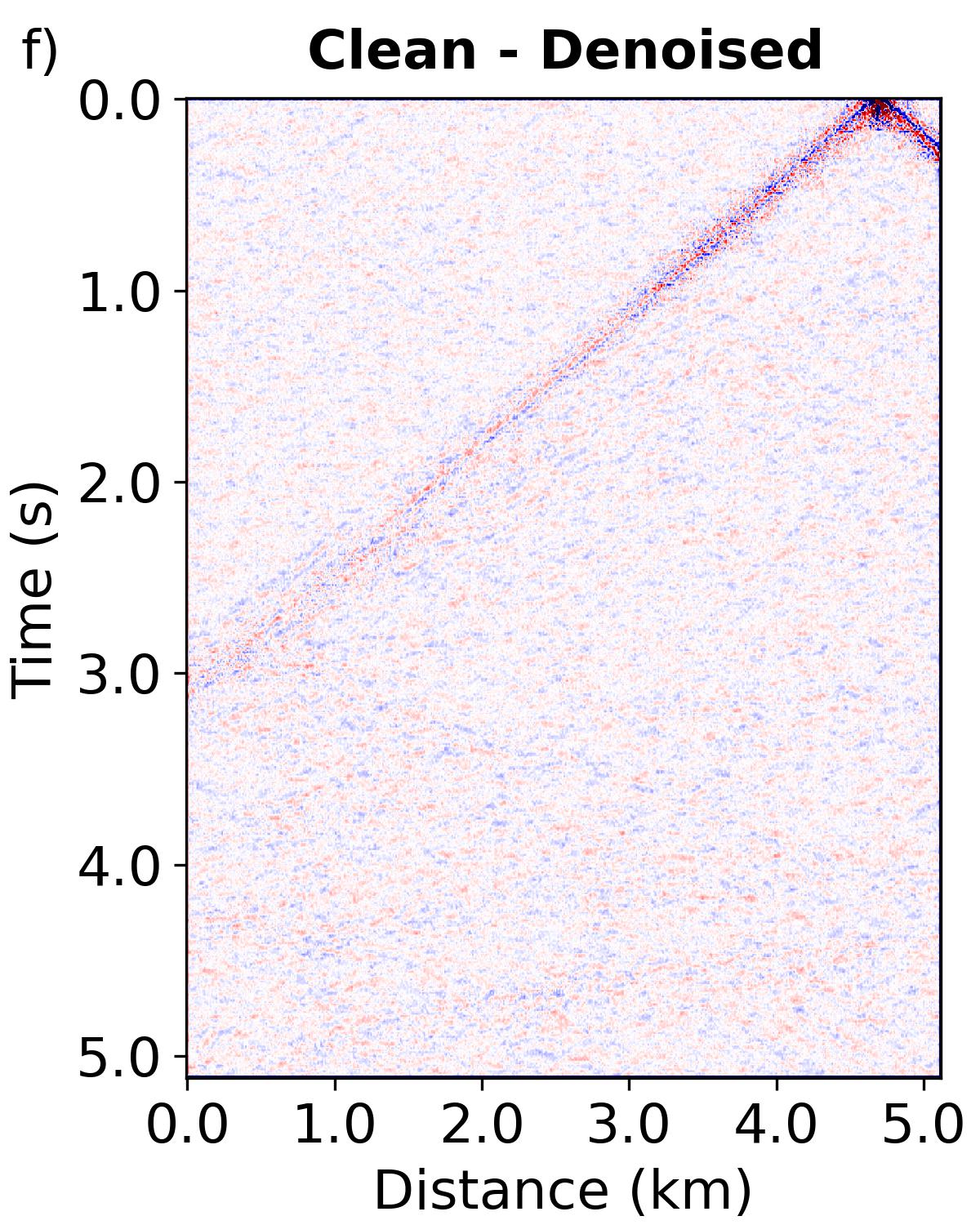}
\includegraphics[width=0.25\textwidth]{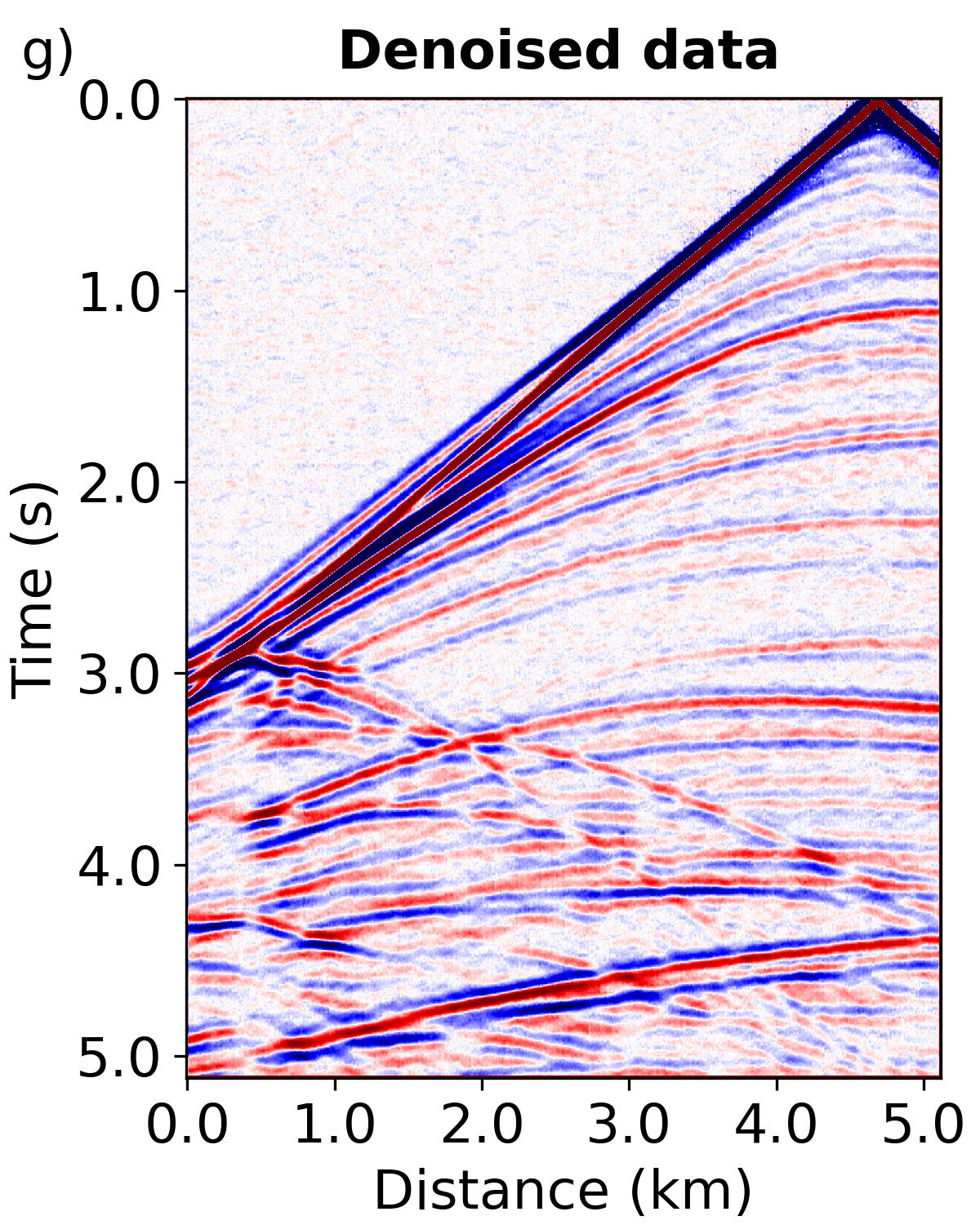}
\hspace{0.3cm}
\includegraphics[width=0.25\textwidth]{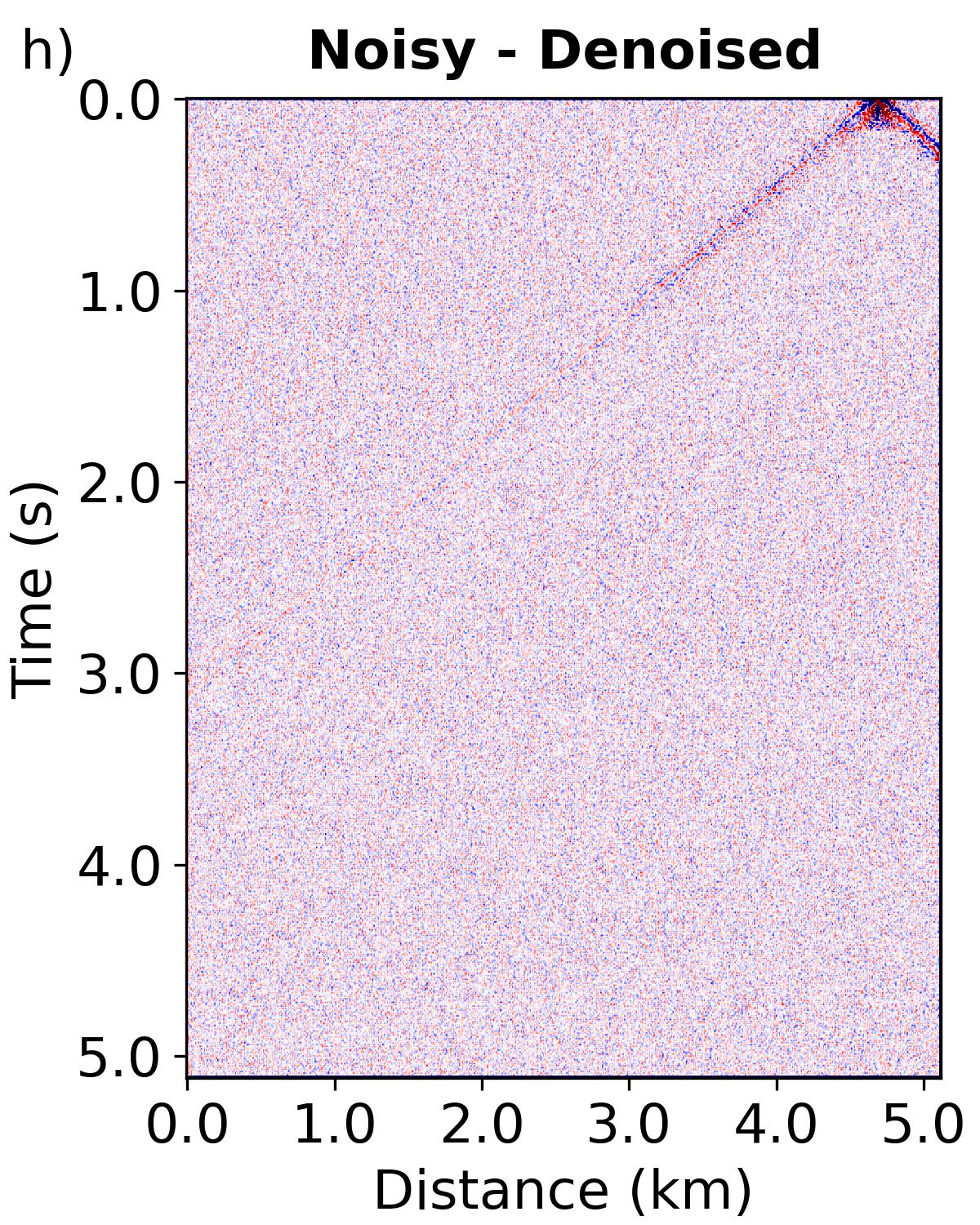}
\hspace{0.3cm}
\includegraphics[width=0.25\textwidth]{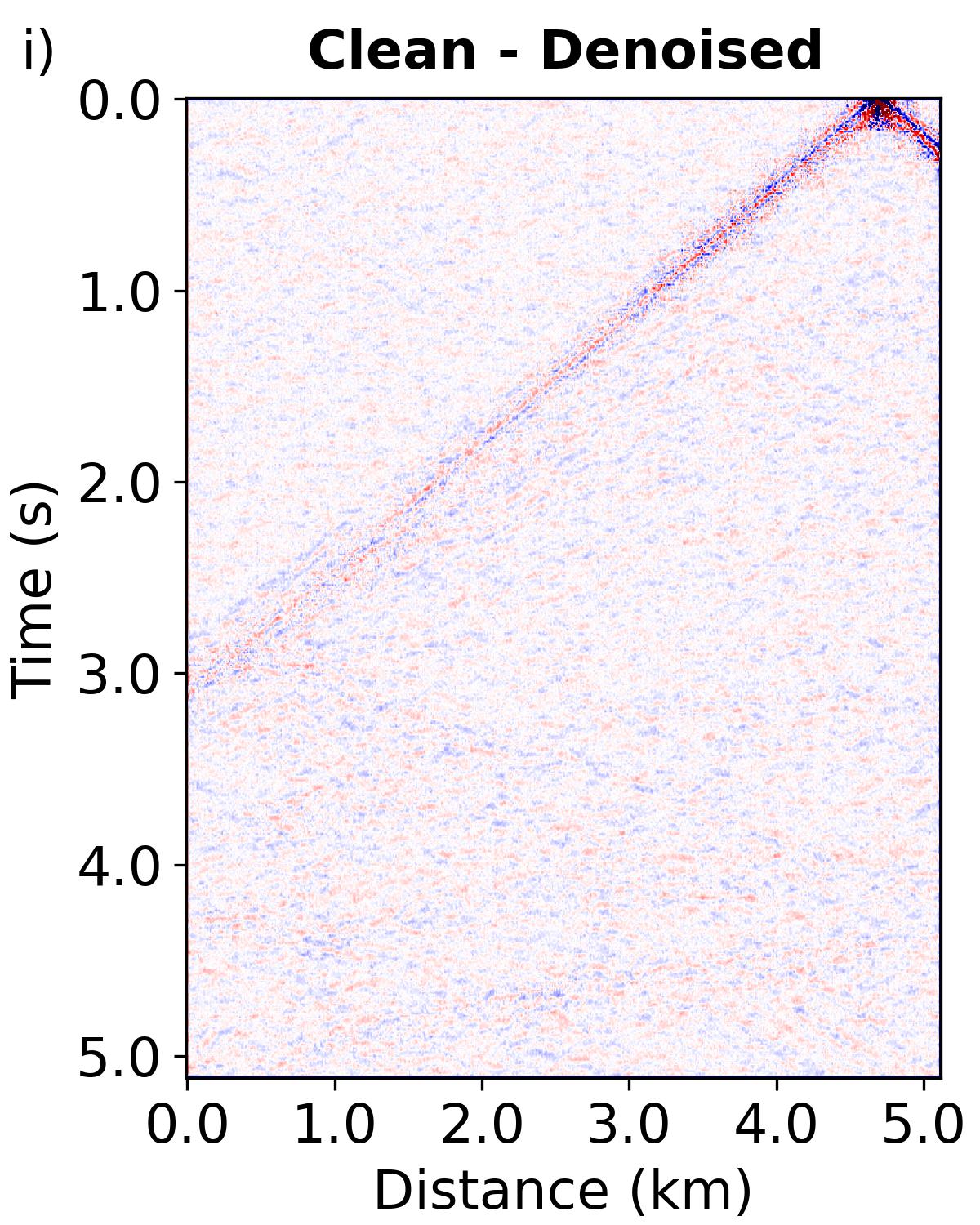}
\includegraphics[width=0.25\textwidth]{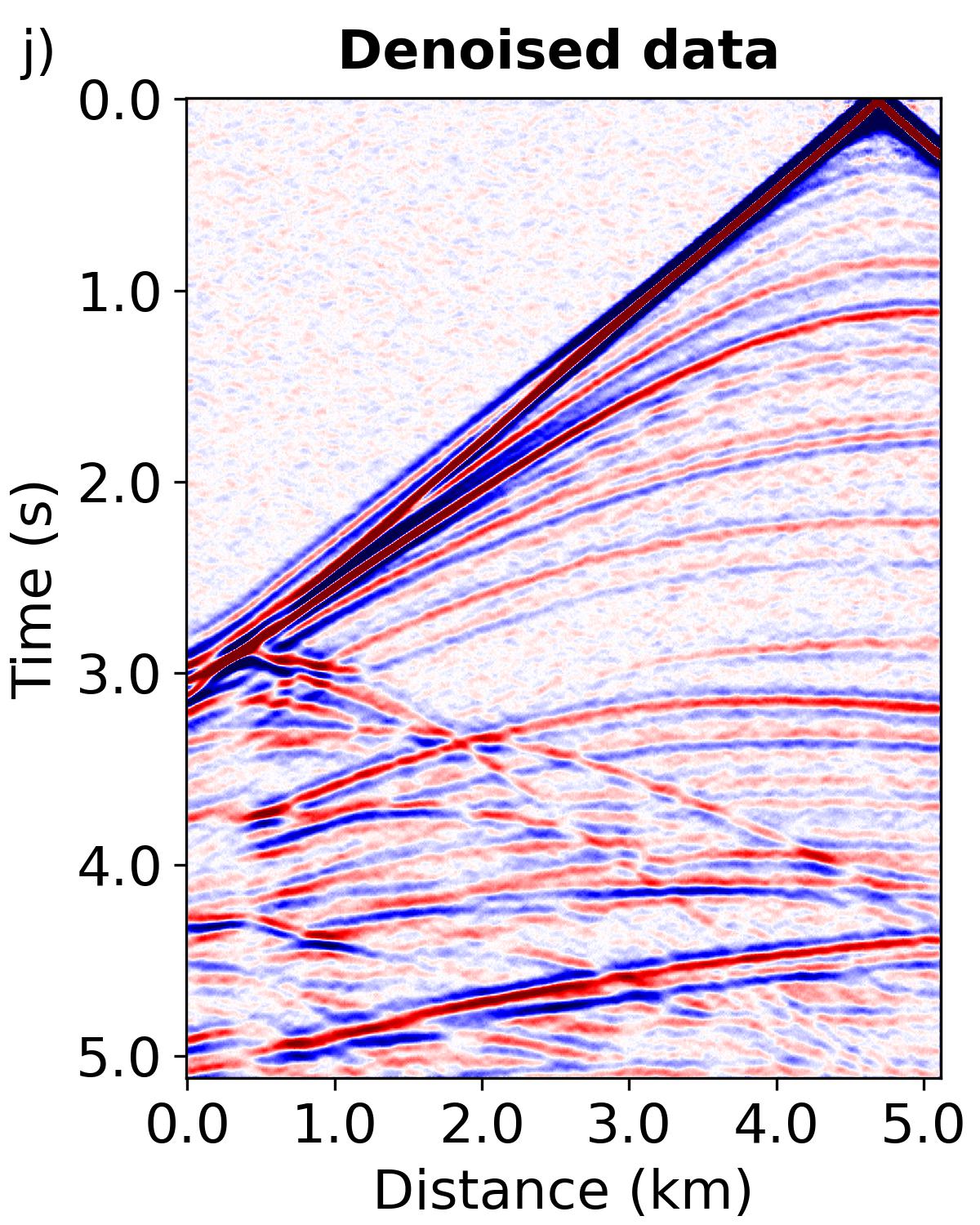}
\hspace{0.3cm}
\includegraphics[width=0.25\textwidth]{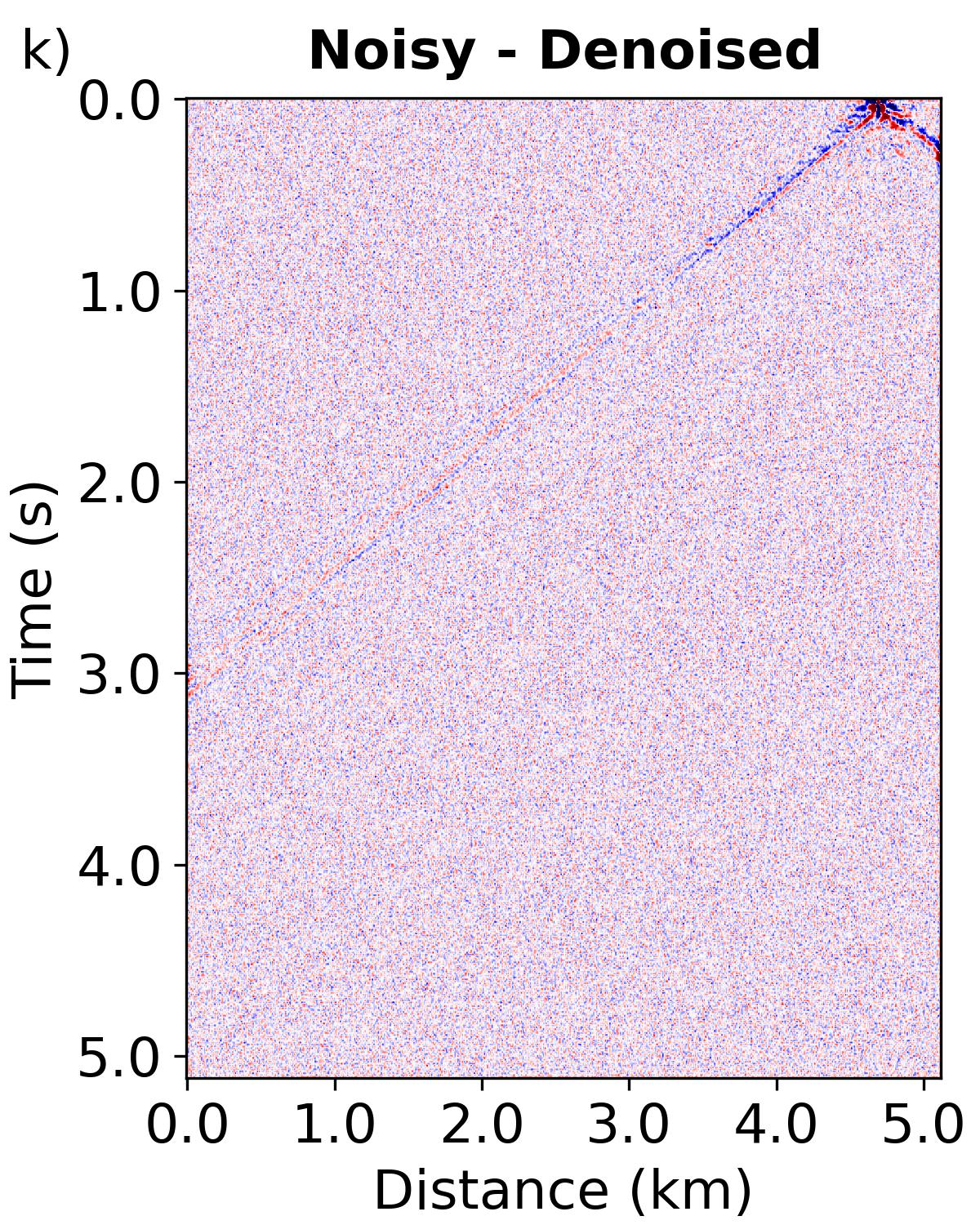}
\hspace{0.3cm}
\includegraphics[width=0.25\textwidth]{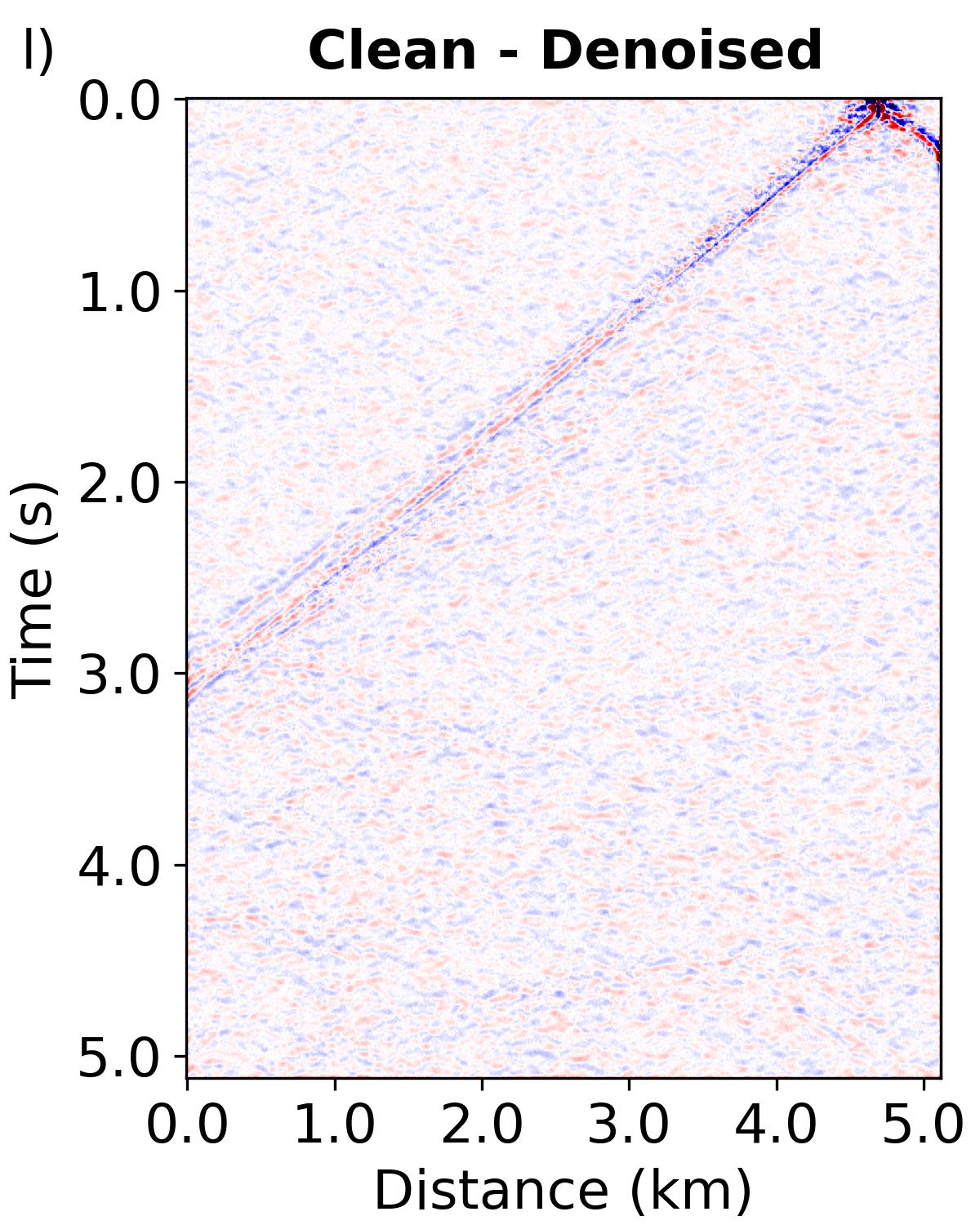}
\caption{Denoising comparisons of neural networks training with meta-learning initialization and random initialization on synthetic data. The first, second, and third rows correspond to random initialization-base neural network with 10, 100, and 300 epochs of training, respectively, and the fourth row corresponds to meta initialization-based neural network with 10 epochs of training.}
\label{fig7}
\end{figure}

The prediction results of MLIN and RIN for unseen test data are shown in Figure \ref{fig7}, with their corresponding input and label data depicted in Figure \ref{fig6}. In Figure \ref{fig7}, the first, second, and third rows correspond to RIN with 10, 100, and 300 epochs training, respectively, while the fourth row corresponds to MLIN with 10 epoch training. As we can see, the proposed Meta-Processing algorithm leads to an ideal denoising performance after only 10 epochs of optimization, achieving an MSE of 1.98e-6 and an MSSSIM of 9.39e-5. In contrast, the RIN with 10 epochs of optimization shows almost no denoising ability, which is attributed to the lengthy optimization process required by RIN. Even with 300 epochs of training, the denoising results of RIN only reach an MSE of 2.93e-5 and an MSSSIM of 8.38e-5. 

\subsubsection{Interpolation}
\begin{figure}[htp]
\centering
\includegraphics[width=0.35\textwidth]{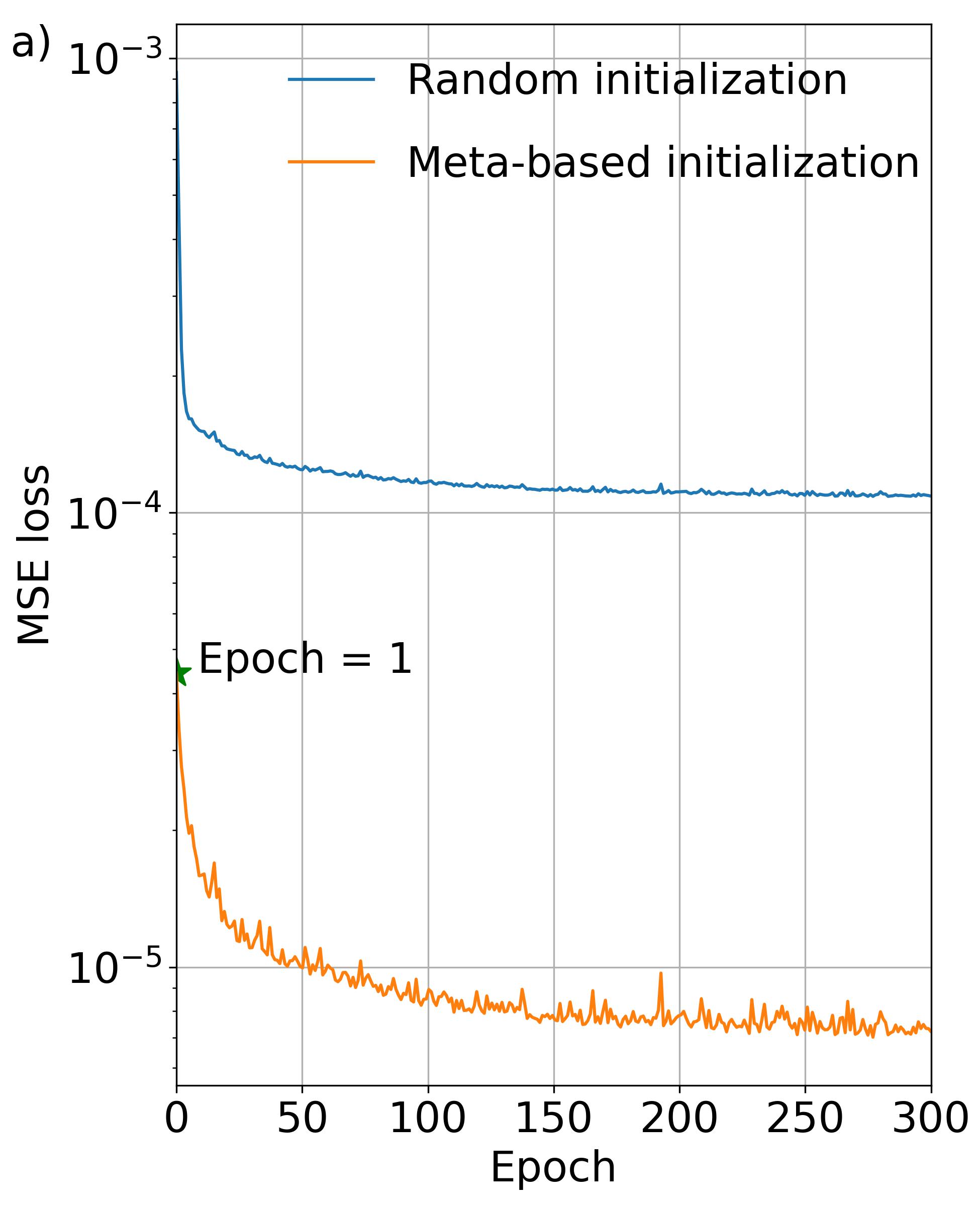}
\hspace{1cm}
\includegraphics[width=0.35\textwidth]{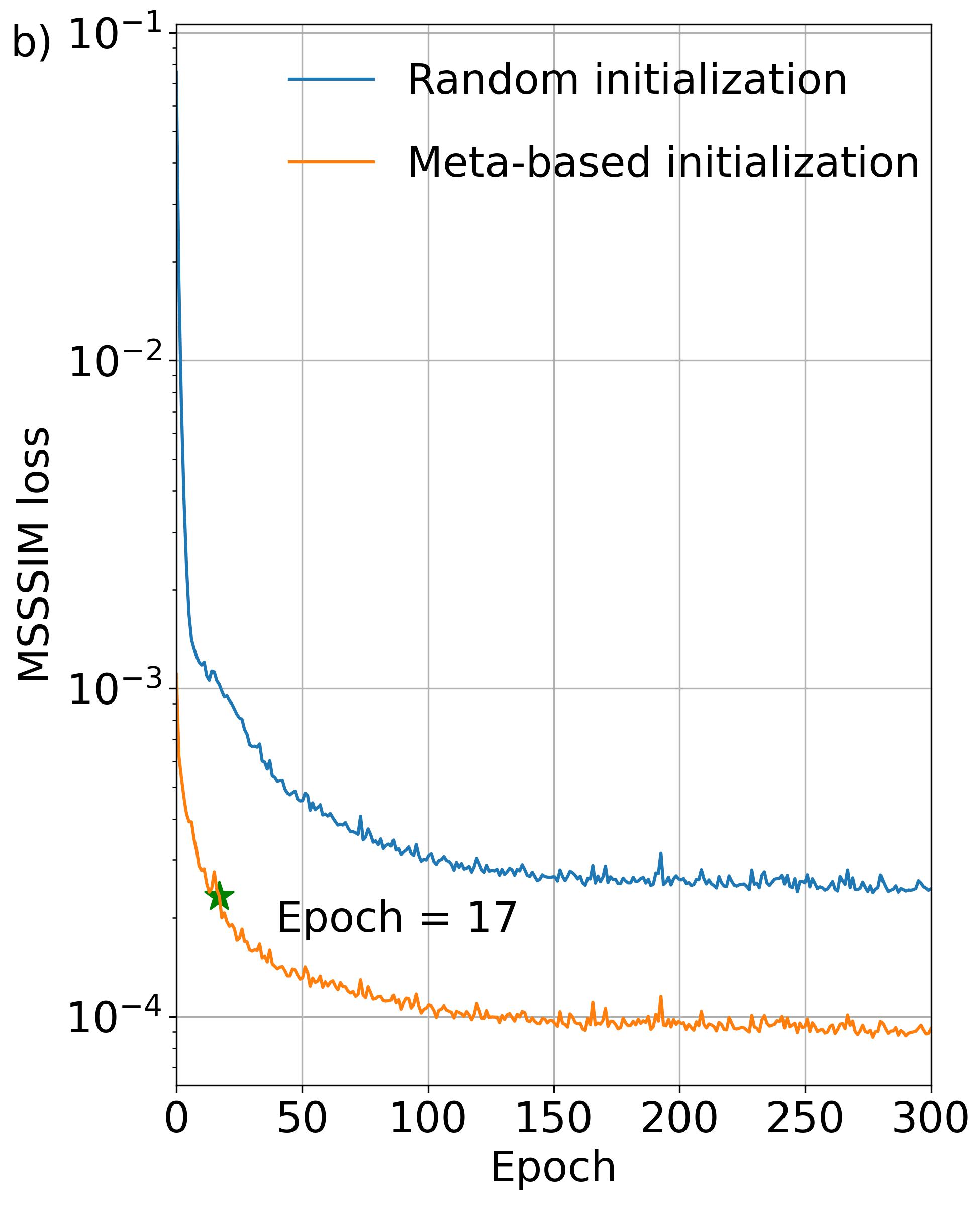}
\caption{The MSE (a) and MSSSIM (b) loss function curves of neural networks training with meta-learning initialization and random initialization of seismic interpolation task. }
\label{fig8}
\end{figure}

\begin{figure}[htp]
\centering
\includegraphics[width=0.25\textwidth]{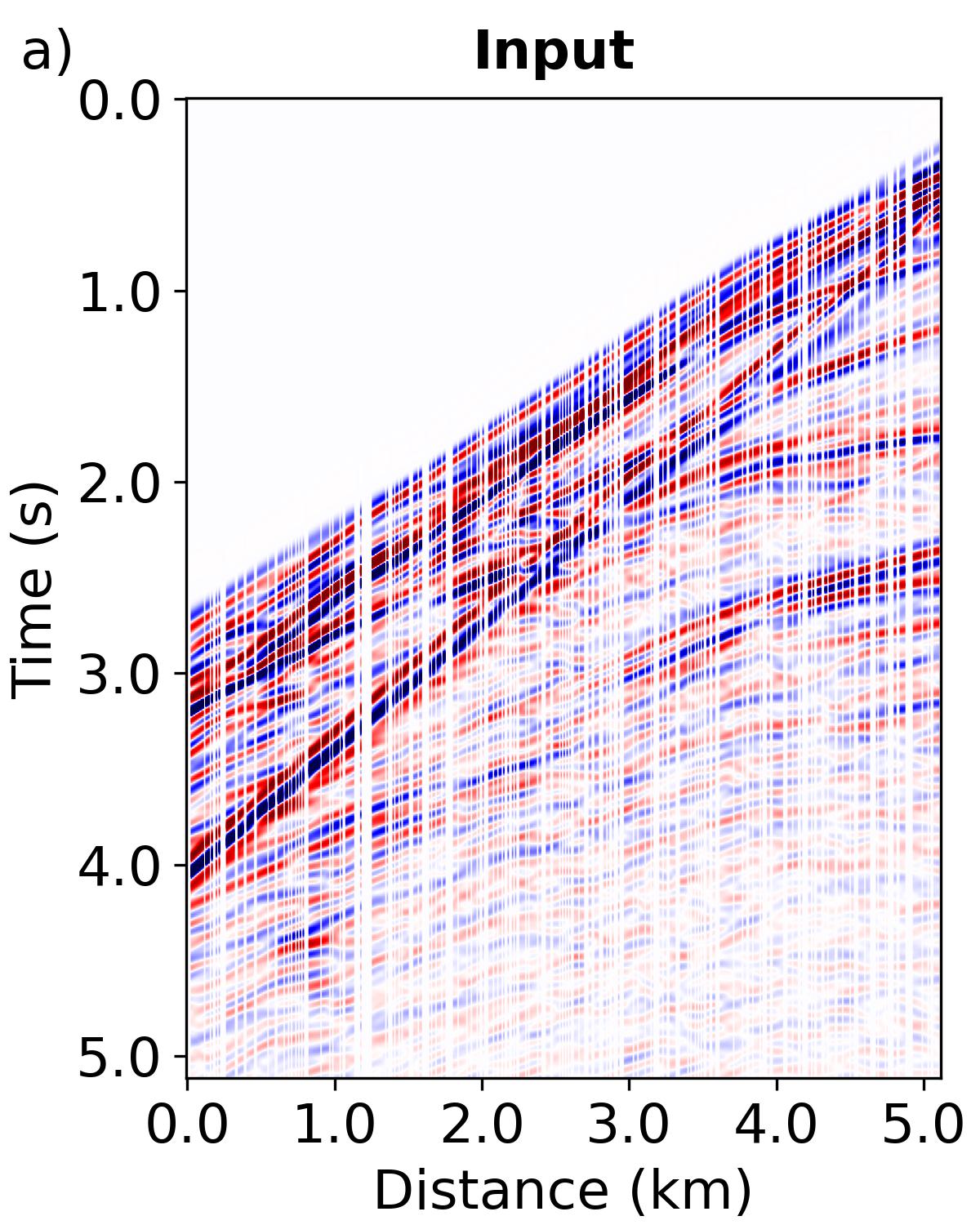}
\hspace{0.3cm}
\includegraphics[width=0.25\textwidth]{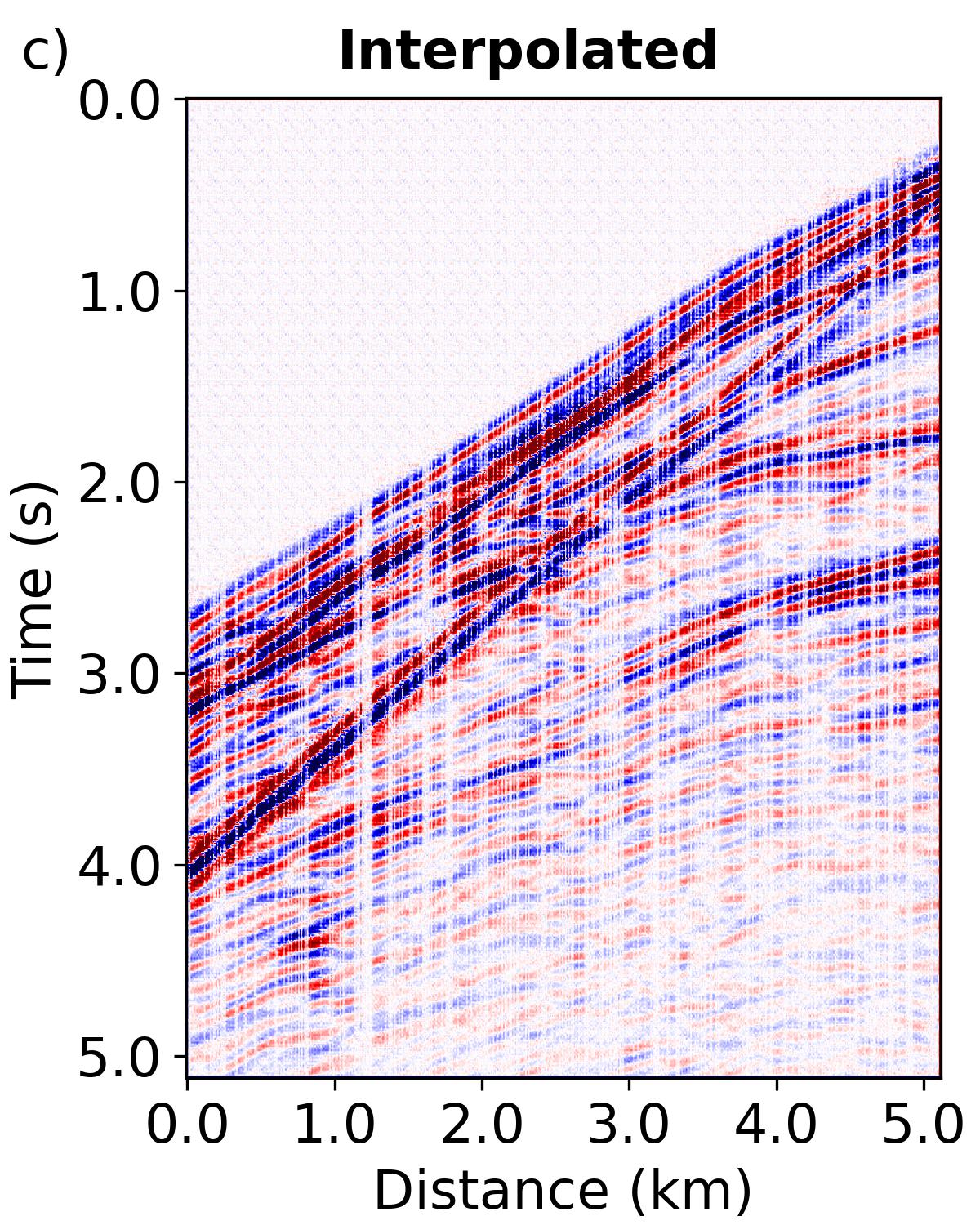}
\hspace{0.3cm}
\includegraphics[width=0.25\textwidth]{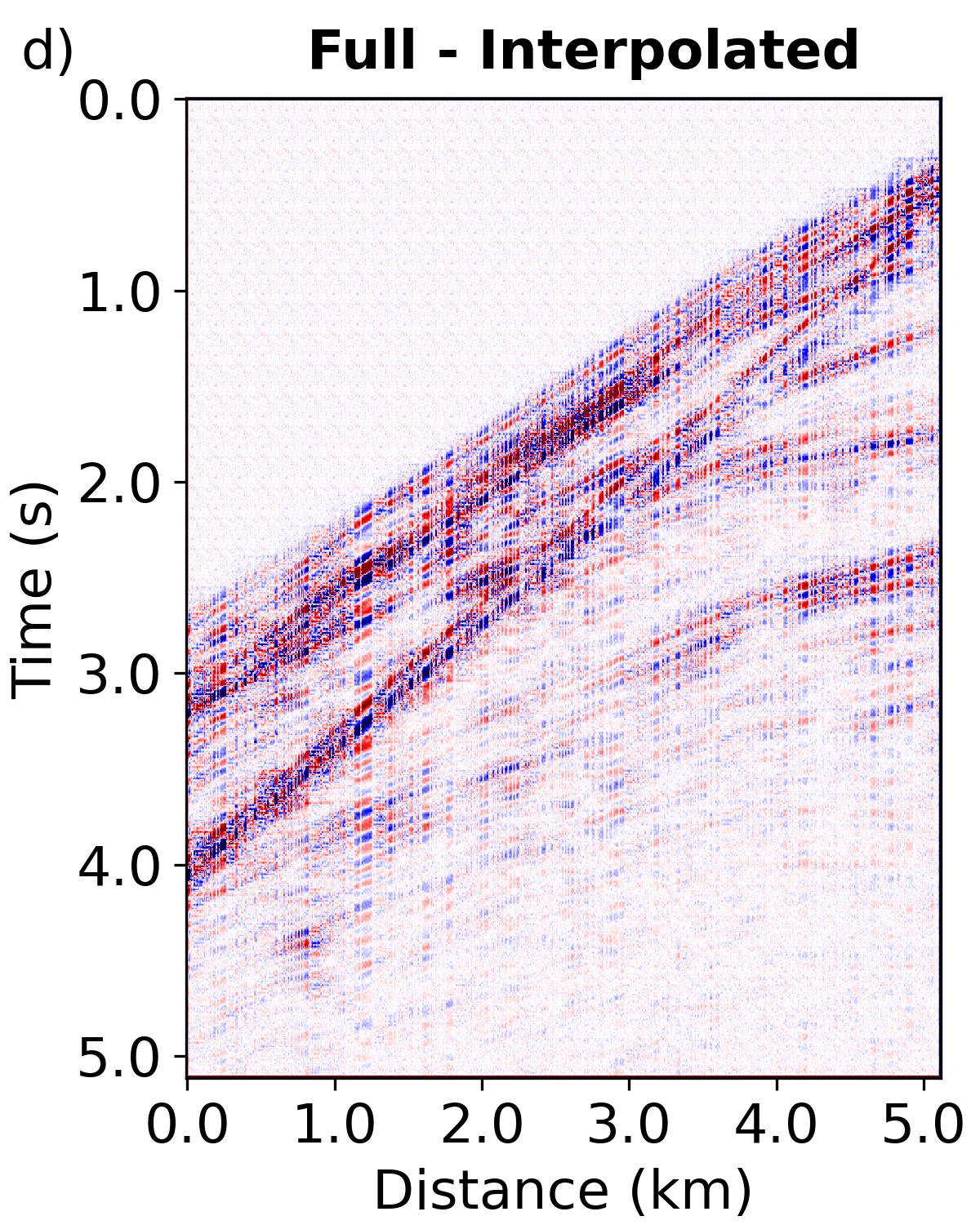}
\includegraphics[width=0.25\textwidth]{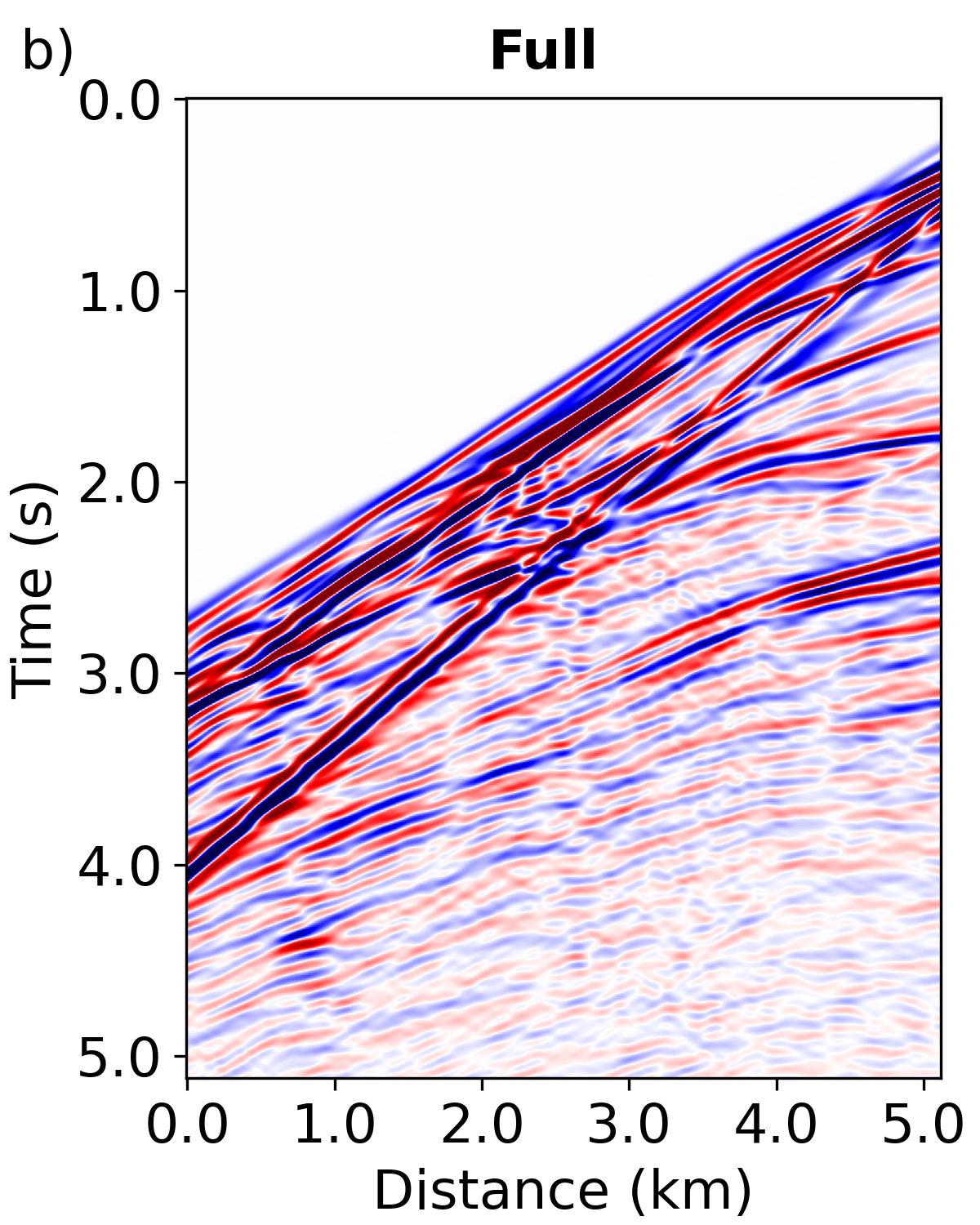}
\hspace{0.3cm}
\includegraphics[width=0.25\textwidth]{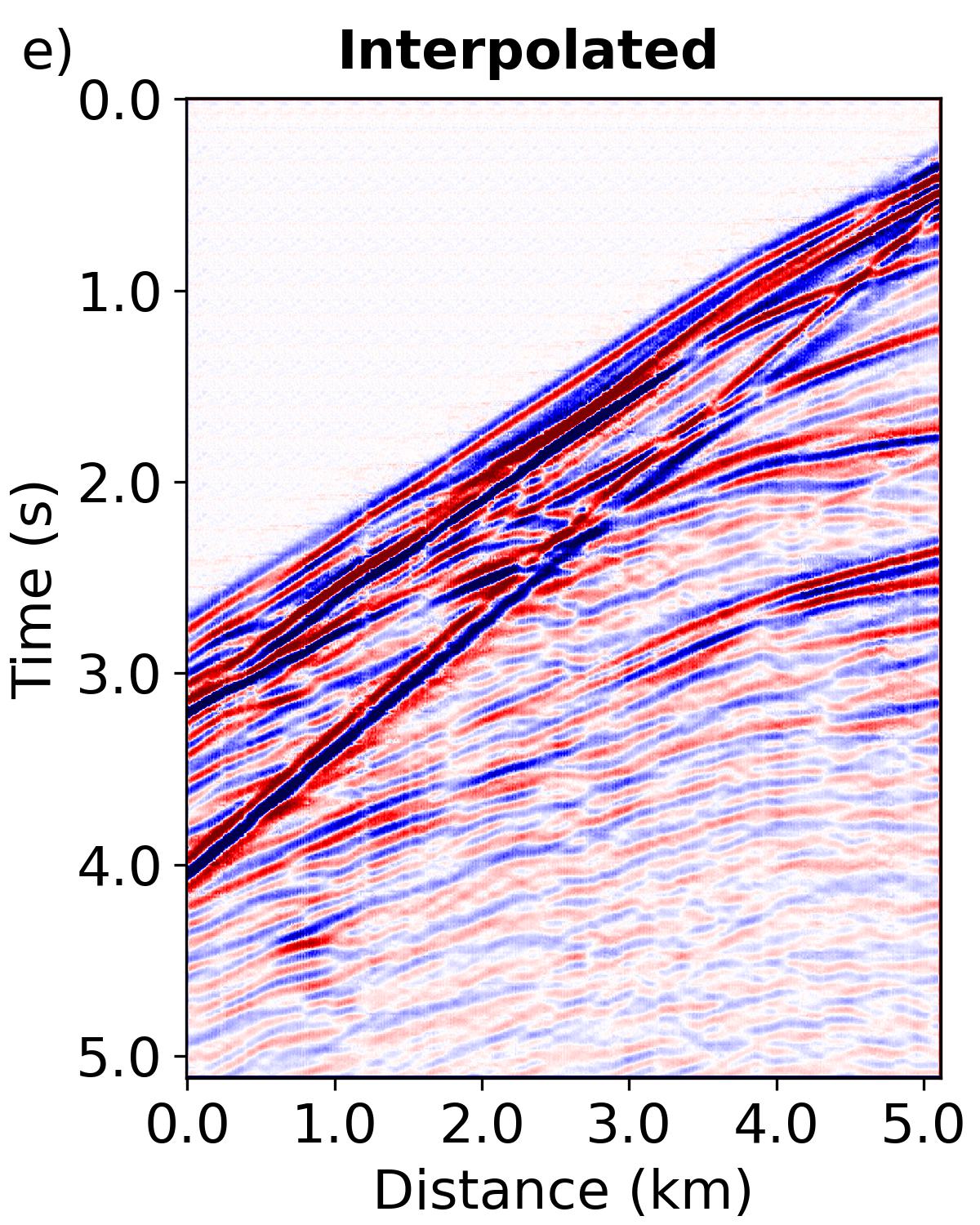}
\hspace{0.3cm}
\includegraphics[width=0.25\textwidth]{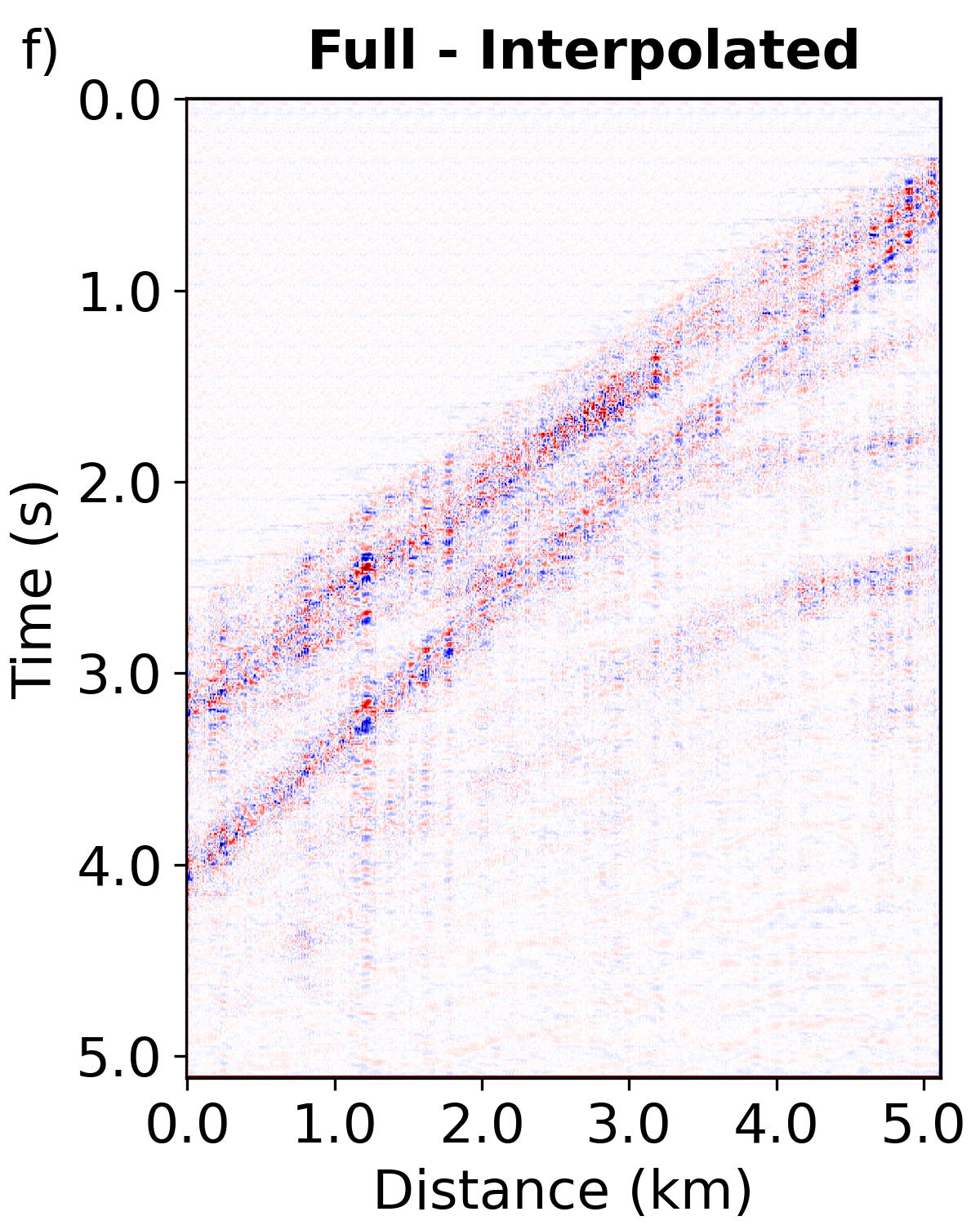} \\
\hspace{4.4cm}
\includegraphics[width=0.25\textwidth]{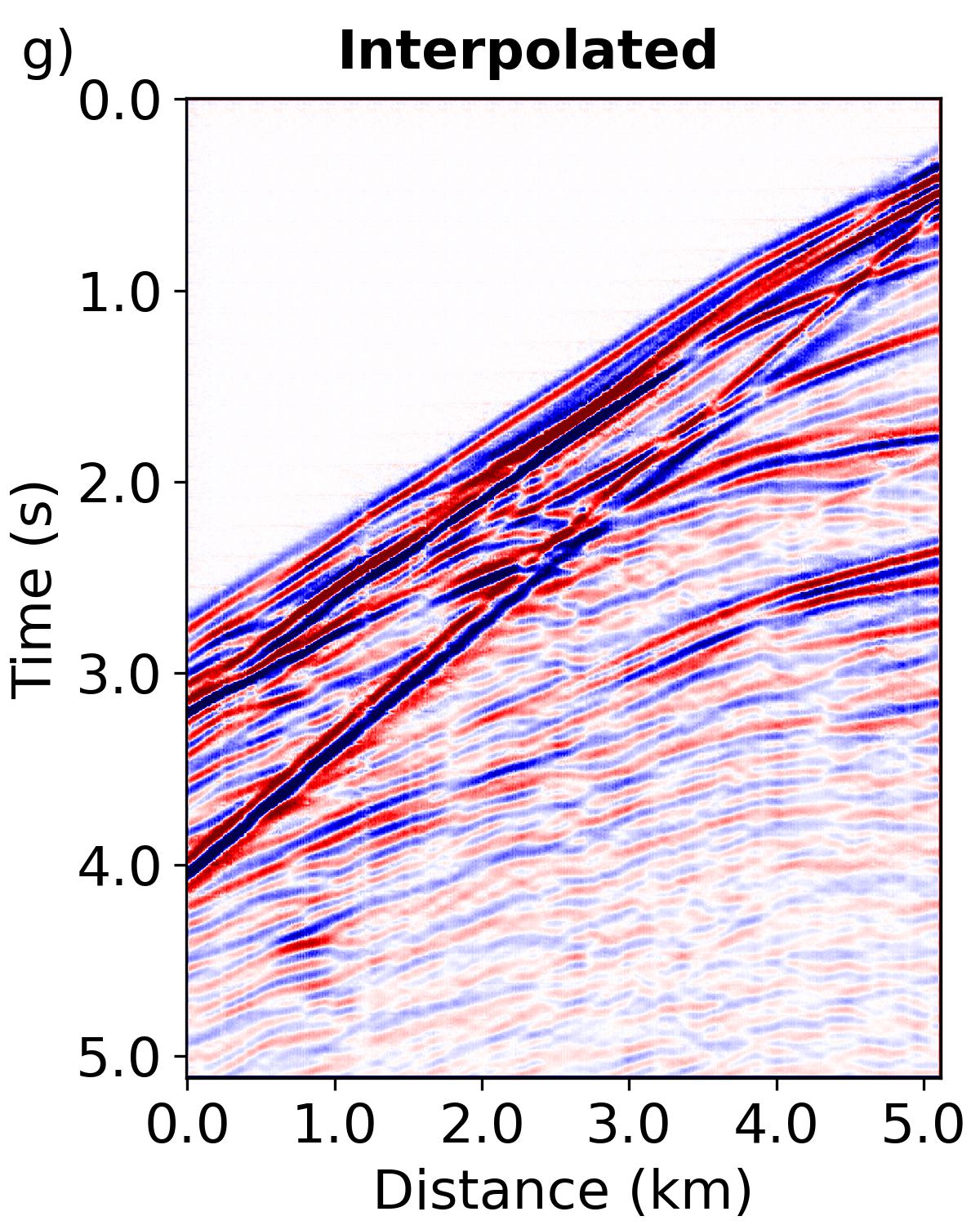}
\hspace{0.3cm}
\includegraphics[width=0.25\textwidth]{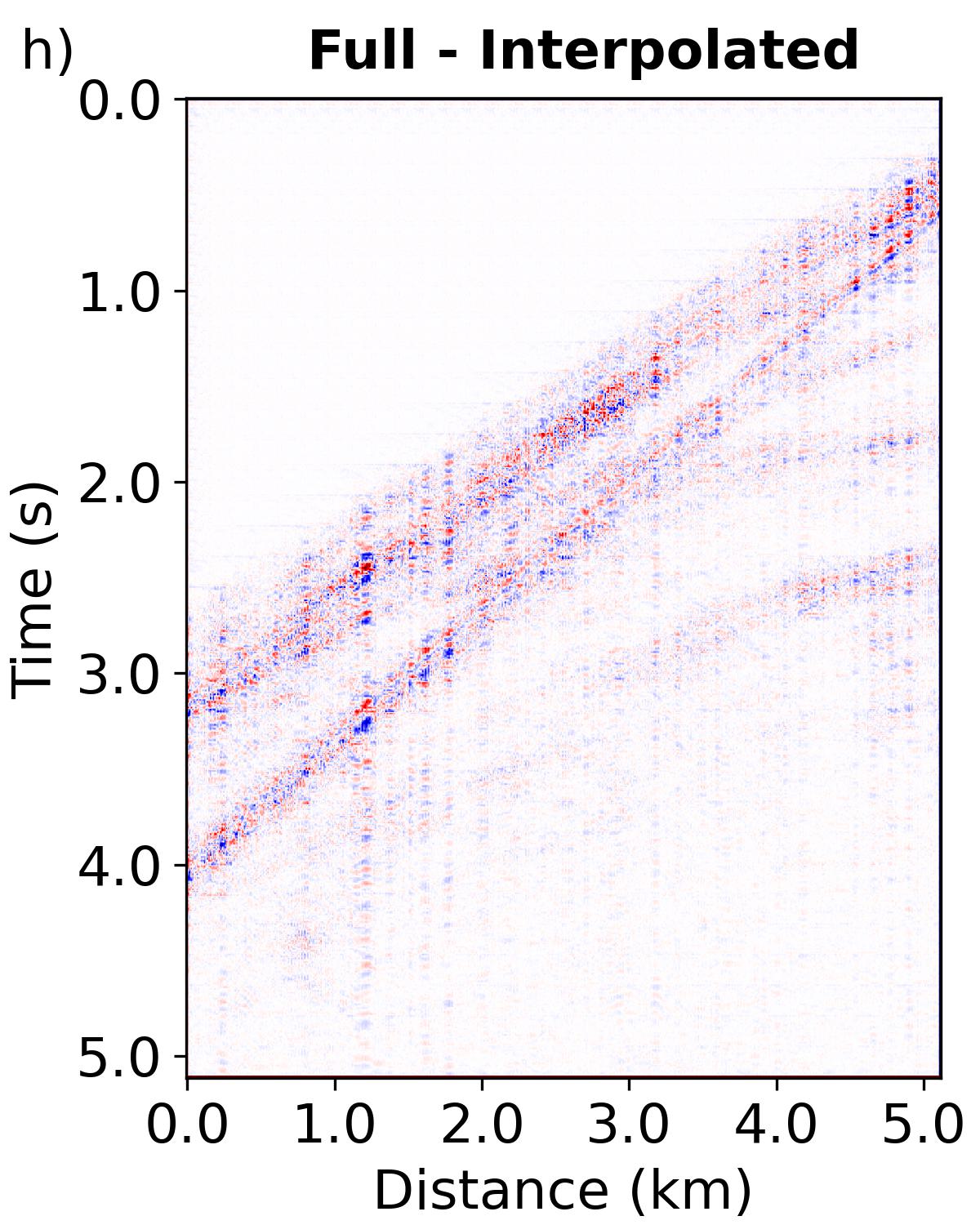} \\
\hspace{4.4cm}
\includegraphics[width=0.25\textwidth]{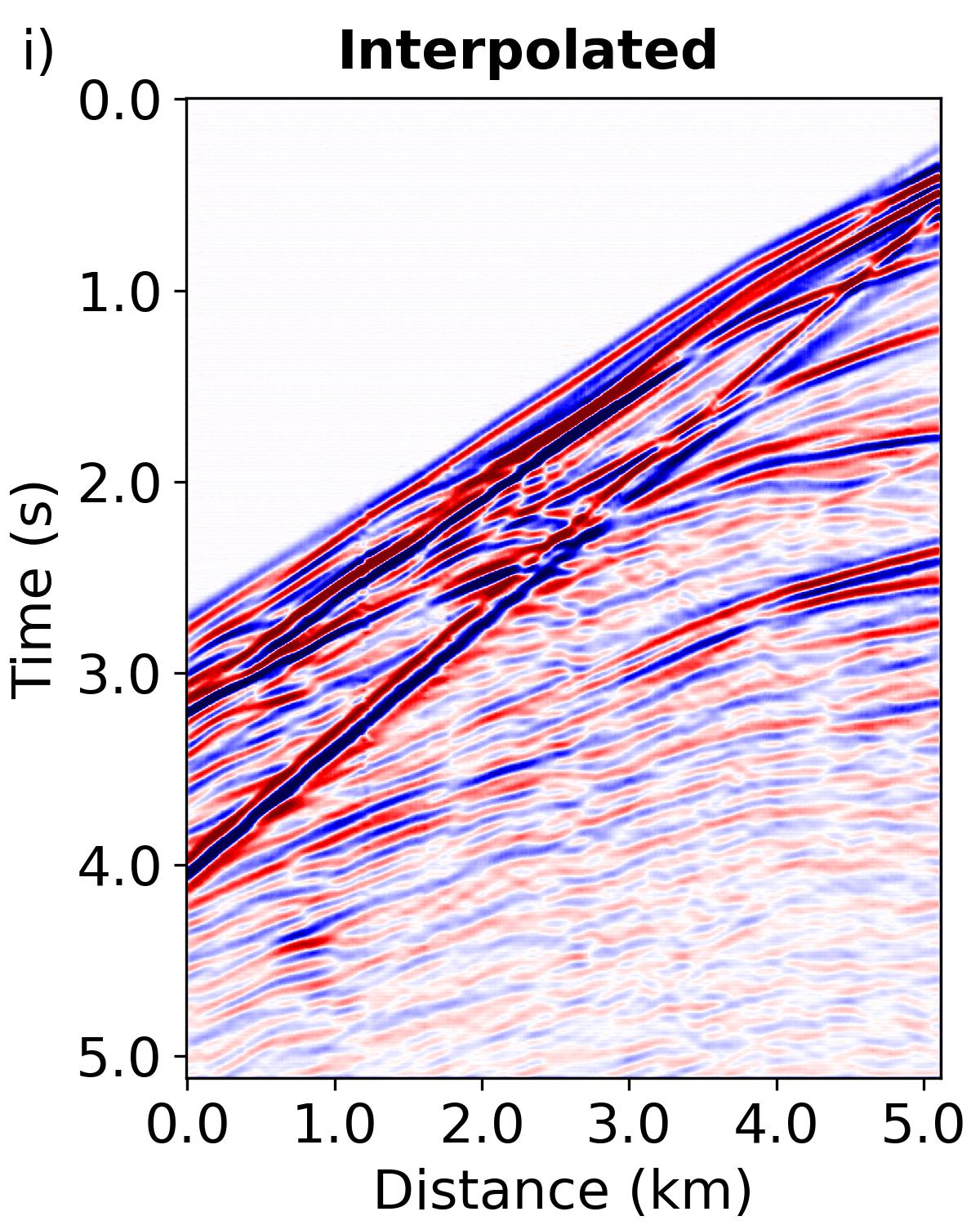}
\hspace{0.3cm}
\includegraphics[width=0.25\textwidth]{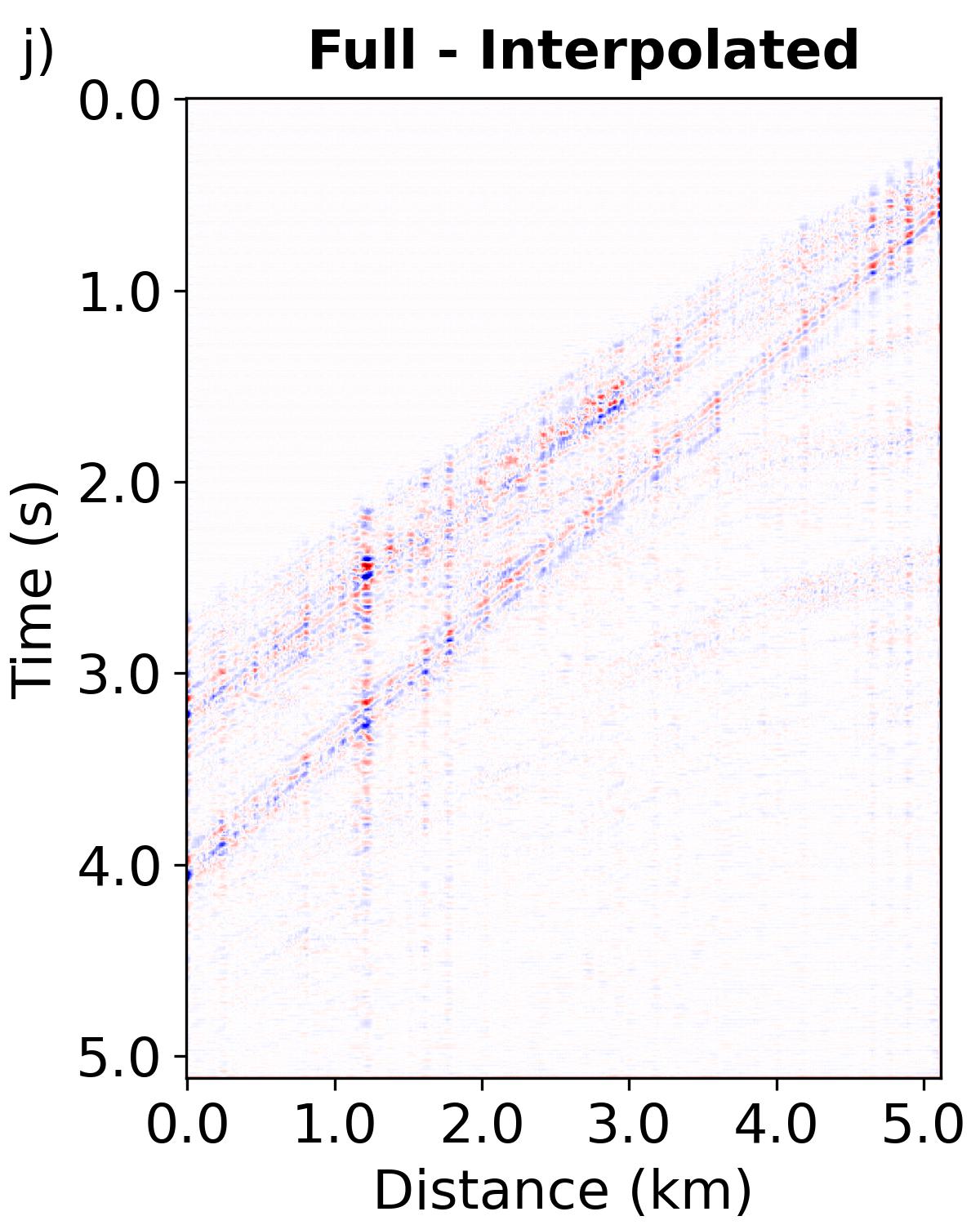}

\caption{Interpolation comparisons of neural networks training with meta-learning initialization and random initialization on synthetic data. (a) Input missing data and (b) the corresponding full data. (c), (e), and (f) are the prediction results from random initialization-base neural network with 10, 100, and 300 epochs of training, respectively, and their corresponding differences with the full data are (d), (f), and (h), respectively. (i) is the prediction results to the meta initialization-based neural network with 10 epochs of training, and (j) is the corresponding error with full data.}
\label{fig9}
\end{figure}

Next, we present the fine-tuning training of MLIN for a seismic interpolation task and its prediction results on the test data set, which are also compared with RIN to demonstrate its performance. Figures \ref{fig8}a and \ref{fig8}b show the MSE and MSSSIM loss curves as a function of the training epochs, respectively. Likewise, compared to RIN, MLIN exhibits a significant superiority in terms of convergence speed and accuracy. Interpolation comparisons of MLIN and RIN to test data set are displayed in Figure \ref{fig9}. The results clearly demonstrate that our method provides a better interpolation performance and less energy leakage compared to the full data, even for RINs trained with 300 epochs. Numerically, our method achieves an MSE of 2.99e-7 and an MSSSIM of 3.70e-5 with 10 epoch training, while the RIN trained for 300 iterations only holds an MSE of 4.78e-5 and an MSSSIM of 9.75e-5. 

\subsubsection{Ground roll attenuation}
Ground roll noise is a type of seismic noise that can severely affect the quality of seismic inversion and imaging. This type of noise is often prevalent in land data, but also present in ocean bottom recording. It is a low-frequency in nature and propagates horizontally near the surface with shear wave velocity speed and can easily overwhelm reflections. In this task, we will try to use the Meta-Processing algorithm to effectively attenuate Ground roll. 

\begin{figure}[htp]
\centering
\includegraphics[width=0.35\textwidth]{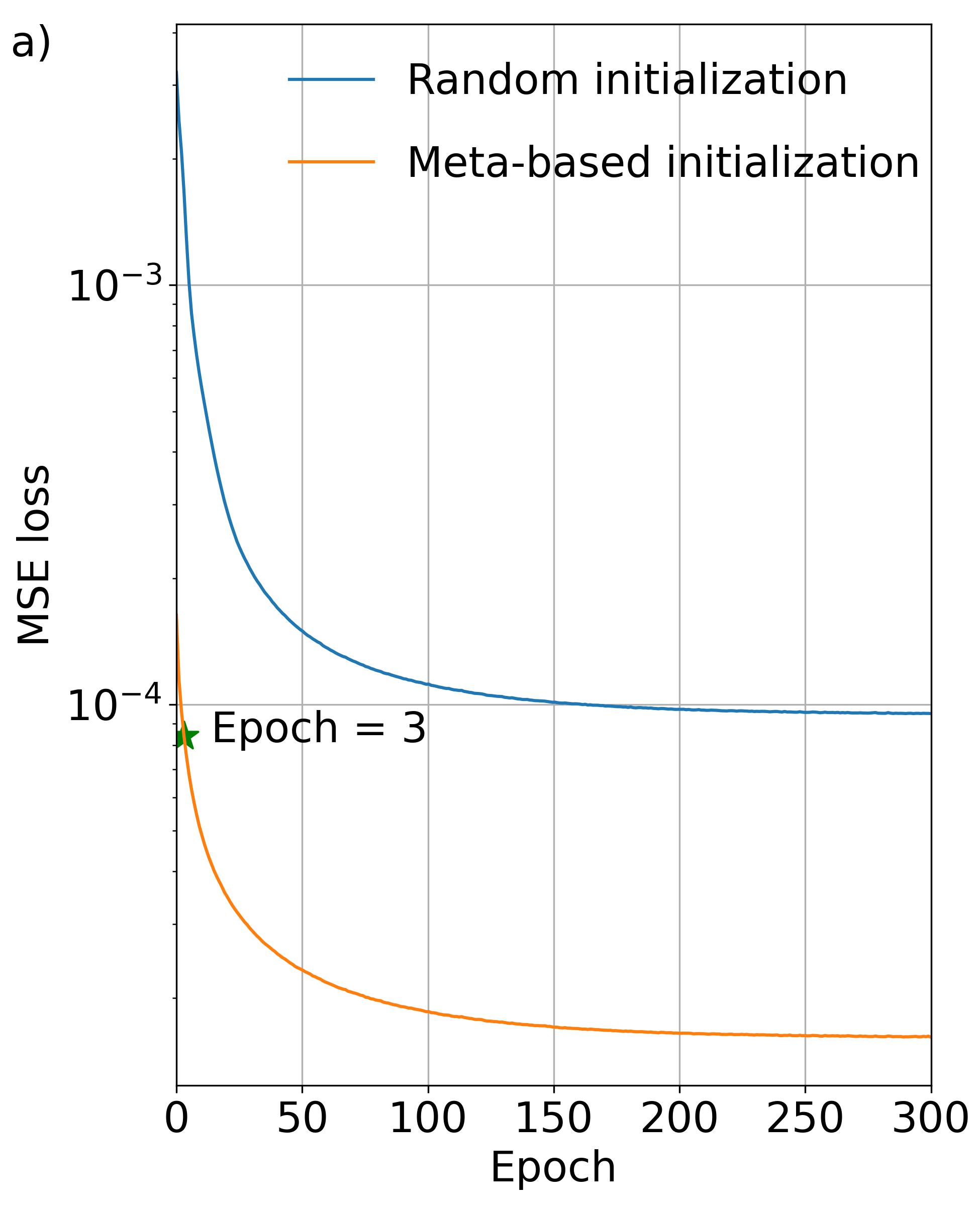}
\hspace{1cm}
\includegraphics[width=0.35\textwidth]{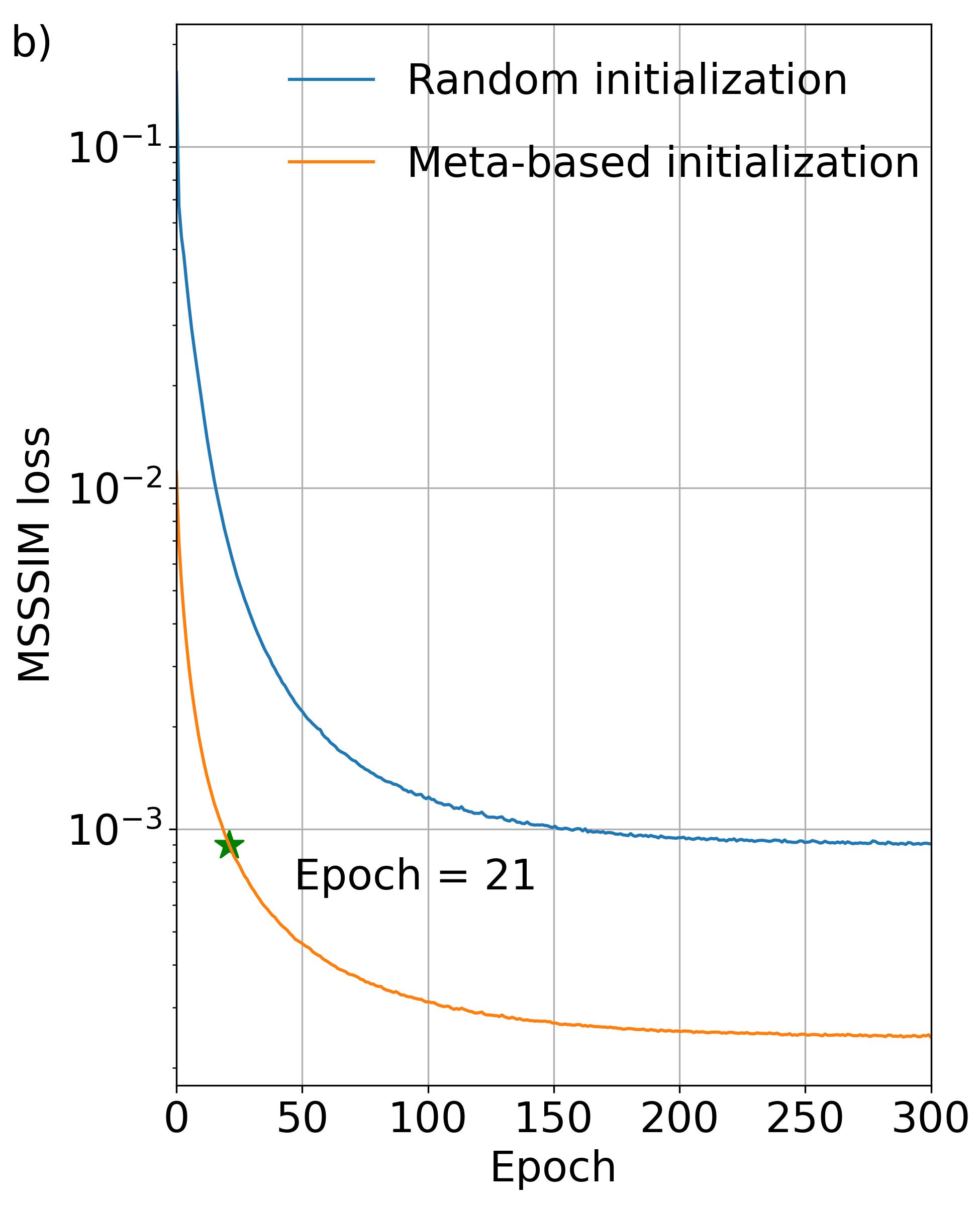}
\caption{The MSE (a) and MSSSIM (b) loss function curves of neural networks training with meta-learning initialization and random initialization of Ground roll attenuation task. }
\label{fig10}
\end{figure}

\begin{figure}[htp]
\centering
\includegraphics[width=0.3\textwidth]{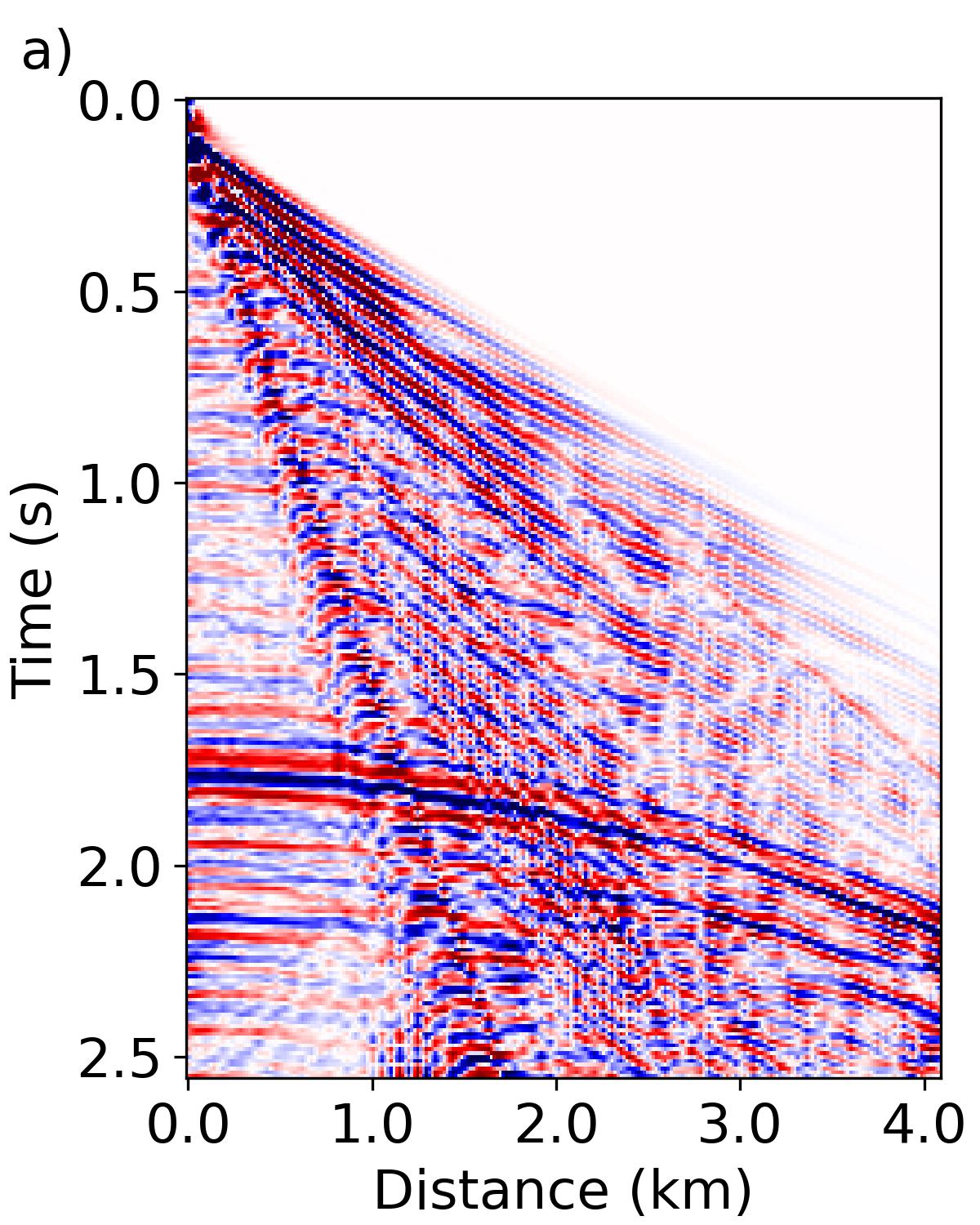}
\hspace{1cm}
\includegraphics[width=0.3\textwidth]{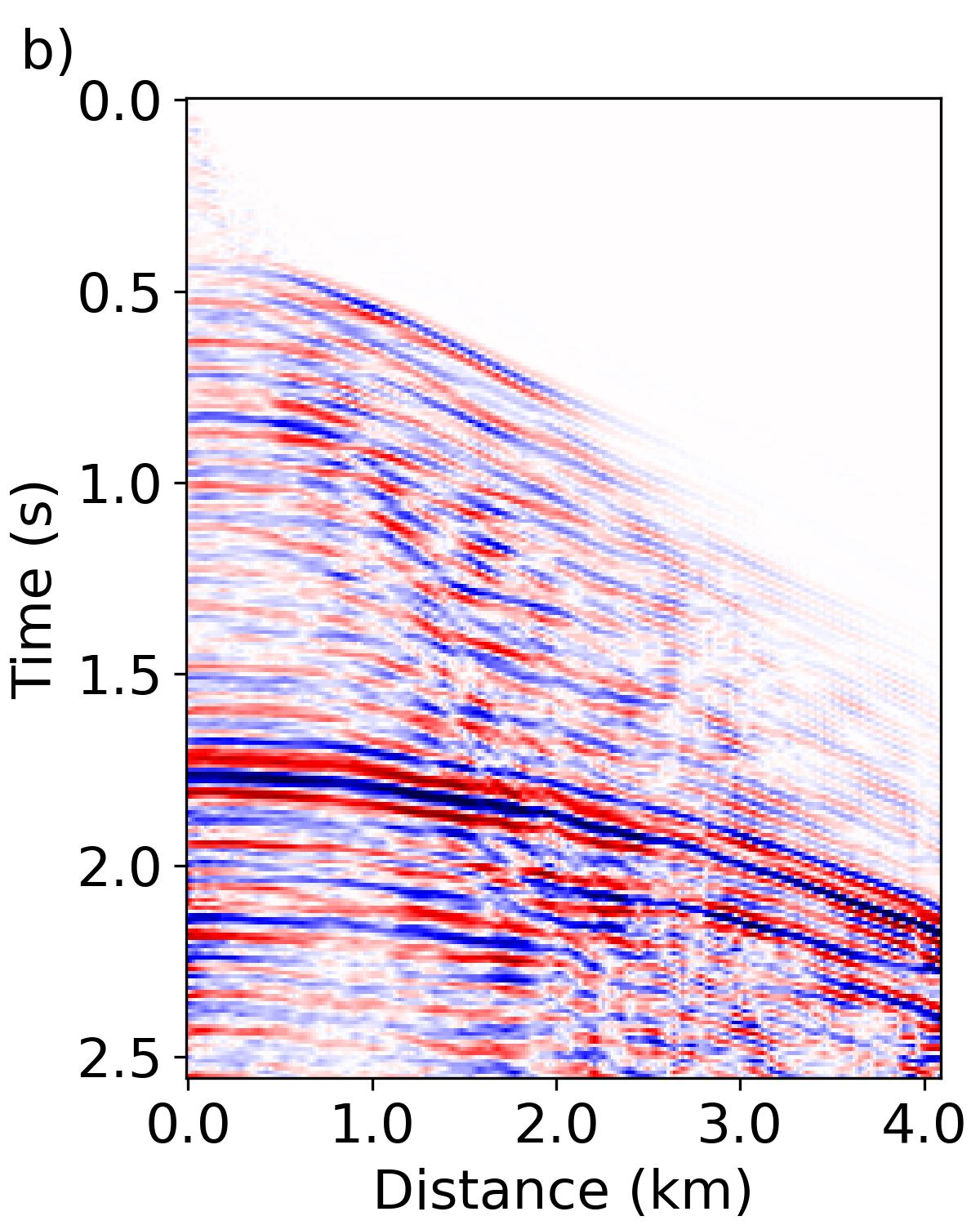}
\caption{The noisy (a) and clean (b) data of synthetic test dataset. }
\label{fig11}
\end{figure}

\begin{figure}[htp]
\centering
\includegraphics[width=0.25\textwidth]{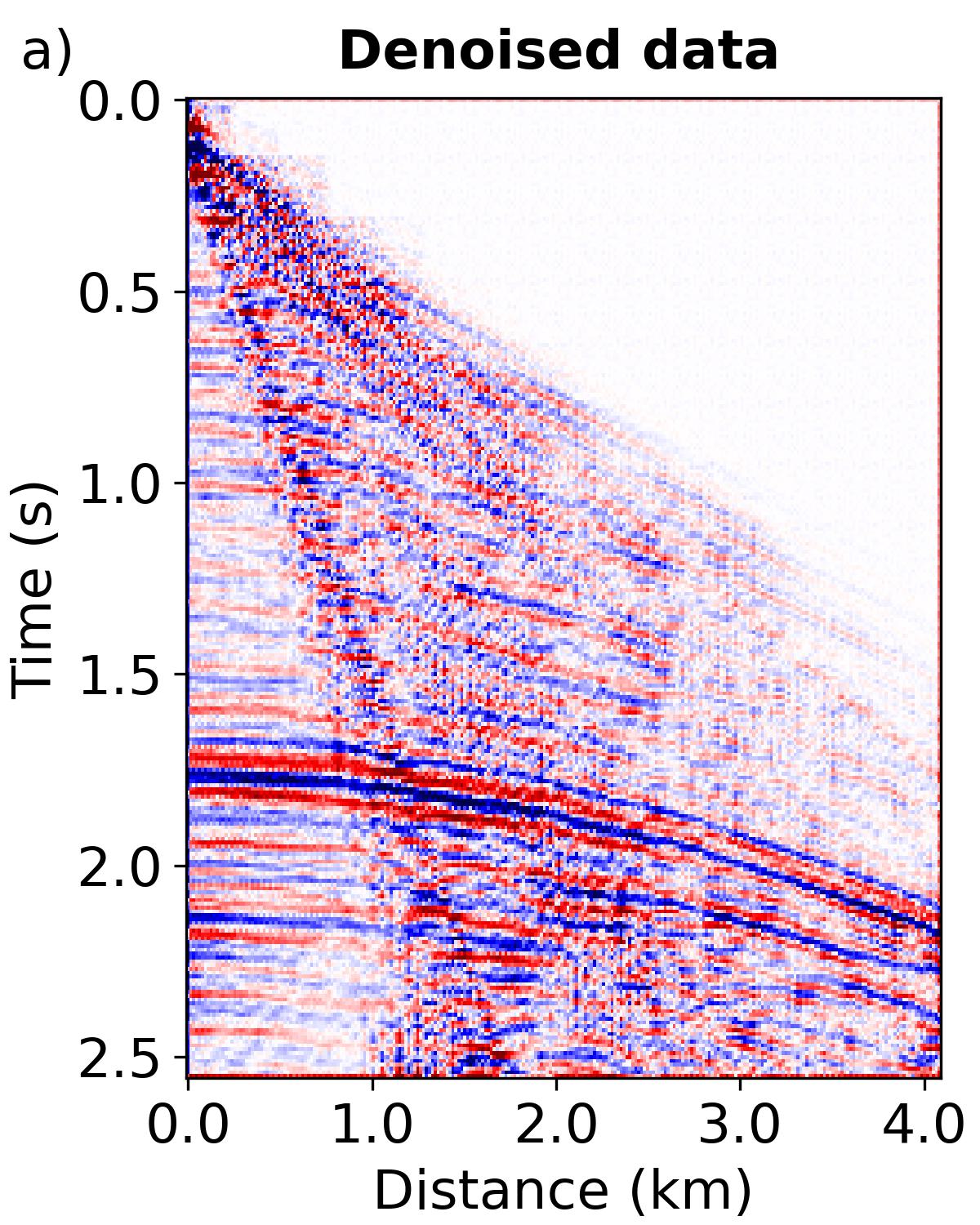}
\hspace{0.3cm}
\includegraphics[width=0.25\textwidth]{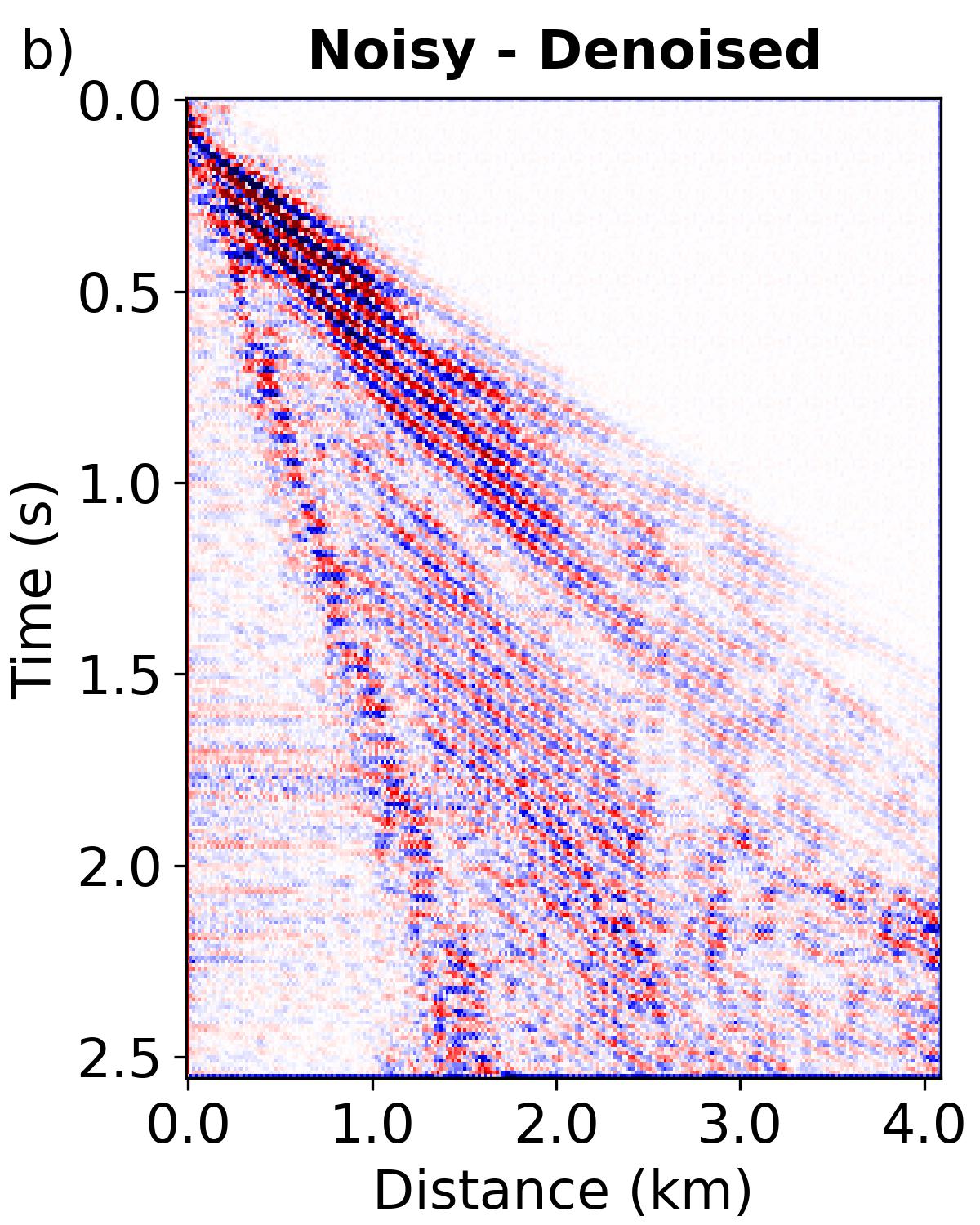}
\hspace{0.3cm}
\includegraphics[width=0.25\textwidth]{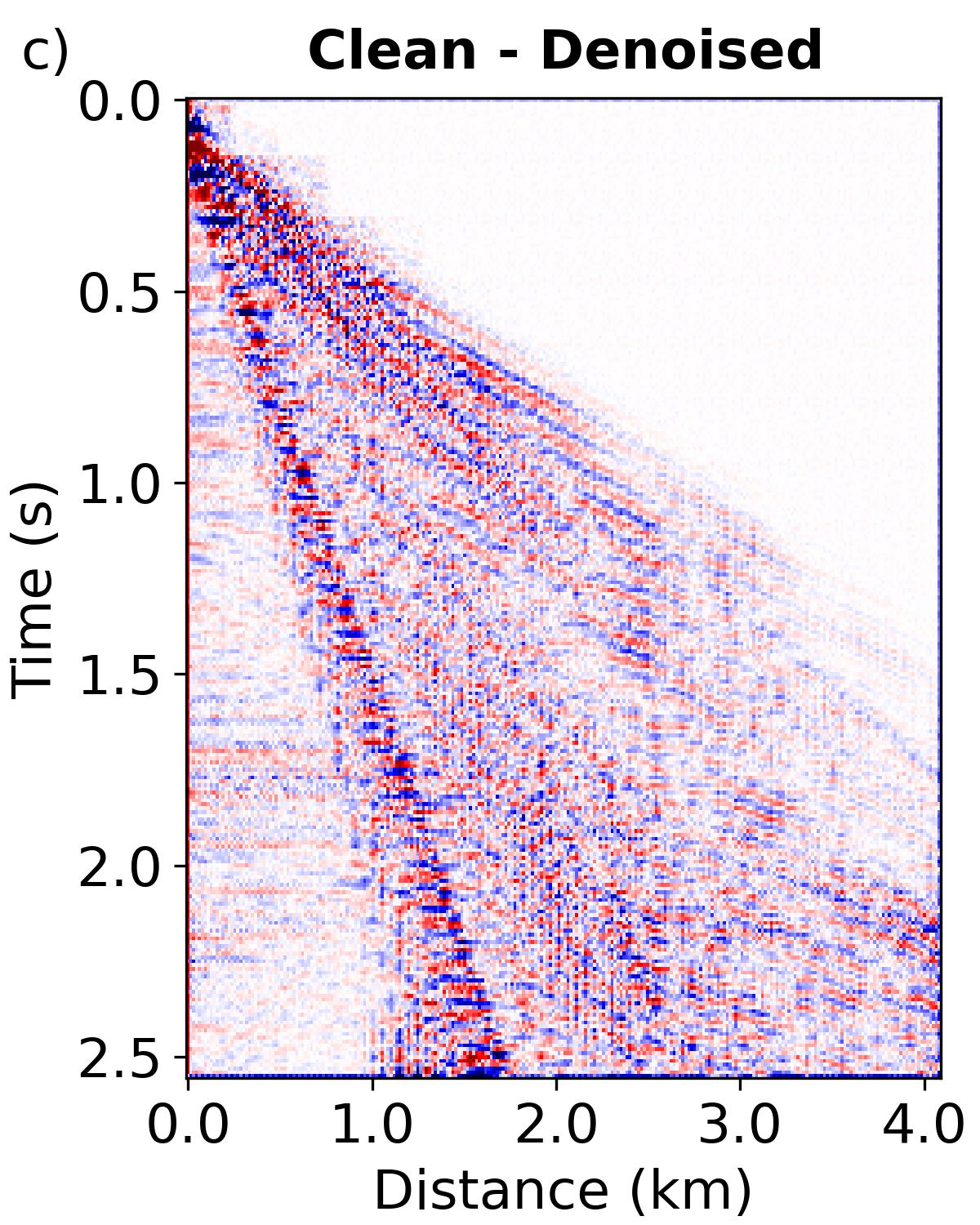}
\includegraphics[width=0.25\textwidth]{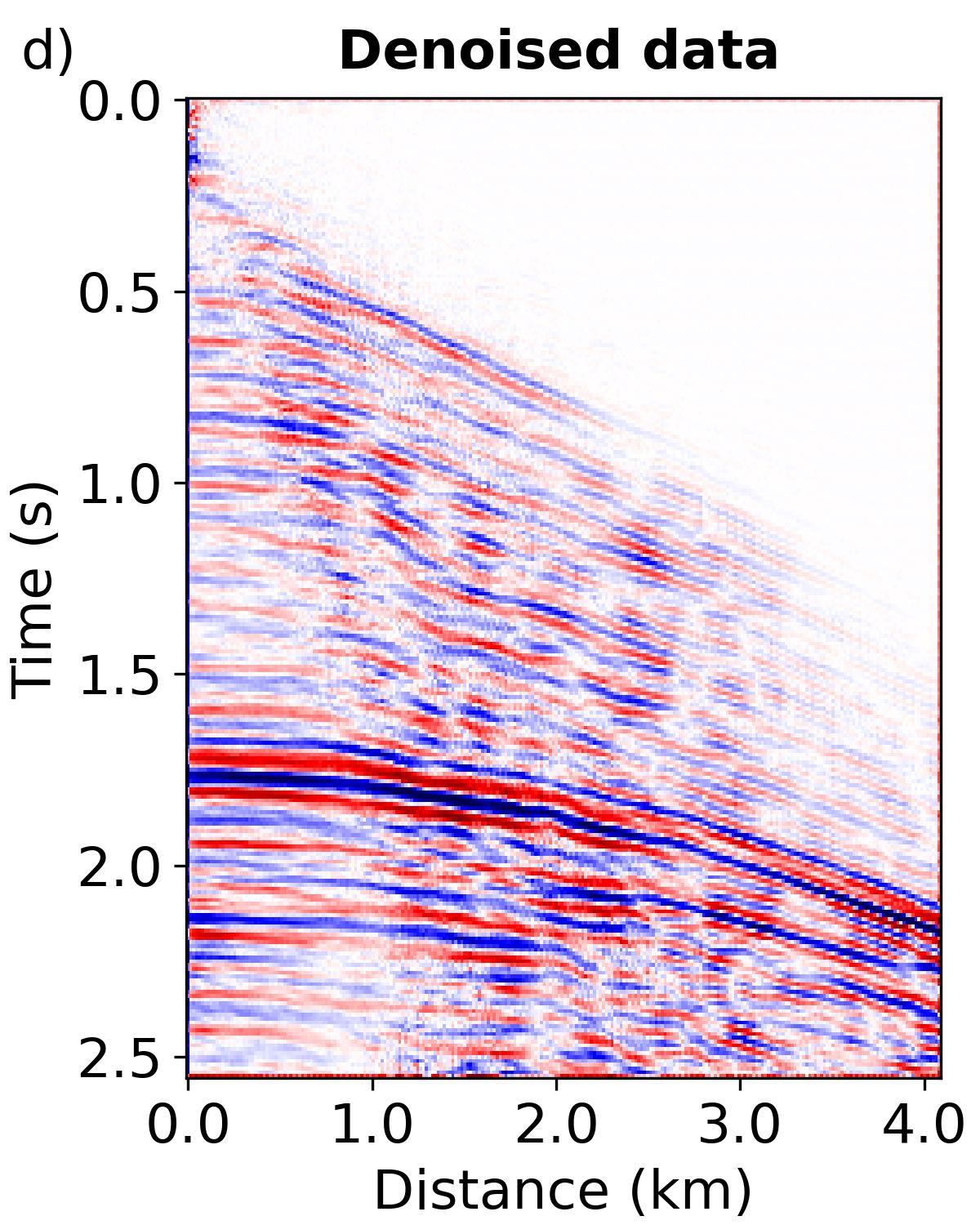}
\hspace{0.3cm}
\includegraphics[width=0.25\textwidth]{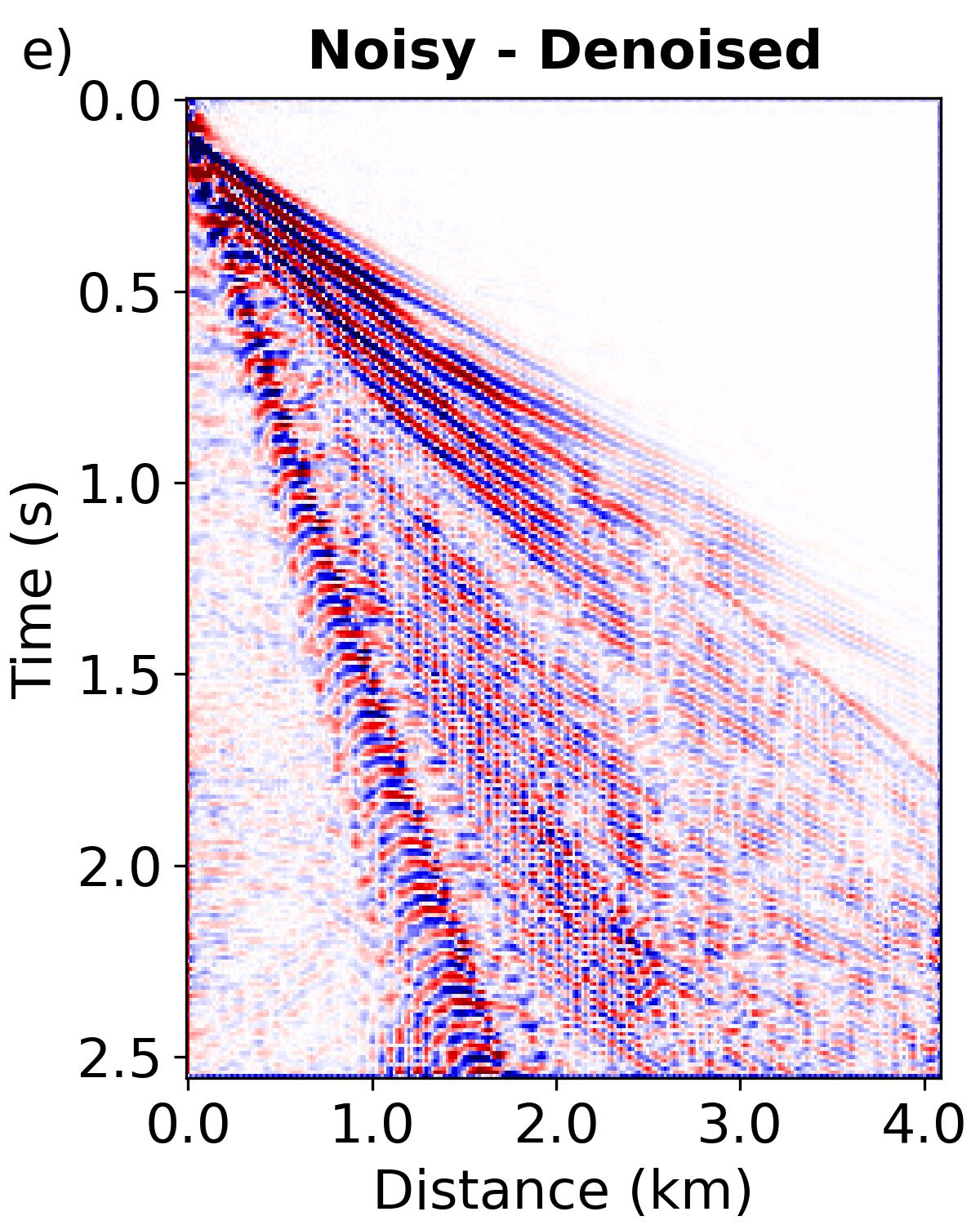}
\hspace{0.3cm}
\includegraphics[width=0.25\textwidth]{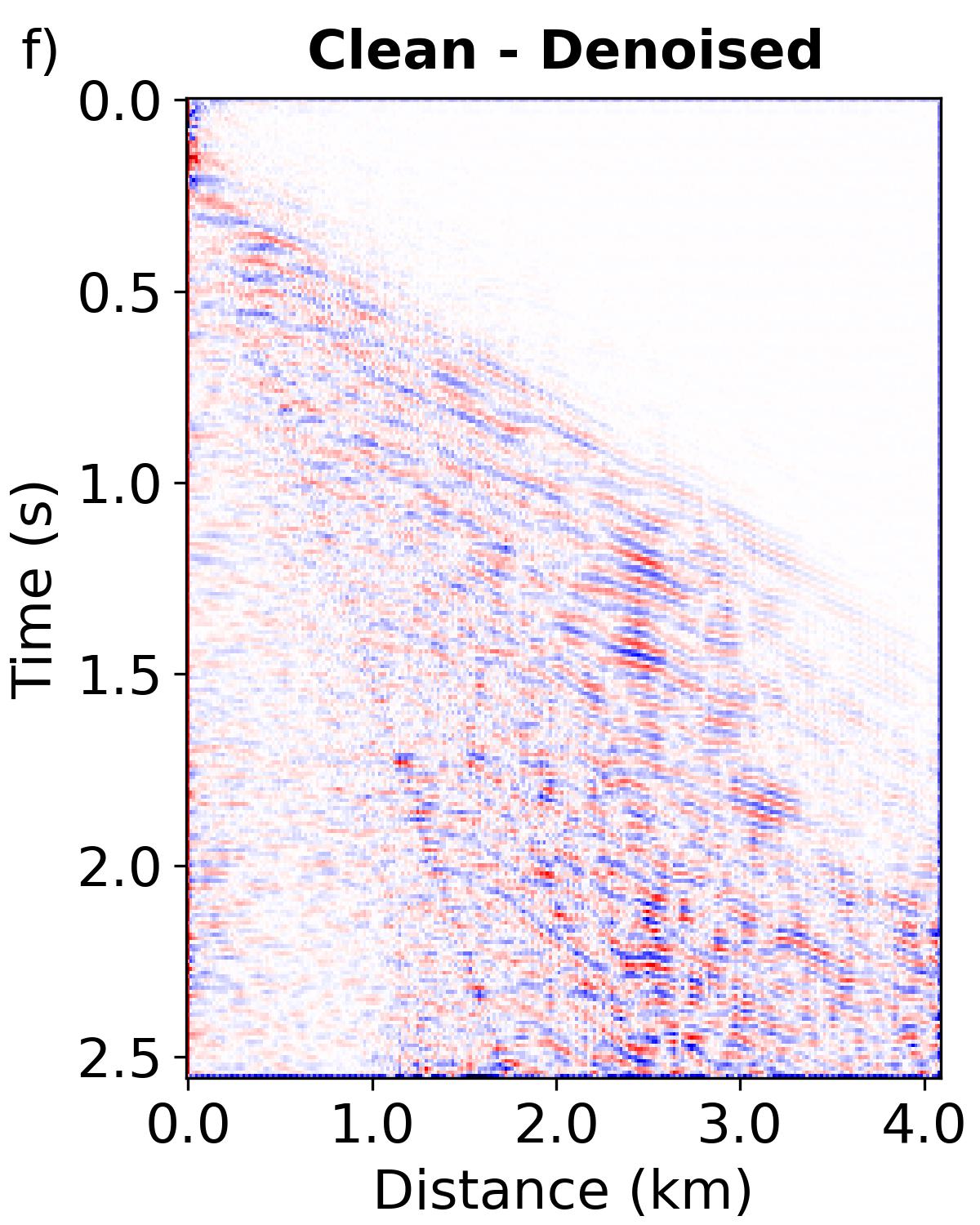}
\includegraphics[width=0.25\textwidth]{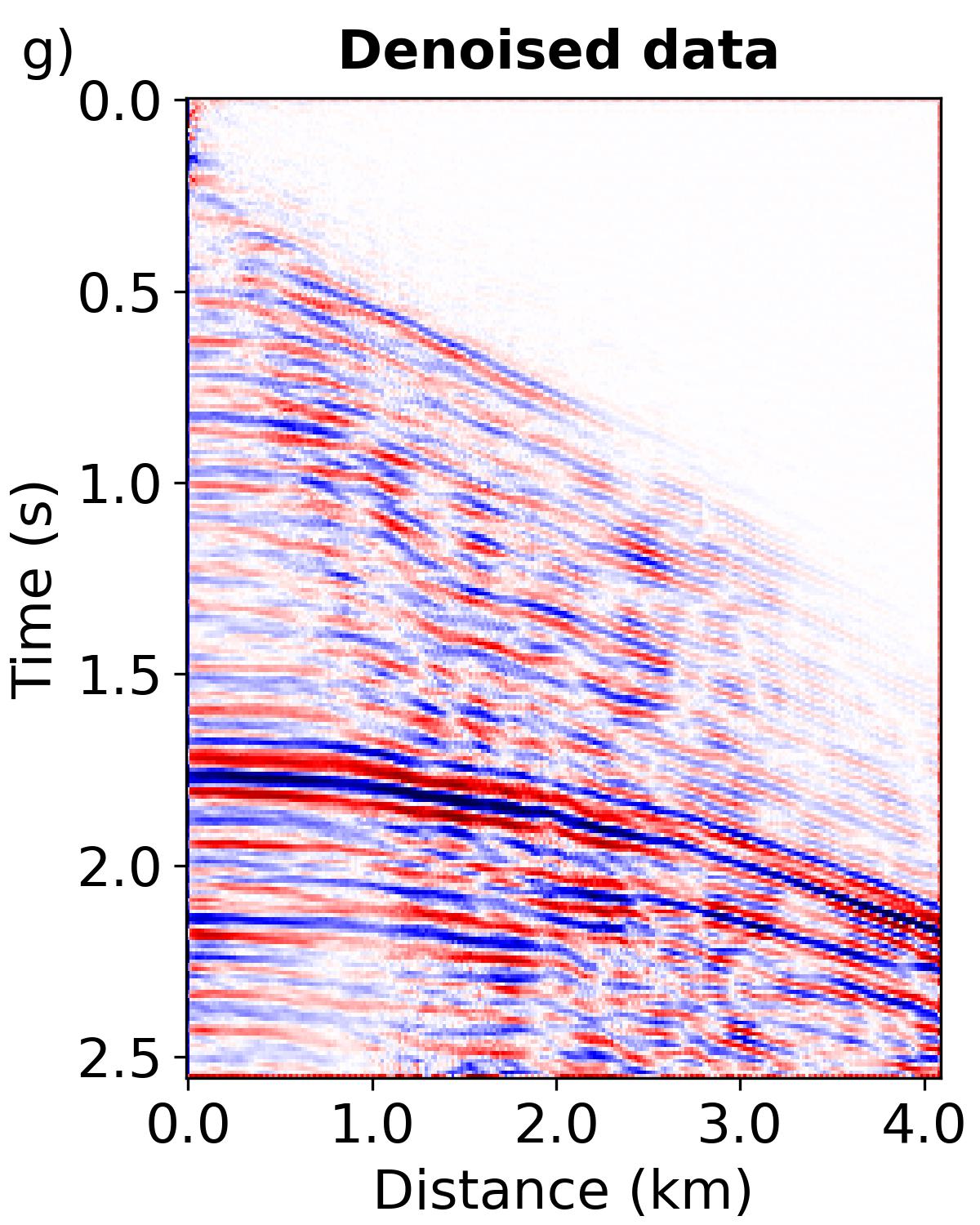}
\hspace{0.3cm}
\includegraphics[width=0.25\textwidth]{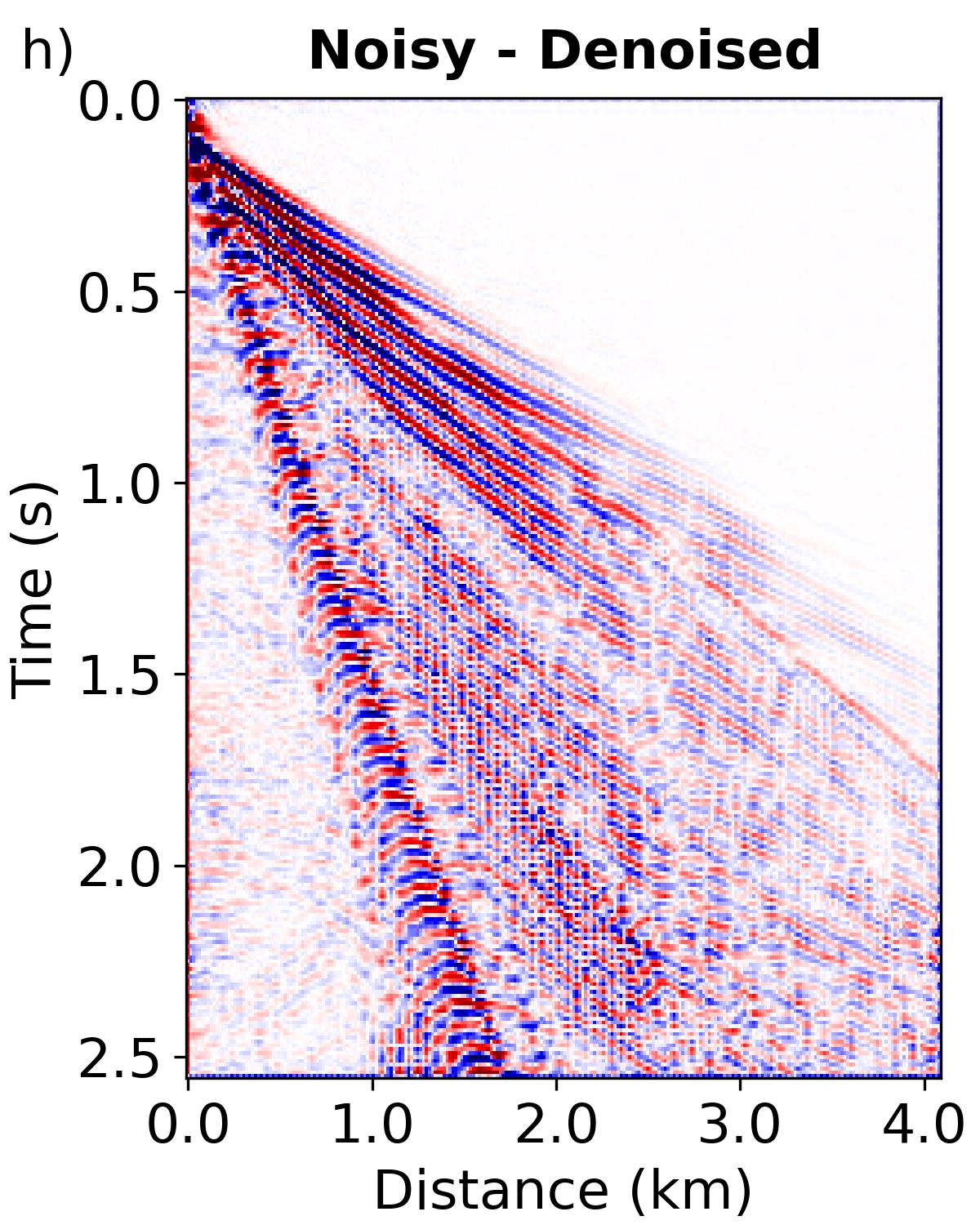}
\hspace{0.3cm}
\includegraphics[width=0.25\textwidth]{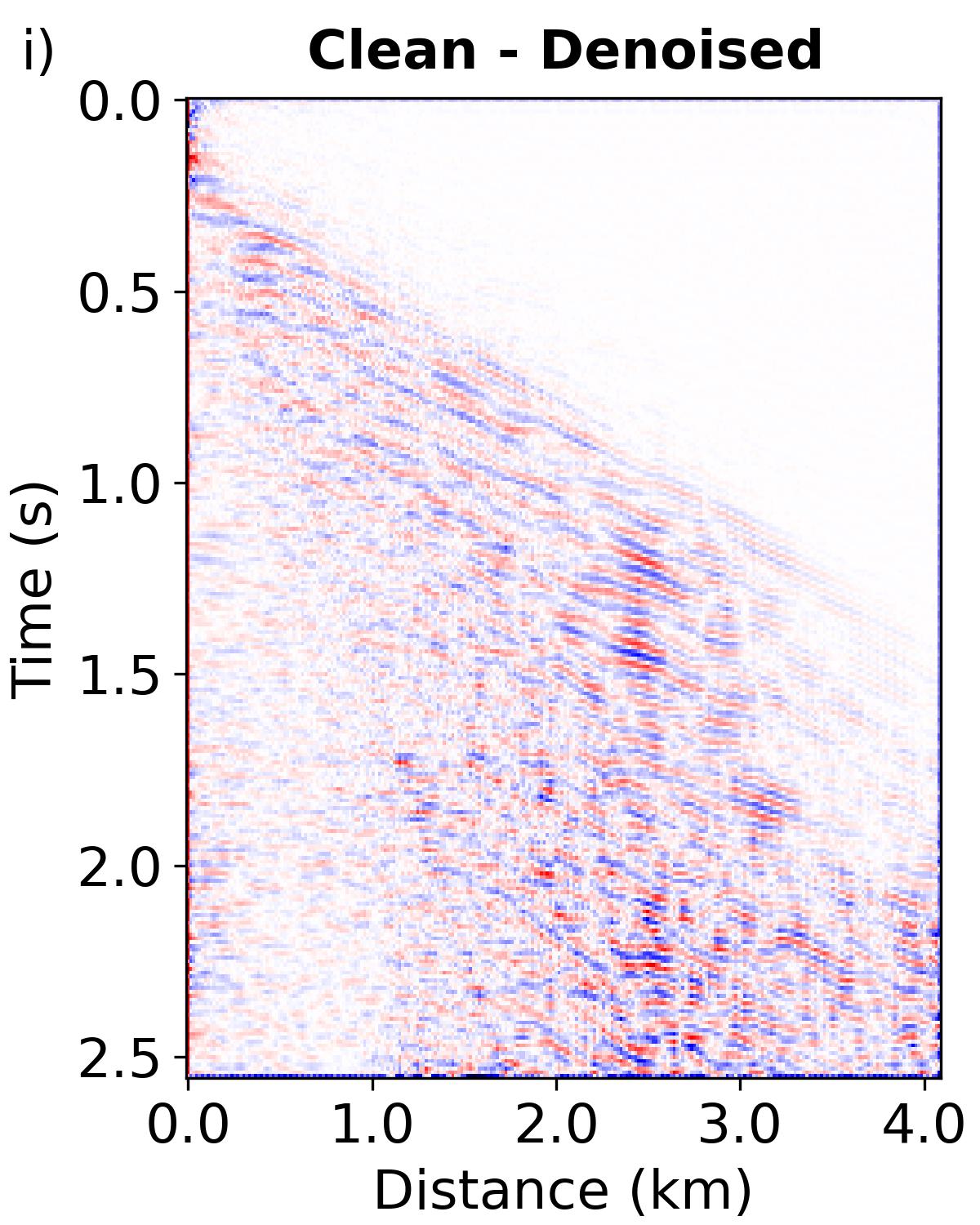}
\includegraphics[width=0.25\textwidth]{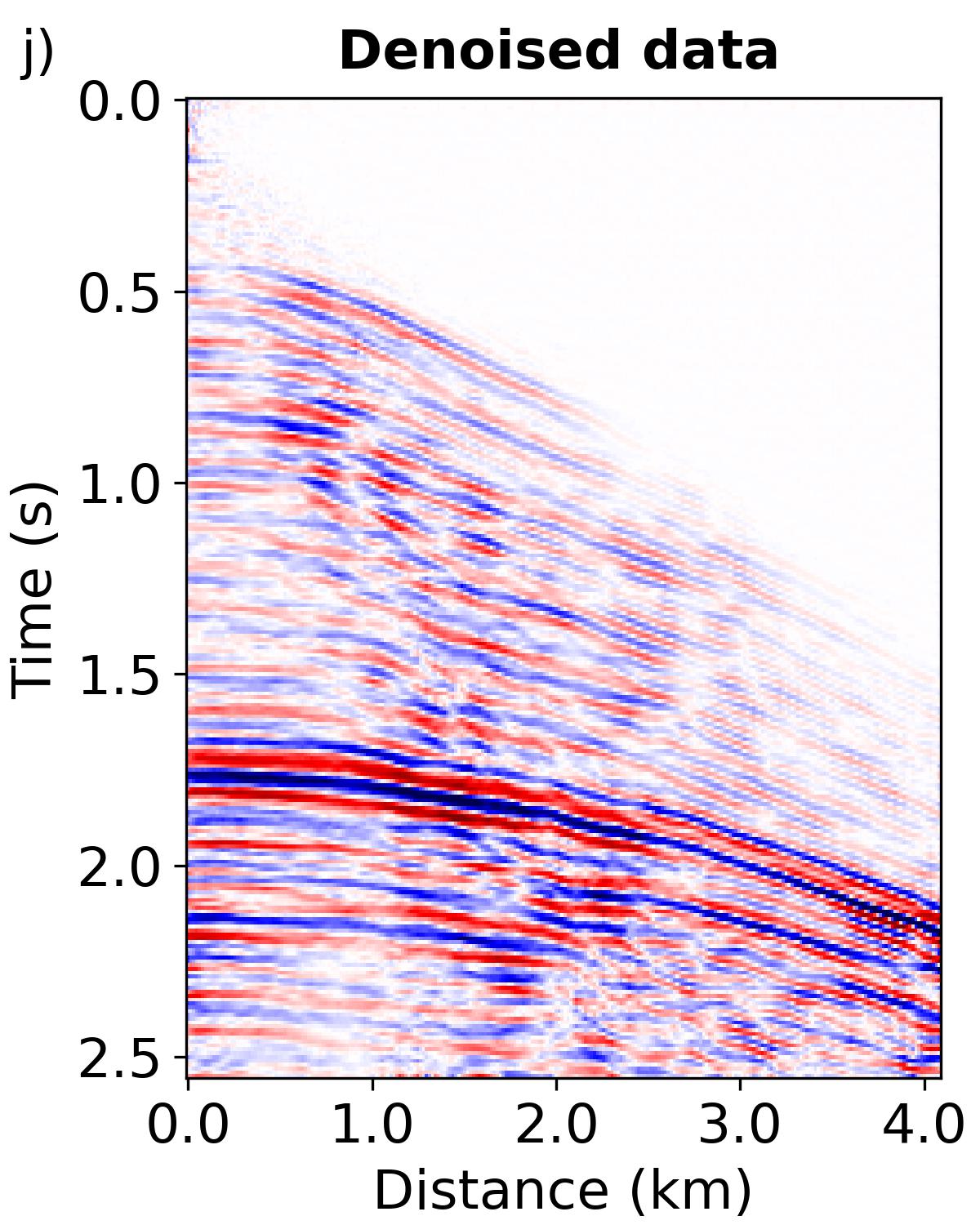}
\hspace{0.3cm}
\includegraphics[width=0.25\textwidth]{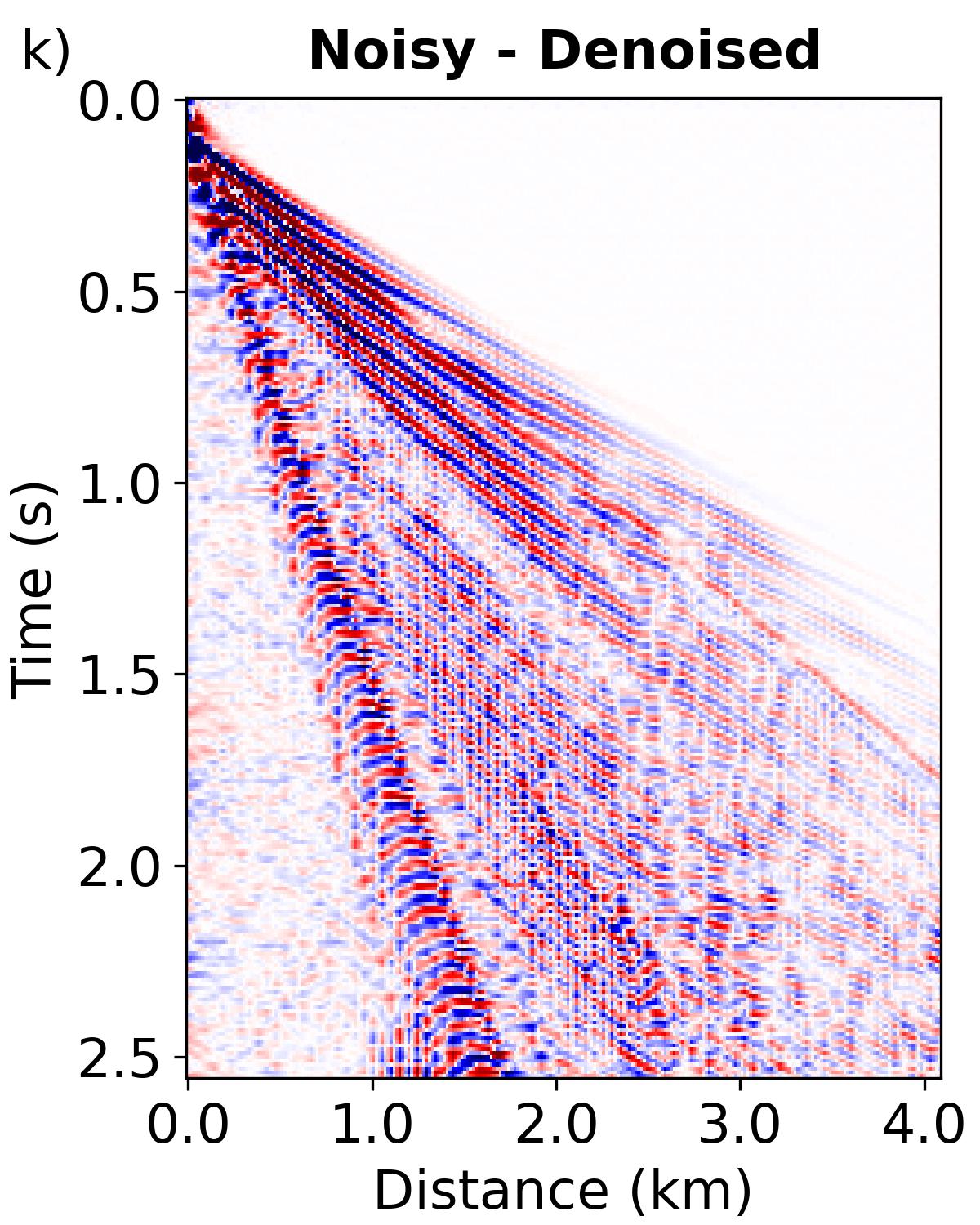}
\hspace{0.3cm}
\includegraphics[width=0.25\textwidth]{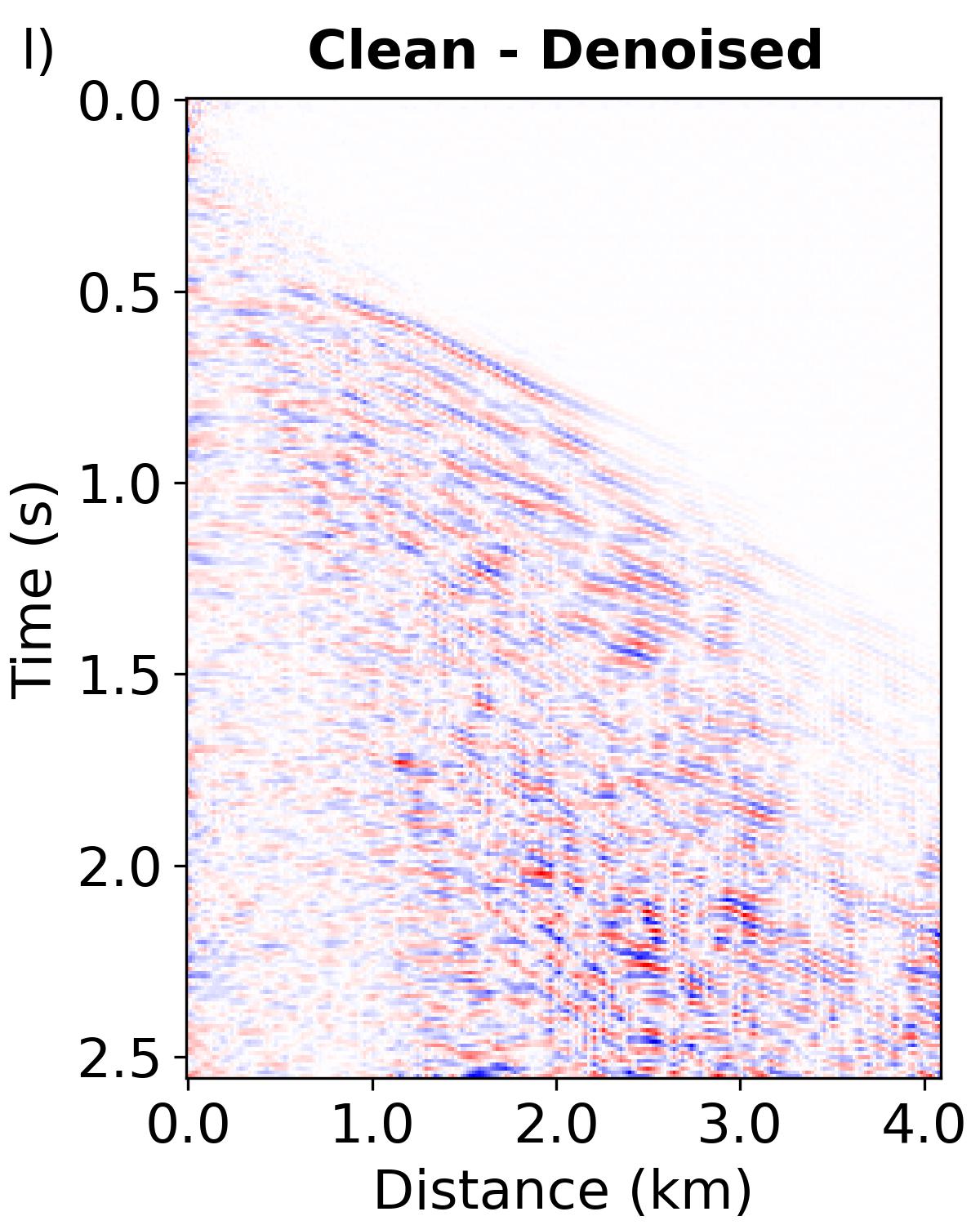}
\caption{Ground roll attenuation comparisons of neural networks training with meta-learning initialization and random initialization on synthetic data. The first, second, and third rows correspond to random initialization-based neural network with 10, 100, and 300 epochs of training, respectively, and the fourth row corresponds to meta initialization-based neural network with 10 epochs of training.}
\label{fig12}
\end{figure}

Figure \ref{fig10} shows the MSE and MSSSIM loss curves of MLIN and RIN with 300 epochs of training. Once again, it verifies that the Meta-Processing algorithm can enable NNs have faster convergence speed and higher accuracy. For the test data set, our method achieves an MSE of 1.25e-4 and an MSSSIM of 8.4e-3 with 10 epochs of training, however, the RIN trained for 300 epochs only to achieve an MSE of 1.84e-4 and an MSSSIM of 8.9e-3. The results are shown in Figure \ref{fig12}, where the first, second, and third rows come from RIN with 10, 100, and 300 epochs of training, respectively, while the fourth row corresponds to MLIN with 10 epochs of training. The corresponding input (noisy) and label (clean) data are displayed in Figure \ref{fig11}. We can see that our method can effectively remove ground roll noise while preserving the signal with only 10 epochs of gradient descent updates. In contrast, the performance of RIN converges slowly as the number of epochs increases, as it is not easy to find the optimal solution from a random set of network parameters. 

\subsubsection{Imaging enhancement}
Seismic imaging is a widely used method for exploring the subsurface structures of the Earth. Several factors can influence the quality of seismic imaging. Specifically, the geometry and spacing of the geophones used during data acquisition can affect the resolution and accuracy of the resulting images. For example, to apply large-scale deepwater surveys in crustal and oceanographic research, ocean bottom node (OBN) seismic acquisition systems typically adopt extremely sparse node spacing to reduce the cost and time consumption in acquisition. The sparse recording, however, leads to poor illumination and reduced continuity of events, posing a huge challenge to imaging. 

In this task, to overcome the challenges brought by the sparse acquisition system in OBN surveys, we propose to train an NN that can map the images from sparse acquisition to dense acquisition. The trained network is expected to directly process the sparse images, improving the continuity and eliminating artifacts. Here, we use a synthetic model of the South China Sea to generate the seismic data. The model consists of a water layer, a series of thin flat transitional layers, and a series of sedimentary rock layers. Dense seismic data are obtained using finite-difference forward modeling with a grid spacing of 3.0 meters vertically and 3.1 meters horizontally, while sparse seismic data are obtained by subsambling the dense data, resulting in an OBN spacing of 310 meters. We employ the common-receiver Gaussian beam migration method \cite{shi2023elastic, cheng2023seismic, cheng2023elastic} to generate the images for training and testing. 

\begin{figure}[htp]
\centering
\includegraphics[width=0.35\textwidth]{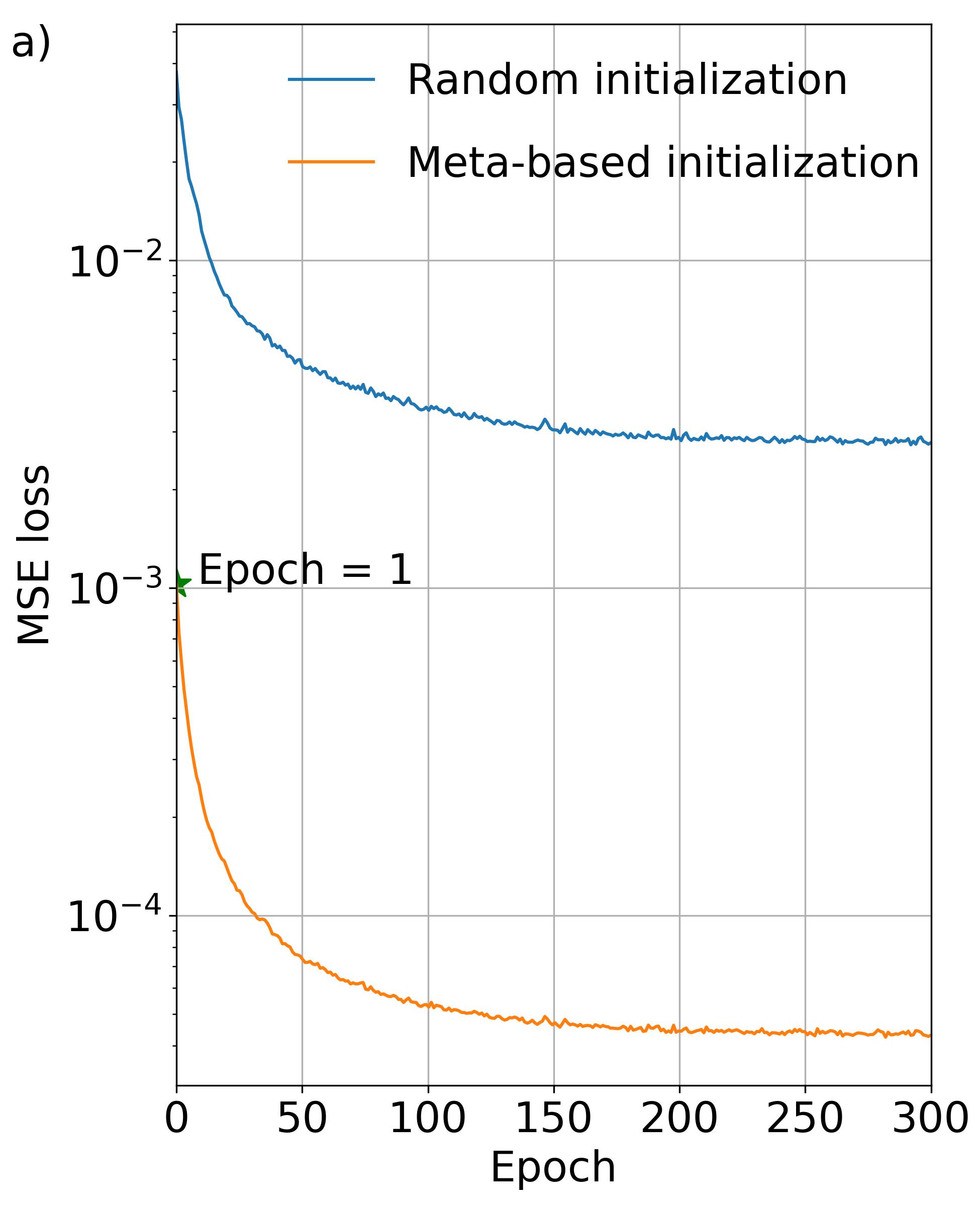}
\hspace{1cm}
\includegraphics[width=0.35\textwidth]{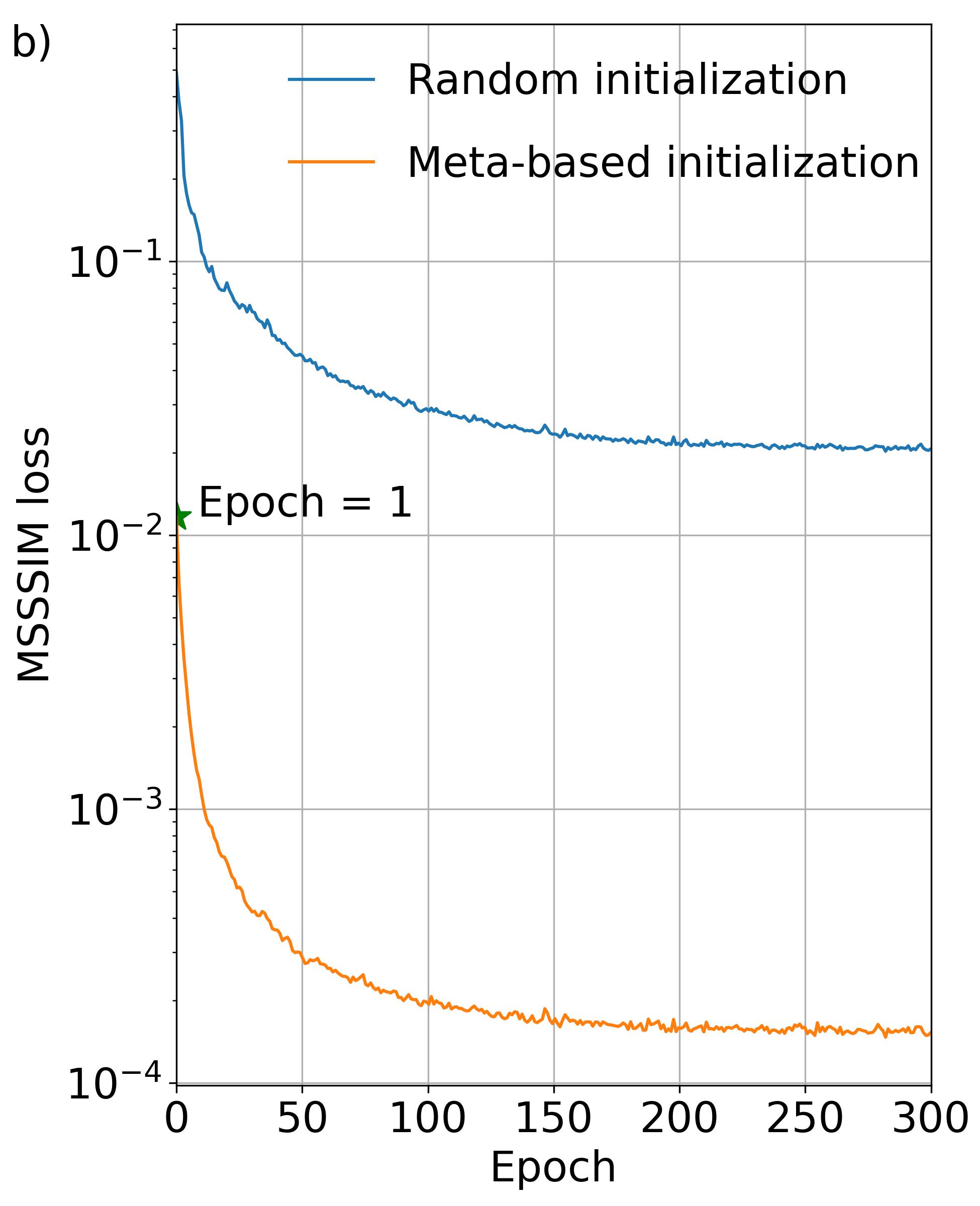}
\caption{The MSE (a) and MSSSIM (b) loss function curves of neural networks training with meta-learning initialization and random initialization of imaging enhancement task. }
\label{fig13}
\end{figure}

Figure \ref{fig13} depicts the MSE and MSSSIM loss curves for the task of imaging enhancement trained by the MLIN and RIN for 300 epochs. Remarkably, after just one epoch of optimization, both MSE and MSSSIM losses of the MLIN are significantly lower than that of the RIN, which undergoes 300 epochs of gradient descent update. This demonstrates that our method, after the Meta-training step, results in significant convergence speed up and accuracy in the imaging enhancement task. 

\begin{figure}[htp]
\centering
\includegraphics[width=0.45\textwidth]{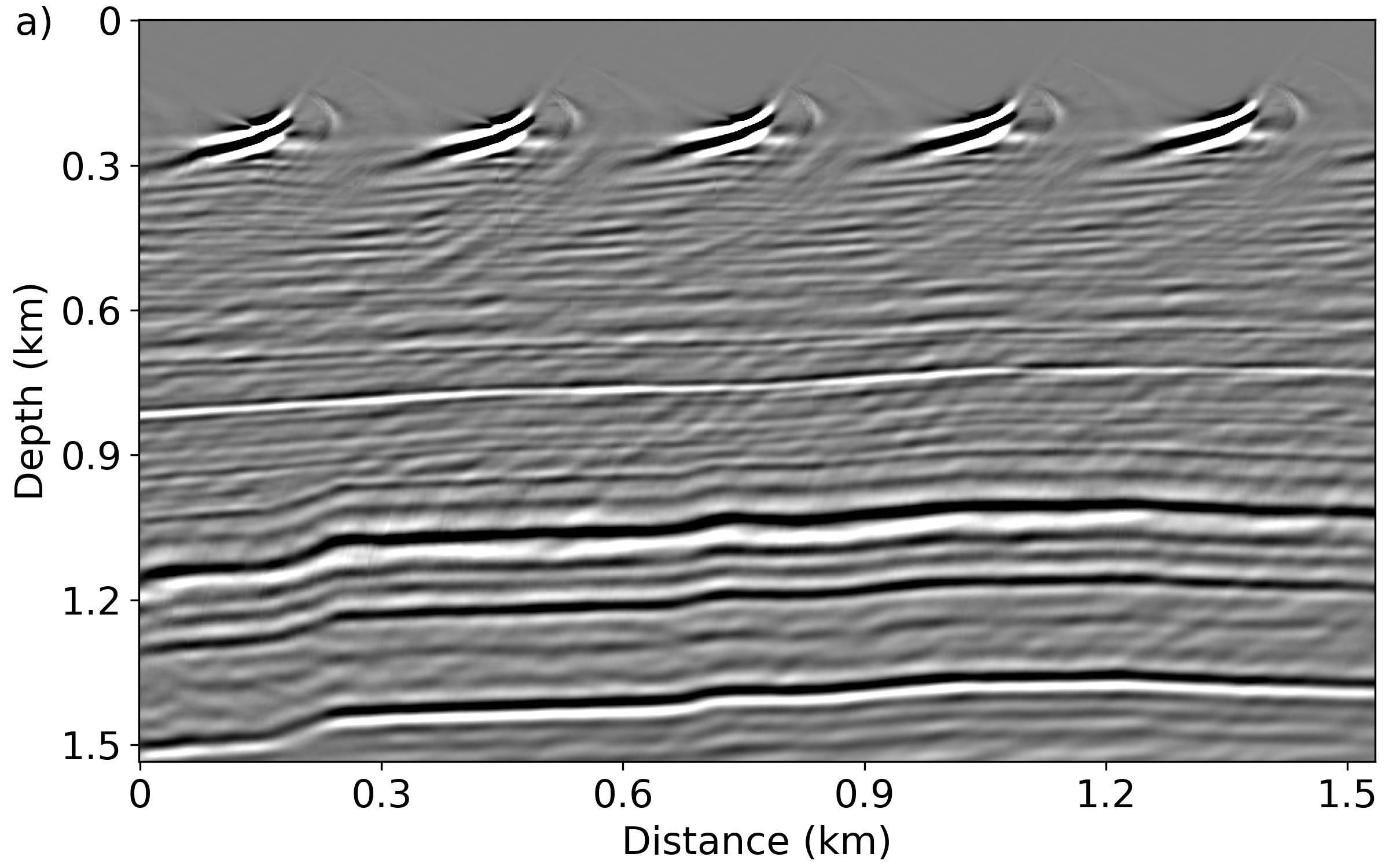}
\hspace{0.3cm}
\includegraphics[width=0.45\textwidth]{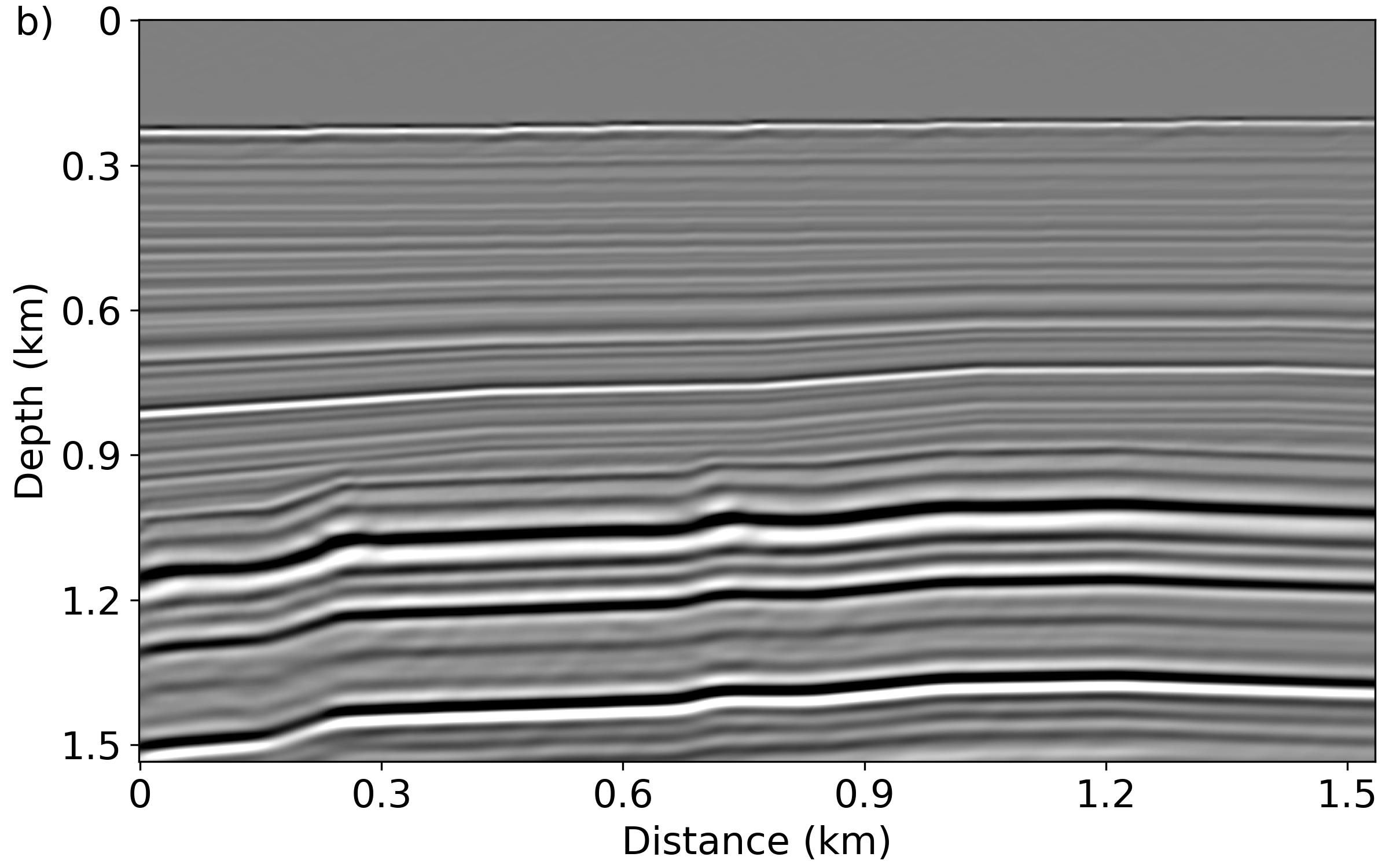}
\vspace{0.3cm}
\includegraphics[width=0.45\textwidth]{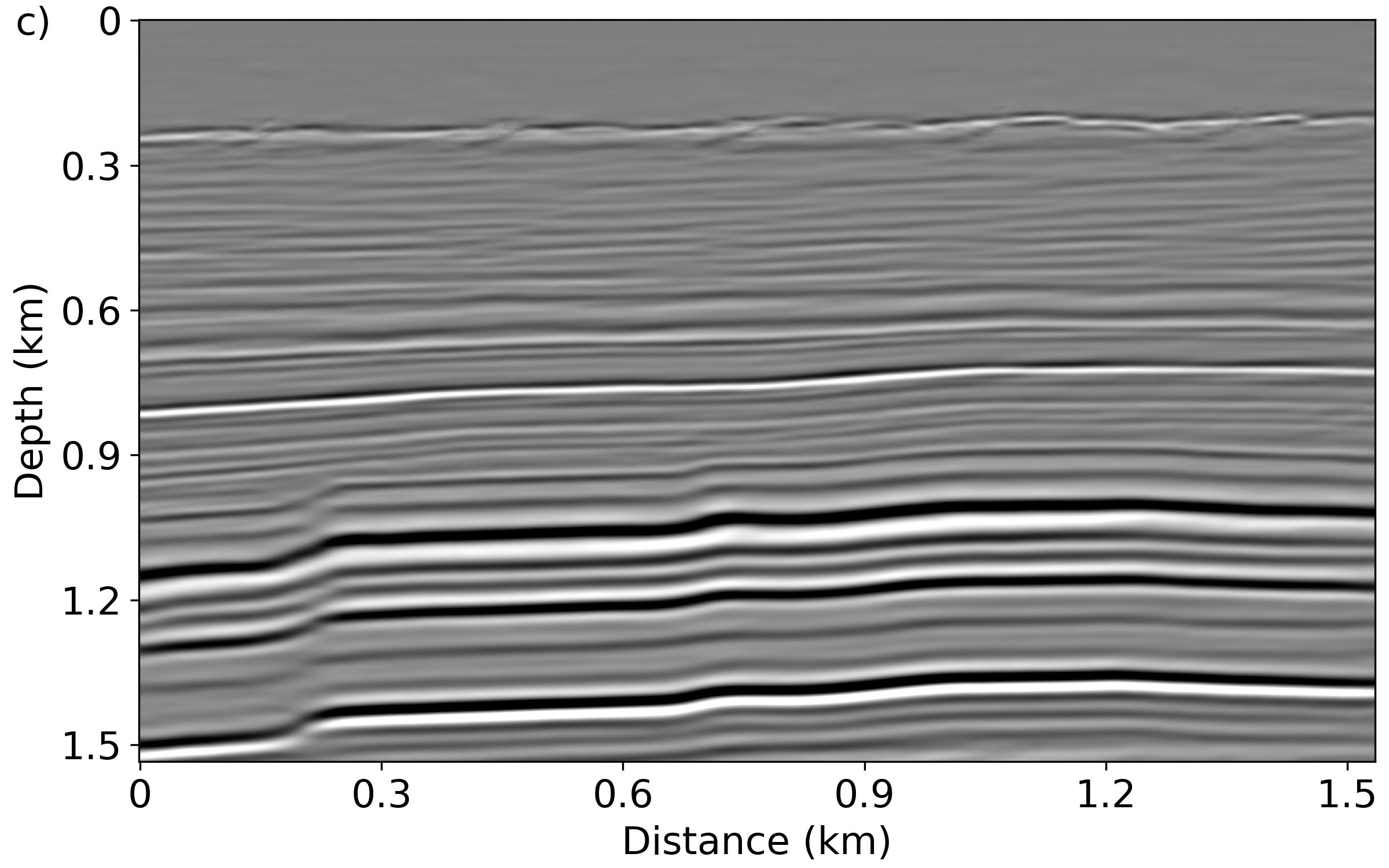}
\hspace{0.3cm}
\includegraphics[width=0.45\textwidth]{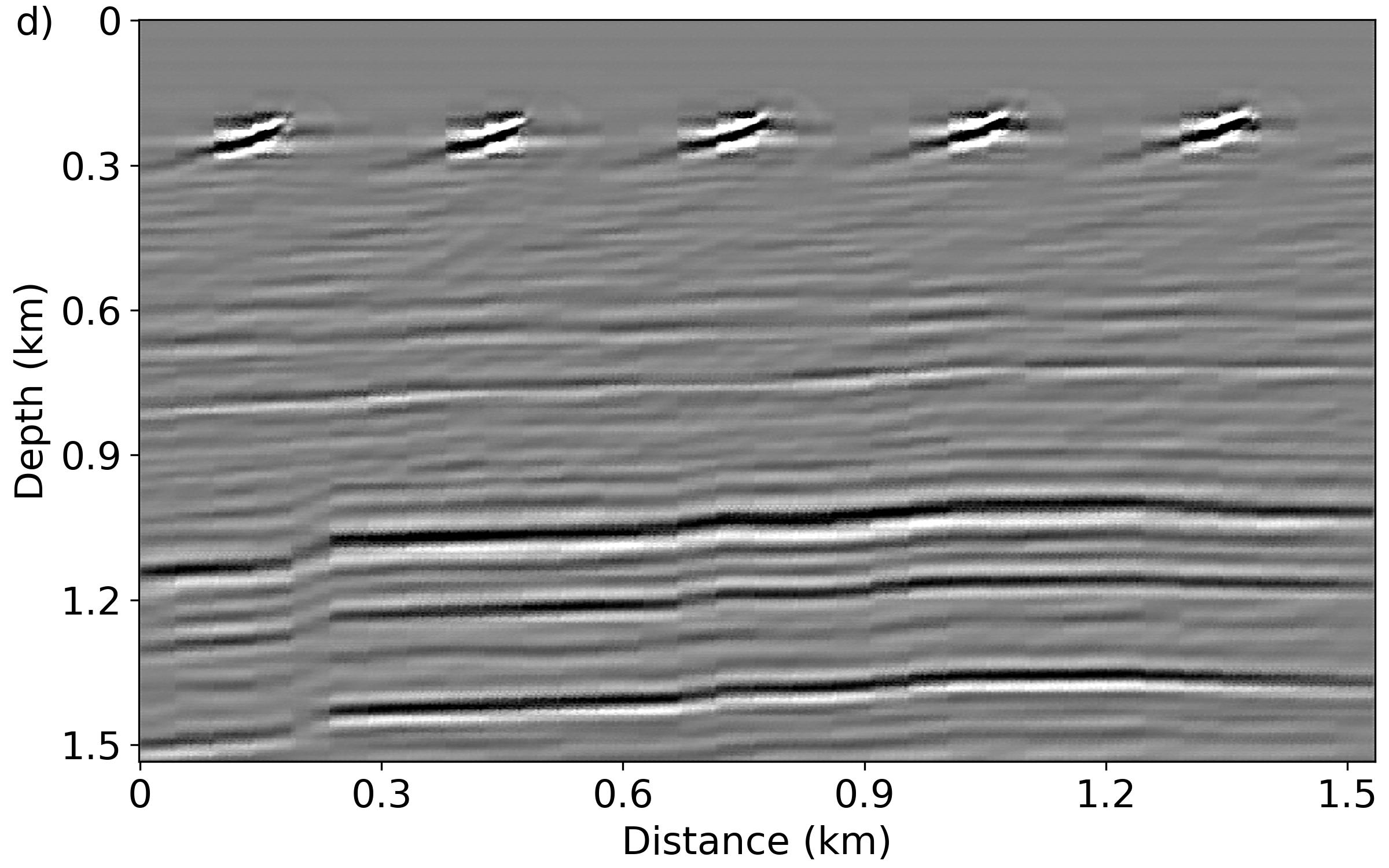}
\vspace{0.3cm}
\includegraphics[width=0.45\textwidth]{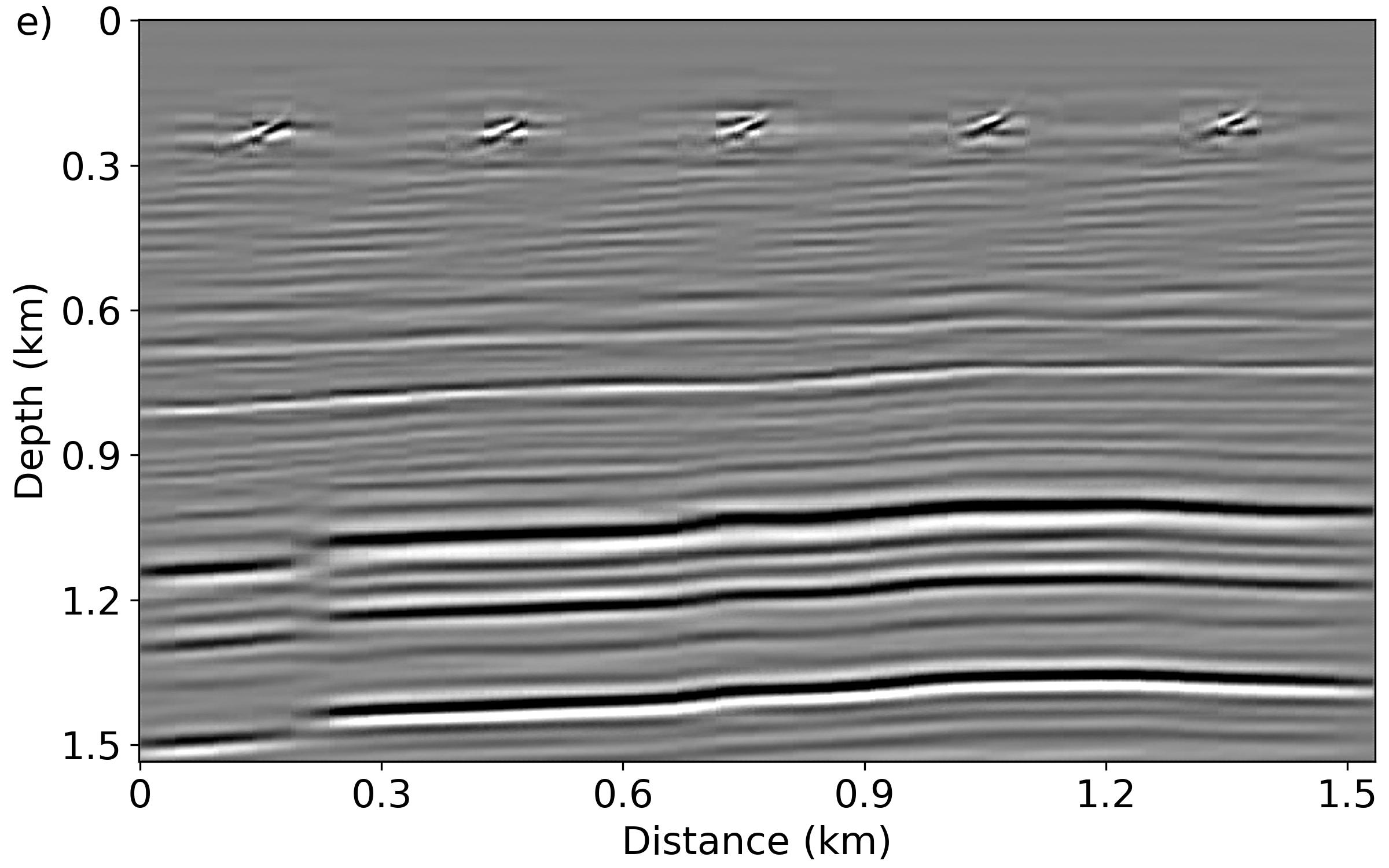}
\hspace{0.3cm}
\includegraphics[width=0.45\textwidth]{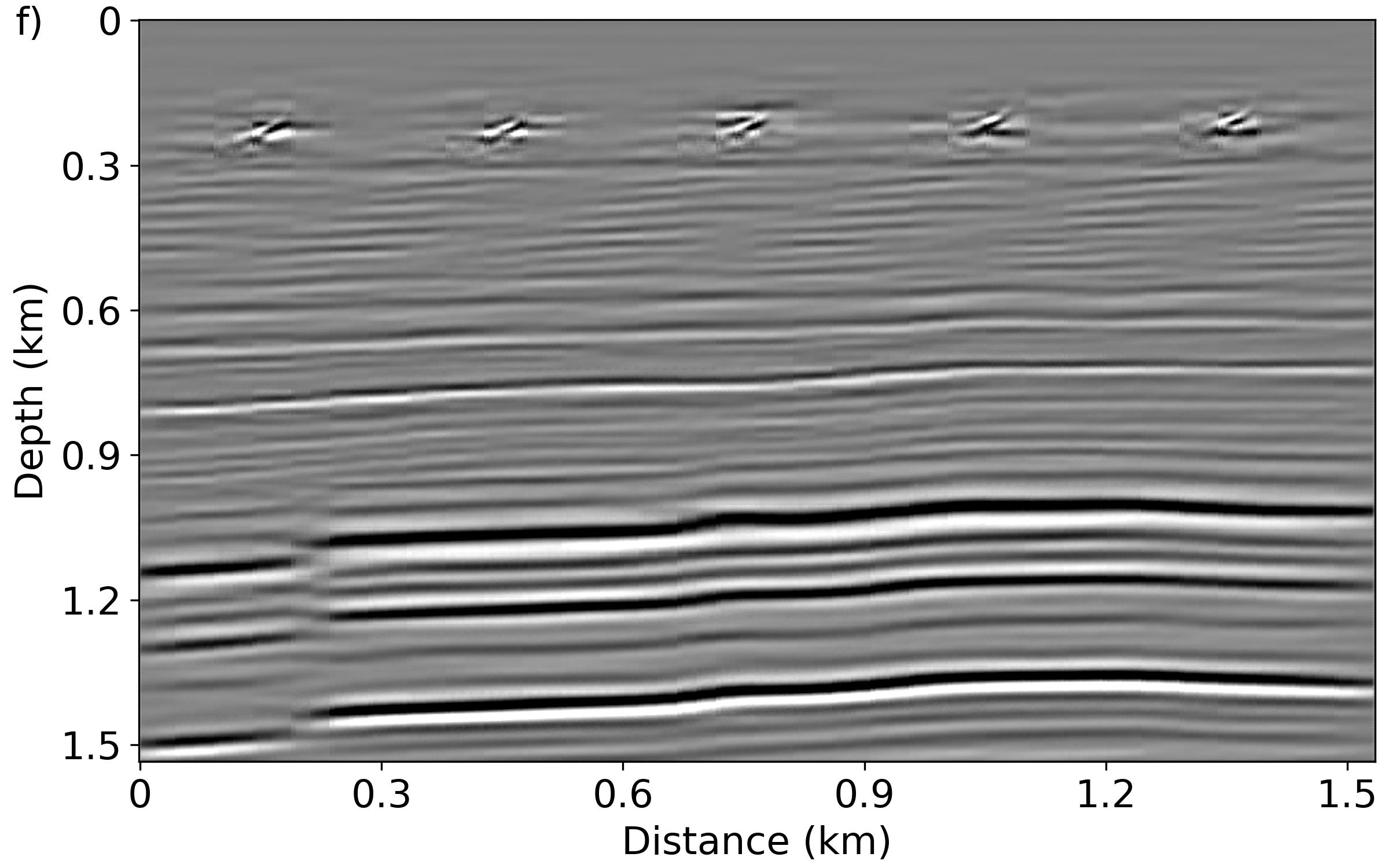}
\caption{Imaging enhancement comparisons of neural networks training with meta-learning initialization and random initialization on synthetic data. (a) Input data and (b) the corresponding label data. (c) is the prediction results to the meta initialization-based neural network with 10 epochs of training. (d), (e), and (f) are the prediction results from random initialization-base neural network with 10, 100, and 300 epochs of training, respectively.}
\label{fig14}
\end{figure}

We further utilize the trained MLIN and RIN to predict the unseen test data, and the results are displayed in Figure \ref{fig14}. As we can see, the original image (see Figure \ref{fig14}a) suffers from poor continuity and is plagued by noise due to the sparse acquisition. In contrast, the image from dense acquisition (see Figure \ref{fig14}b) has high imaging quality. As demonstrated in Figure \ref{fig14}c, the image produced by the MLIN with 10 epochs of training, which achieves an MSE of 2.2e-3 and an MSSSIM of 5.61e-2, exhibits a significant improvement in both events continuity and noise attenuation. These improvements are crucial for accurately interpreting the subsurface structure. However, RIN does not bring noticeable imaging enhancements (see Figure \ref{fig14}d-f). Instead, it disrupts some event continuity in deeper layers. Even after undergoing 300 epochs of gradient descent updates, RIN only reach an MSE of 4.1e-3 and an MSSSIM of 6.7e-2.

\subsubsection{Velocity estimation}
Finally, we evaluate the performance of the Meta-Processing algorithm in a velocity estimation task. Specifically, for each shot gather, we will utilize the trained NNs to predict a root-mean-square velocity (referred to as $V_{rms}$), which is usually measured directly from the seismic data and is often used for normal moveout (NMO) correction. That is, the input to NNs is a single shot gather, and the output is the $V_{rms}$. However, it is almost impossible to directly predict the $V_{rms}$ of the entire model size from a single shot gather. Therefore, we follow Harsuko and Alkhalifah \cite{harsuko2022storseismic} suggestions and extract the predicted $V_{rms}$ laterally from the shot position to half the maximum offset as our result, which is more reliable. Here, we need to emphasize that, in order to guarantee the same dimensions of input and output, we refer Ovcharenko  et al. \cite{ovcharenko2022multi} approach of stretching each model along the depth axis to match the size of the shot-gather data along the temporal dimension. This operation allows the NN architecture to be extended to arbitrary model depths. Furthermore, we randomly sample noise from field data (as we will see later), and inject noise into the synthetic data, since we hope the trained NNs can be better generalize to field data testing. 

\begin{figure}[htp]
\centering
\includegraphics[width=0.35\textwidth]{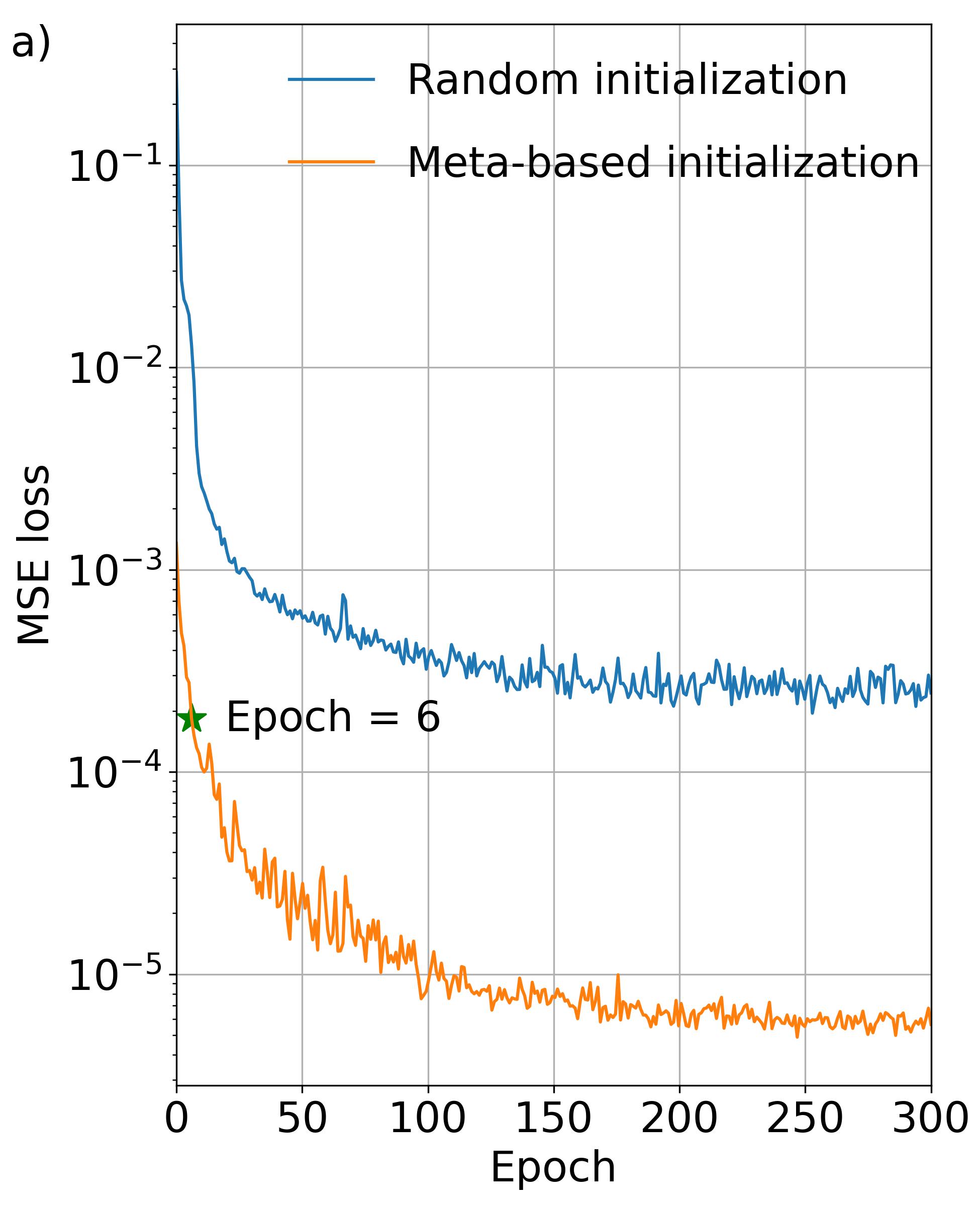}
\hspace{1cm}
\includegraphics[width=0.35\textwidth]{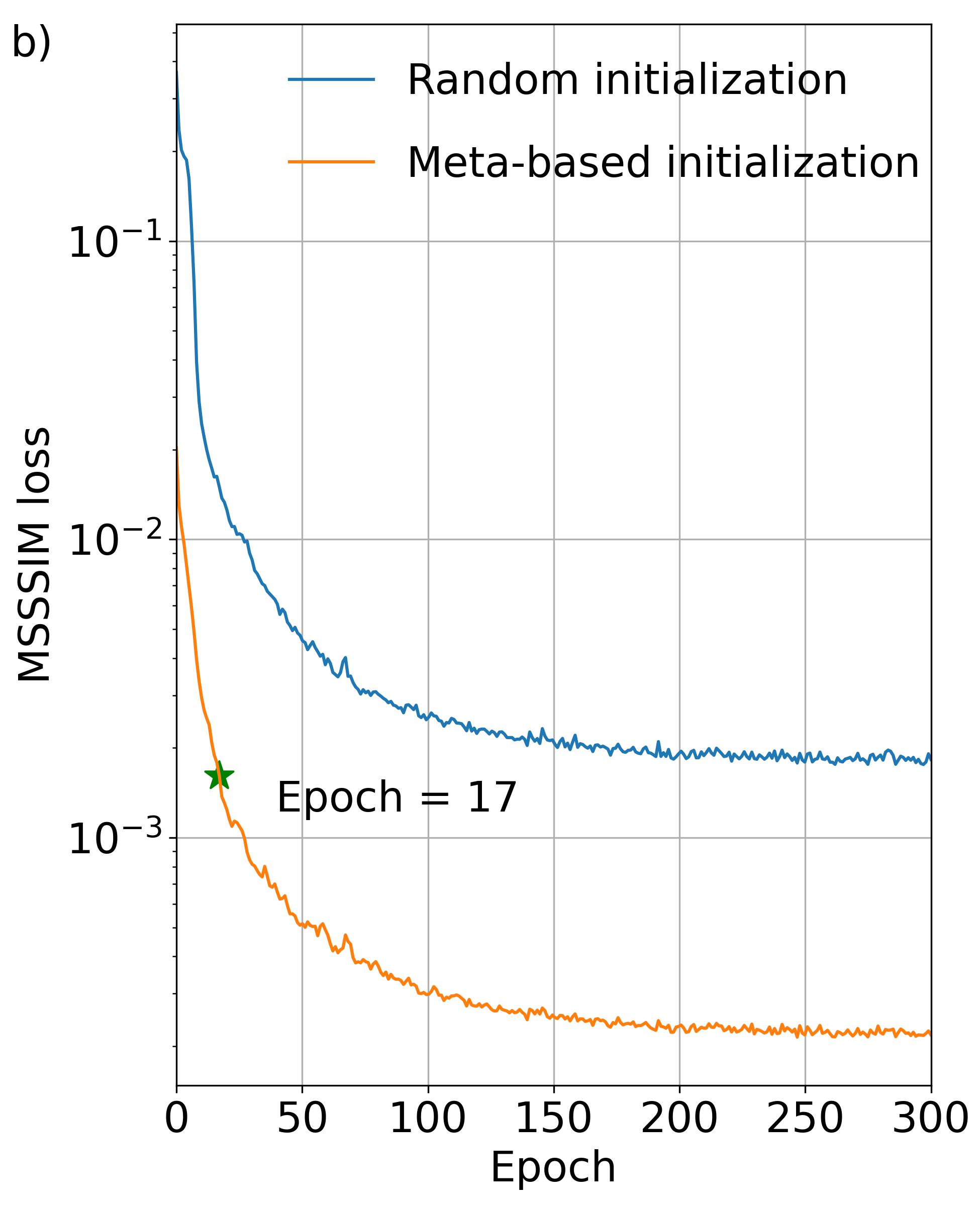}
\caption{The MSE (a) and MSSSIM (b) loss function curves of neural networks training with meta-learning initialization and random initialization of velocity estimation task. }
\label{fig15}
\end{figure}

Figure \ref{fig15} illustrates the MSE and MSSSIM loss curves of the MLIN and RIN trained for 300 epochs in the velocity estimation task. It is evident that our method-driven MLIN results in superior performance over the RIN in terms of convergence speed and accuracy. This outcome is of great significance for practical applications, as we can fine-tune the MLIN with minimal time investment to obtain a reasonably accurate $V_{rms}$, which can be quickly applied to other SPTs such as NMO correction. 

\begin{figure}[htp]
\centering

\includegraphics[width=0.45\textwidth]{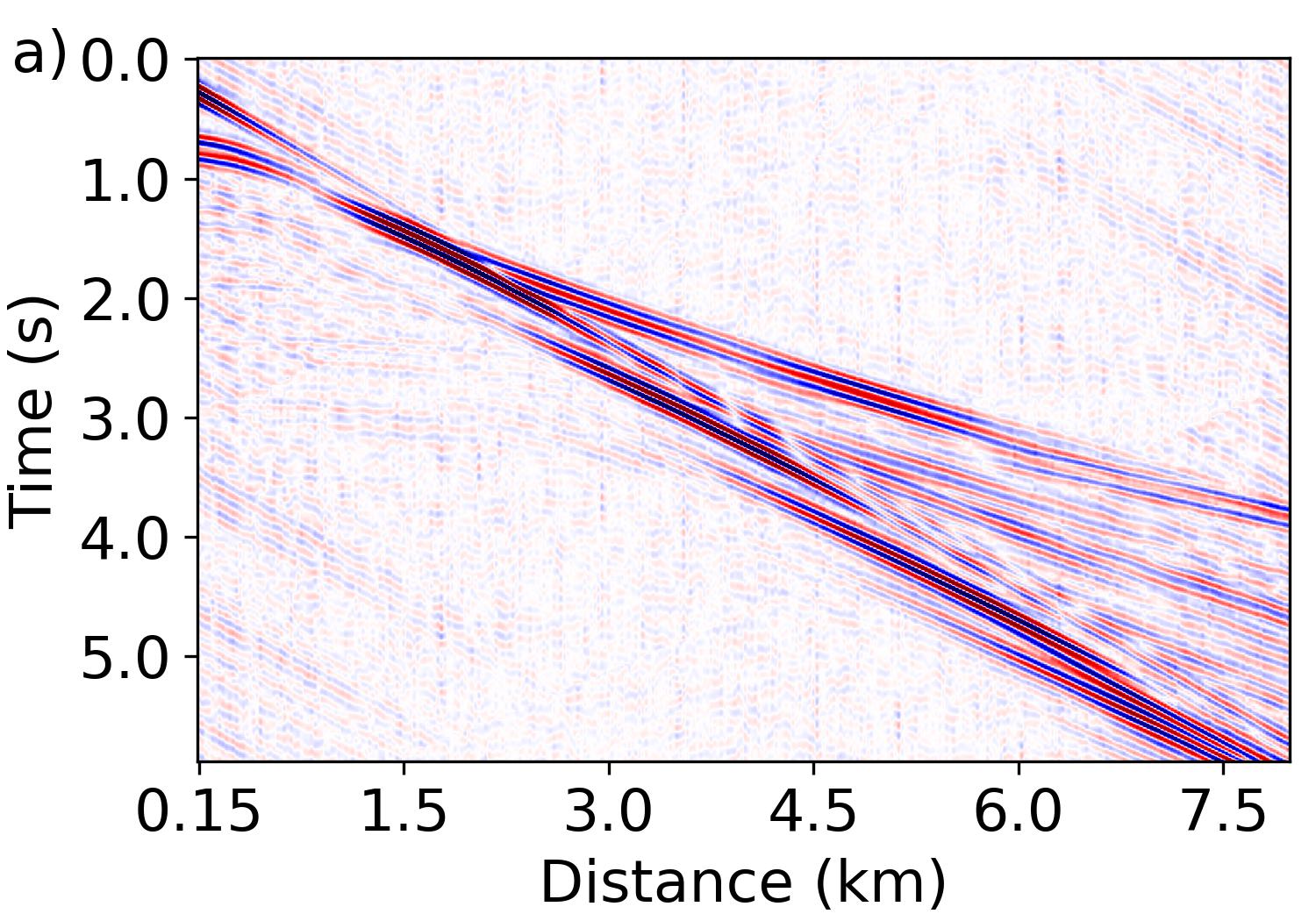}
\hspace{0.3cm}
\includegraphics[width=0.45\textwidth]{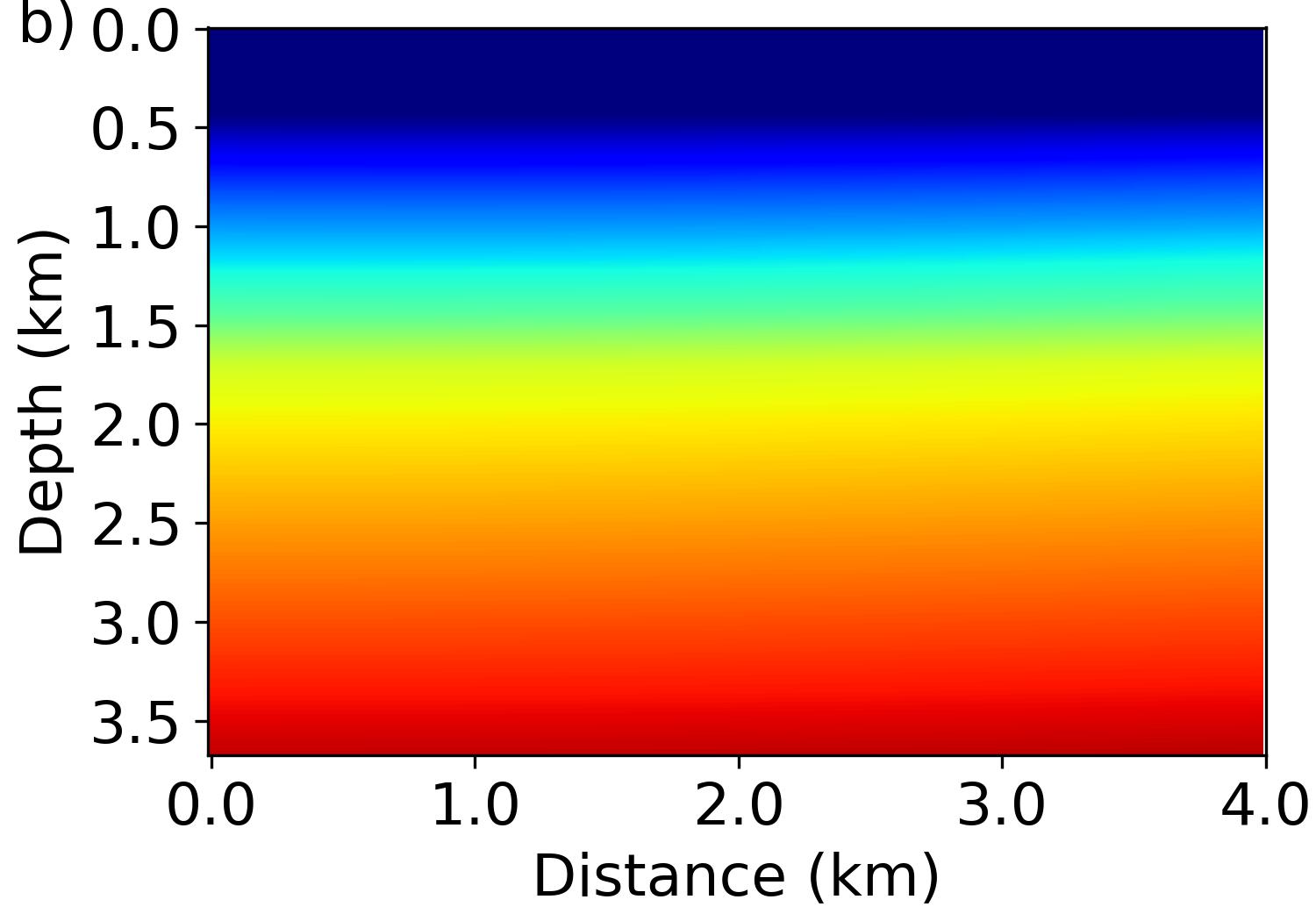}
\includegraphics[width=0.45\textwidth]{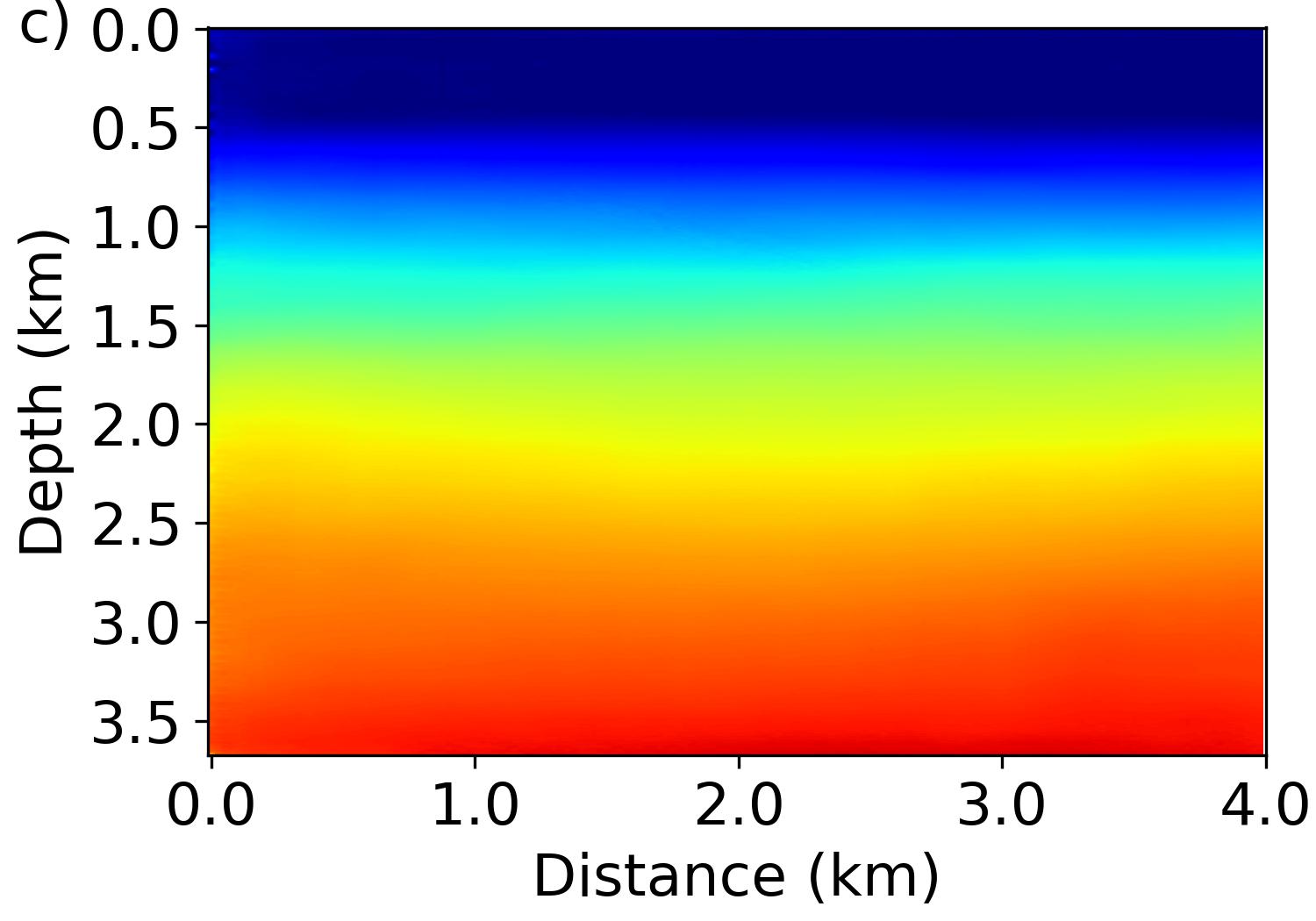}
\hspace{0.3cm}
\includegraphics[width=0.45\textwidth]{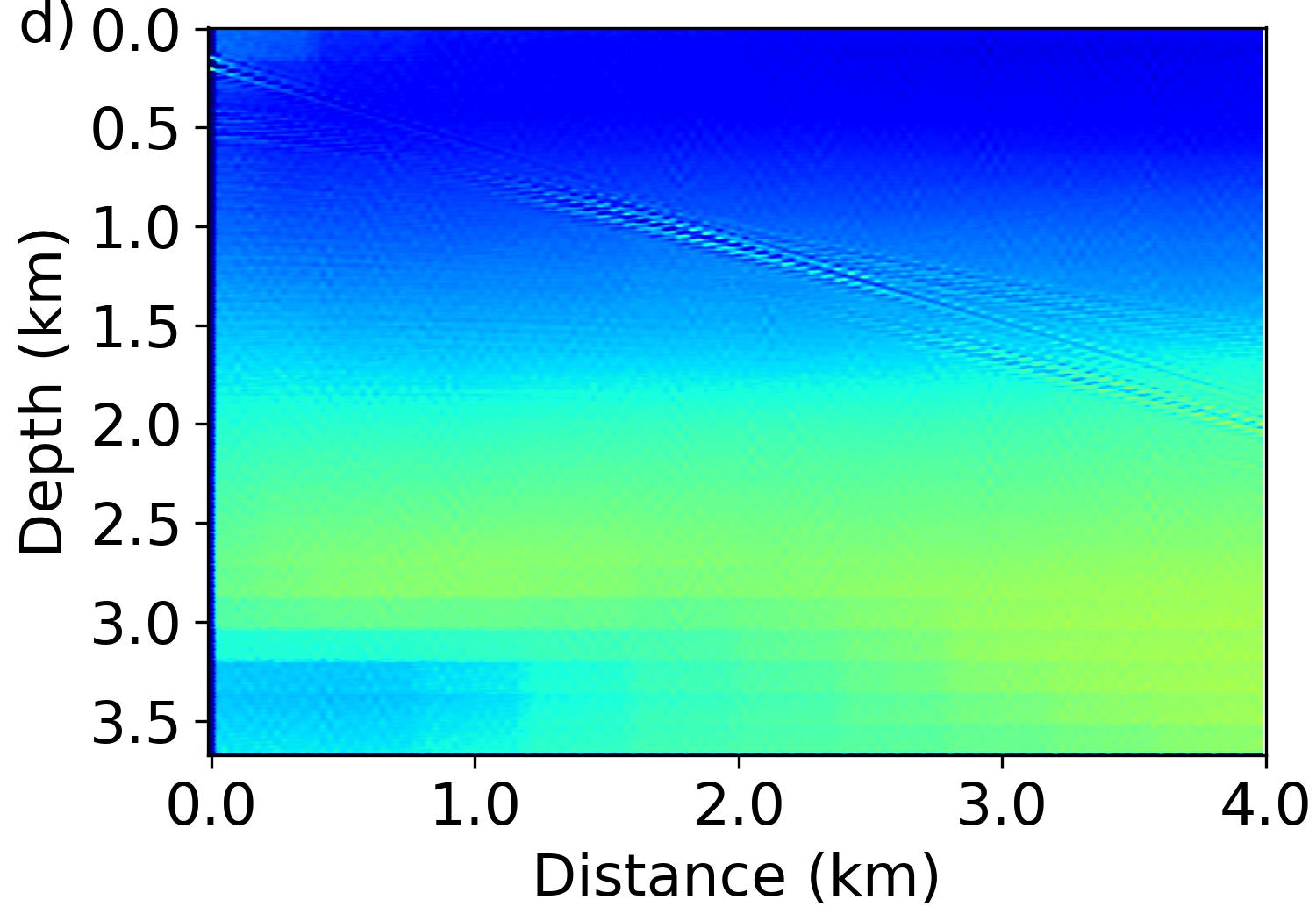}
\includegraphics[width=0.45\textwidth]{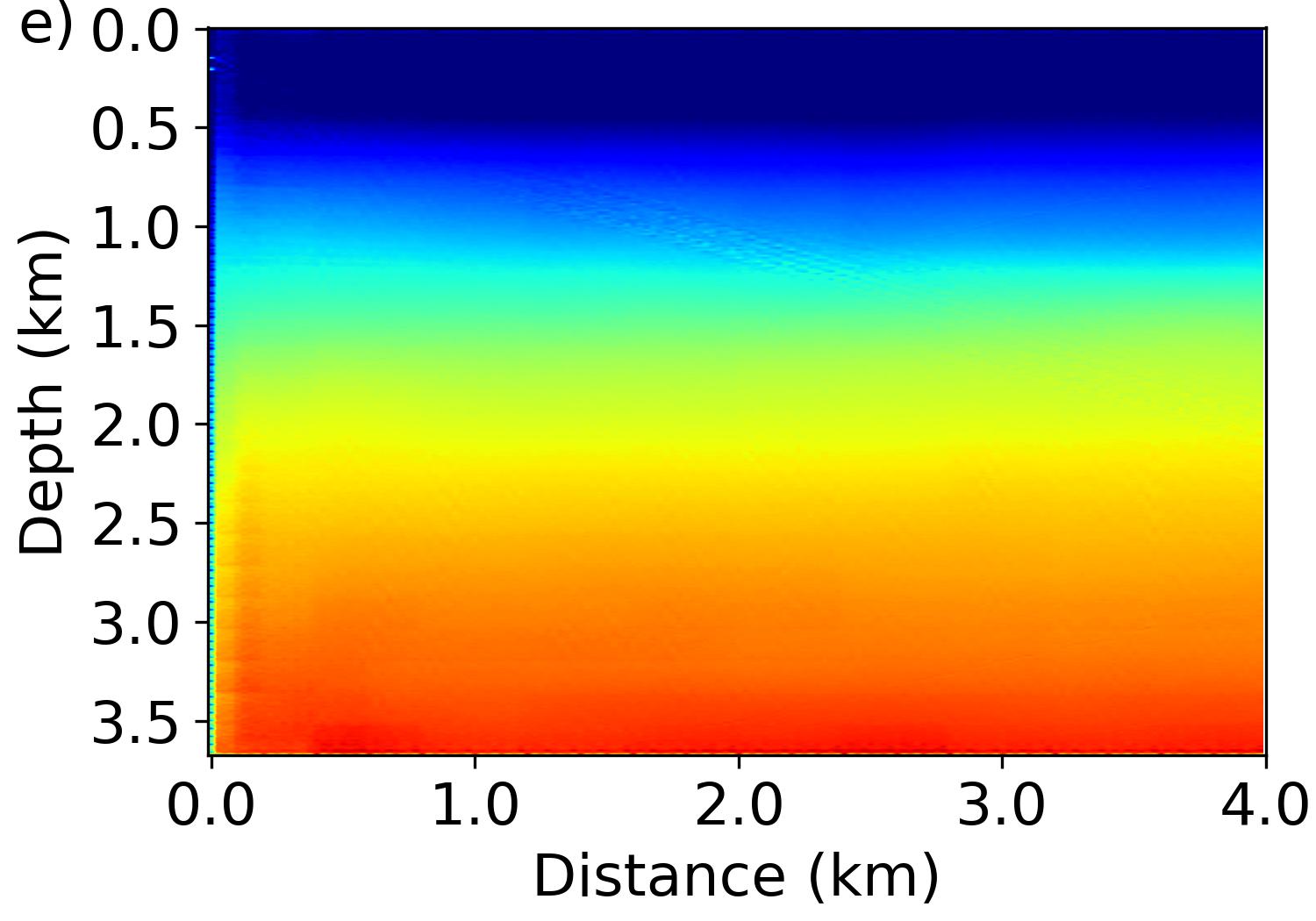}
\hspace{0.3cm}
\includegraphics[width=0.45\textwidth]{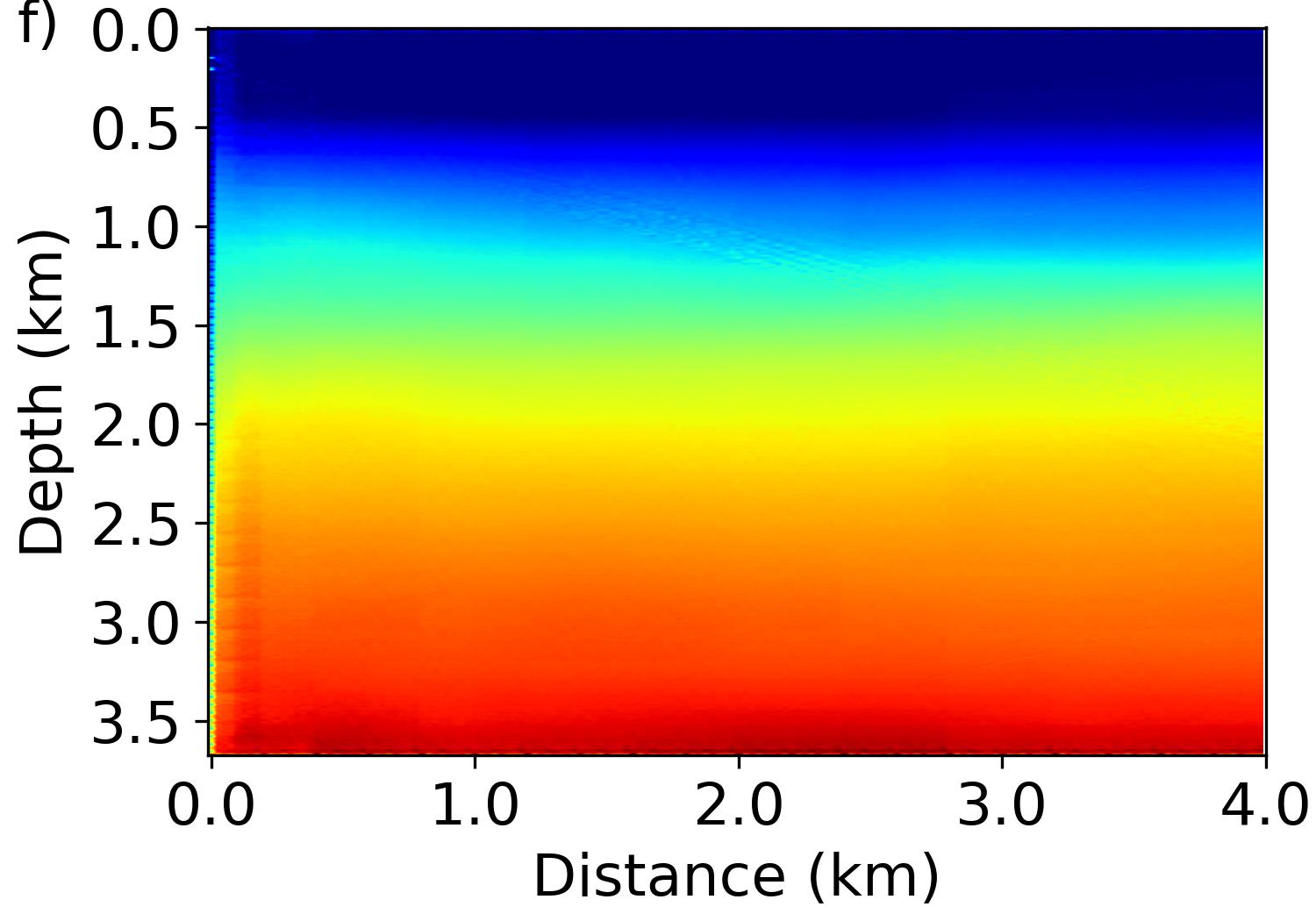}
\includegraphics[width=0.35\textwidth]{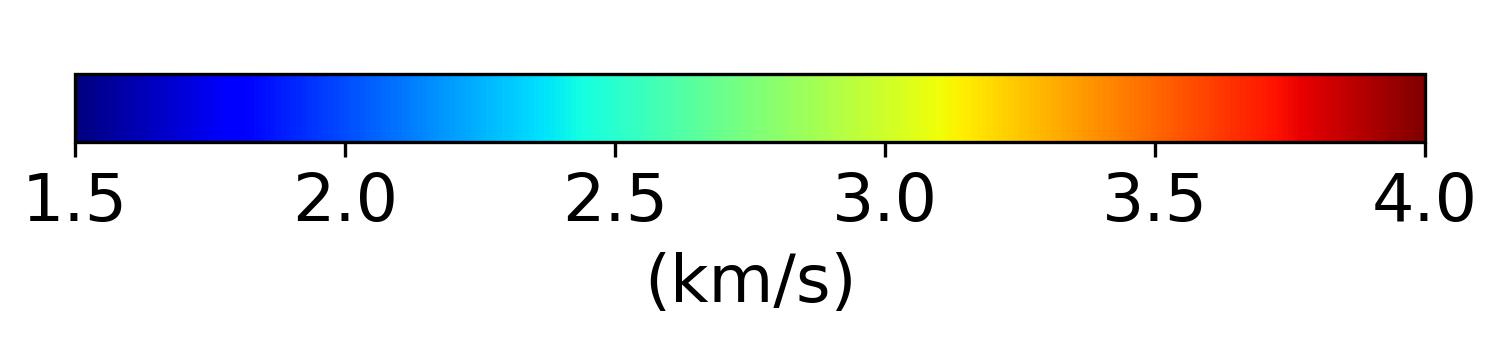}
\caption{Velocity estimation comparisons of neural networks training with meta-learning initialization and random initialization on synthetic data. (a) Input seismic data and (b) the corresponding $V_{rms}$ model. (c) is the predicted $V_{rms}$ from the meta initialization-based neural network with 20 epochs of training. (d), (e), and (f) are the predicted $V_{rms}$ from random initialization-base neural network with 20, 100, and 300 epochs of training, respectively.}
\label{fig16}
\end{figure}

The prediction results of the MLIN and RIN on the test set are presented in Figure \ref{fig16}, where panels (a) and (b) are the input and ground truth, respectively, (c) comes from the prediction result of the MLIN after 20 epochs of training, while (d), (e), and (f) correspond to the prediction results of the RIN trained for 20, 100, and 300 epochs, respectively. We can observe that the MLIN, which only requires 20 epochs of optimization, achieves very close prediction results to the ground truth, with a mean absolute error (MAE) of 54.5 m/s. However, the prediction results of the RIN trained for 20 epochs have large errors and contain a lot of signal artifacts. Although the accuracy of the RIN gradually improves with the increase in the number of epochs, there are still signal artifacts and noise, as seen in Figures \ref{fig16}e-f. Even with 300 epochs of gradient descent updates, the RIN only achieves an MAE of 68.5 m/s, which is still higher than the error of MLIN trained for only 20 epochs.

\subsection{Field data}
In the following, we will utilize the MLIN and RIN, previously trained on synthetic data, to directly predict field data, including seismic denoising, imaging enhancement, and velocity estimation tasks. To improve the networks' generalization capability to the field data, we normalize the amplitude of the field data using mean amplitude normalization instead of the traditional maximum value normalization method. Visually, this means that the normalized field data and the synthetic data have similar amplitudes for most events within the same display range. This step helps in bringing the feature distribution of the field data closer to that of the training synthetic data set. 

\subsubsection{Denoising}
\begin{figure}[htp]
\centering
\includegraphics[width=0.23\textwidth]{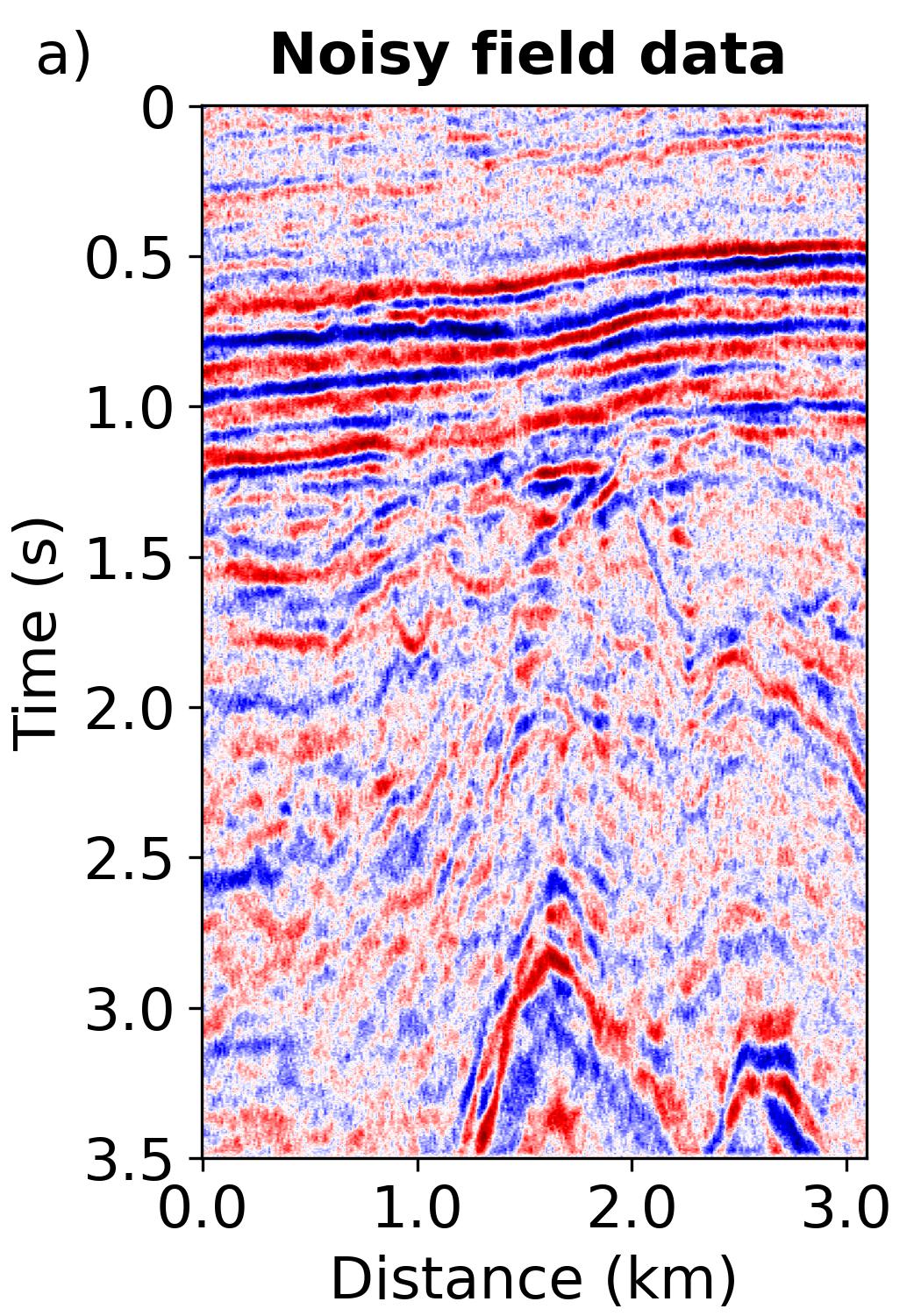} \\
\includegraphics[width=0.23\textwidth]{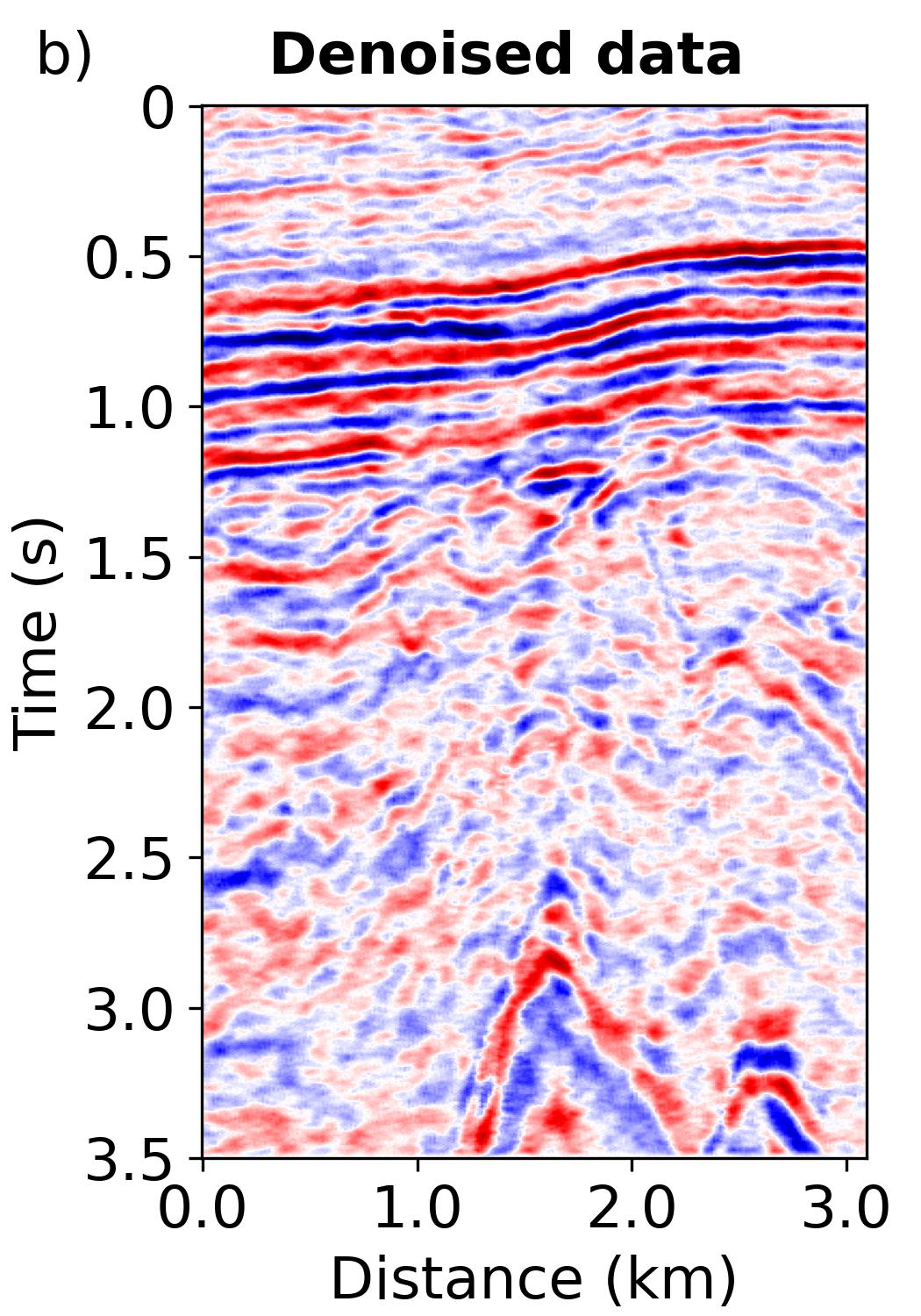}
\includegraphics[width=0.23\textwidth]{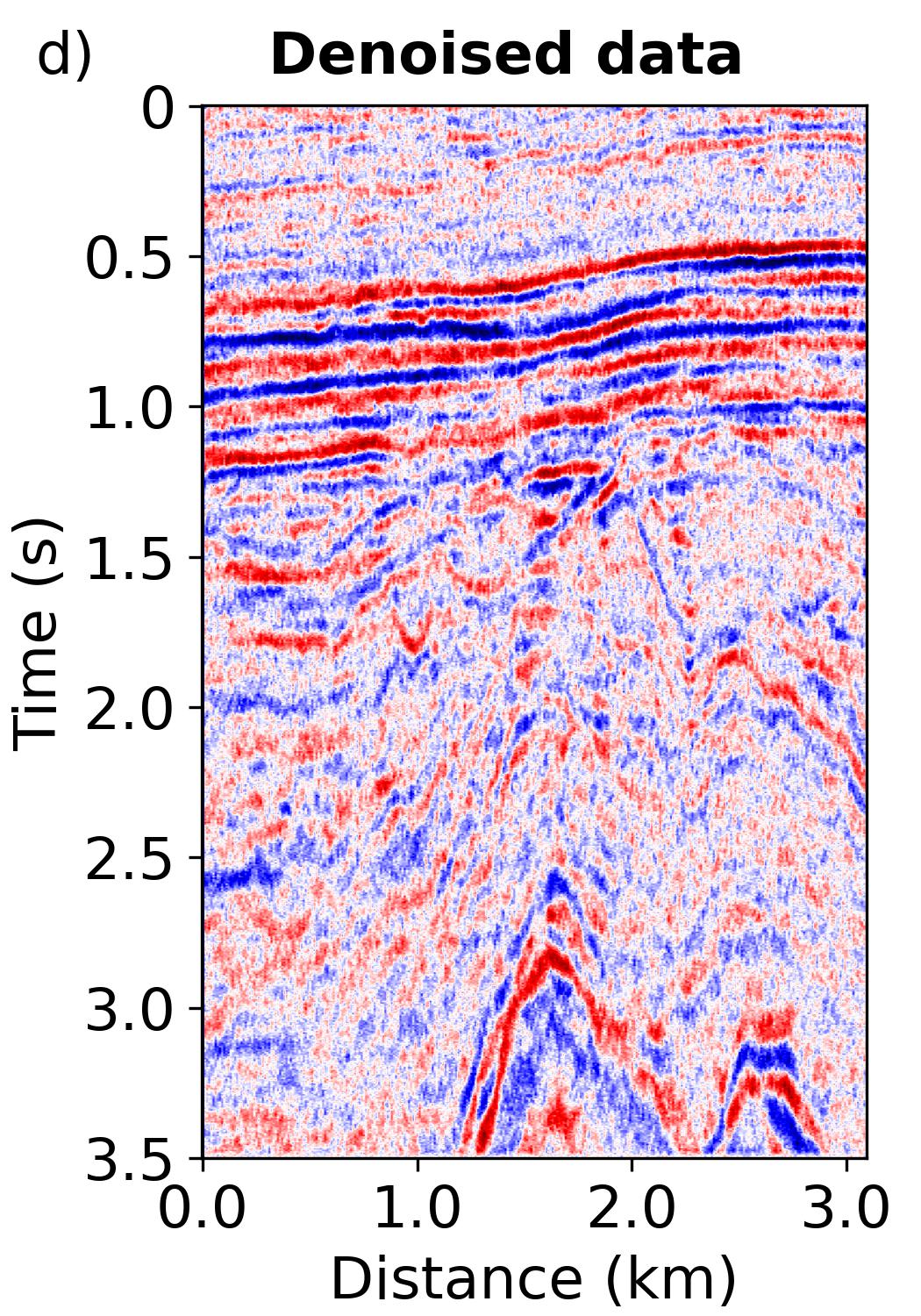}
\includegraphics[width=0.23\textwidth]{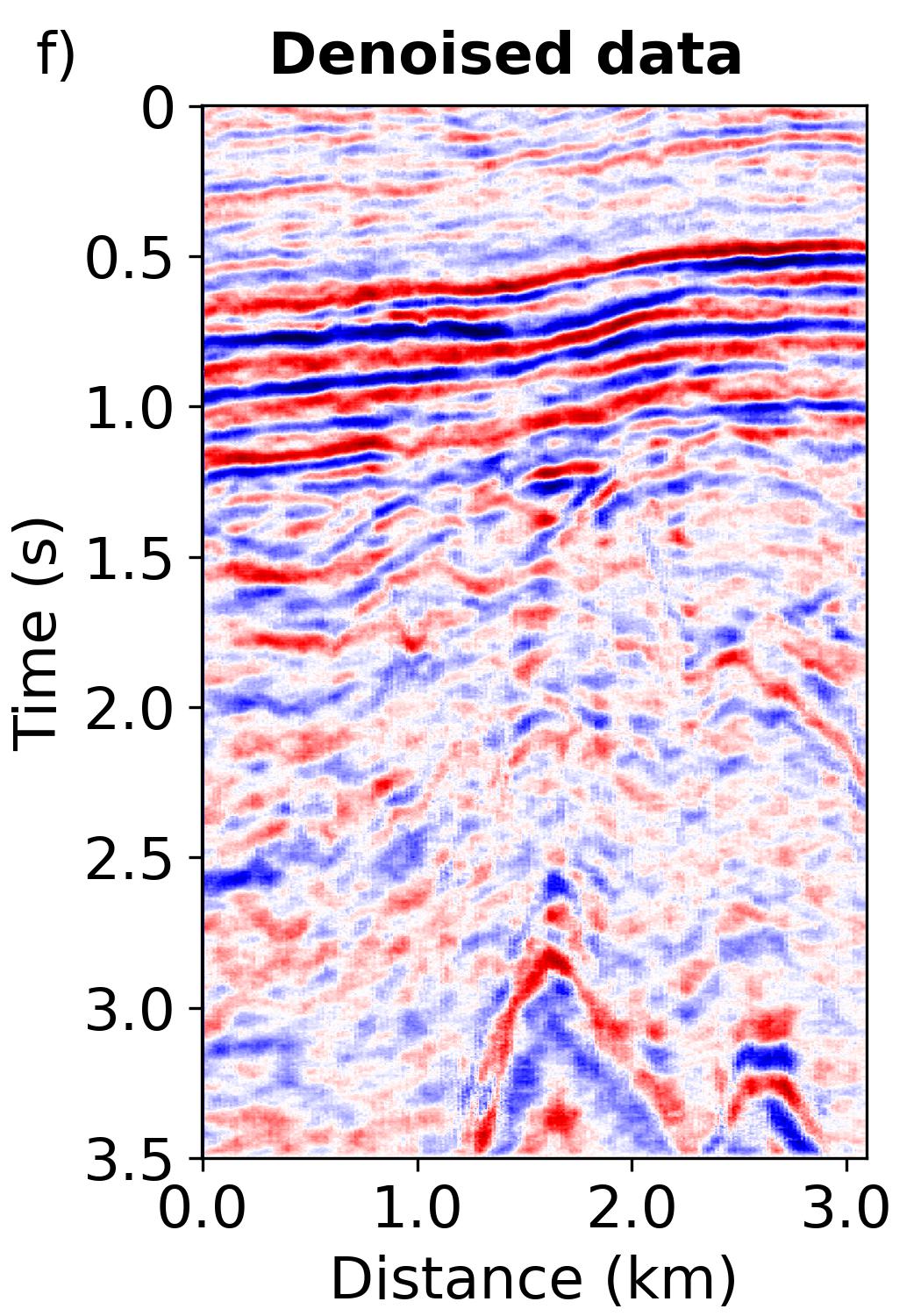} 
\includegraphics[width=0.23\textwidth]{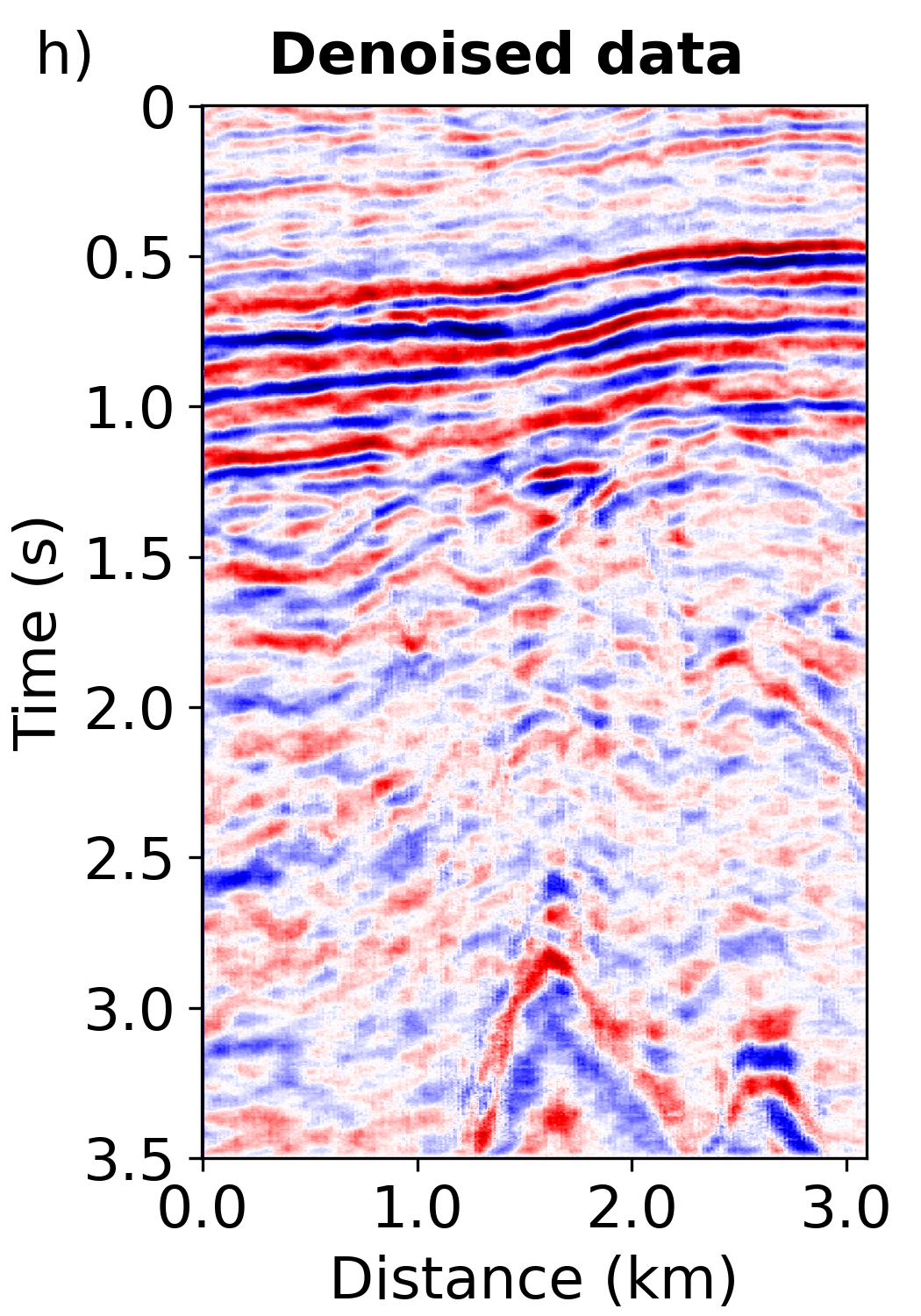} \\
\includegraphics[width=0.23\textwidth]{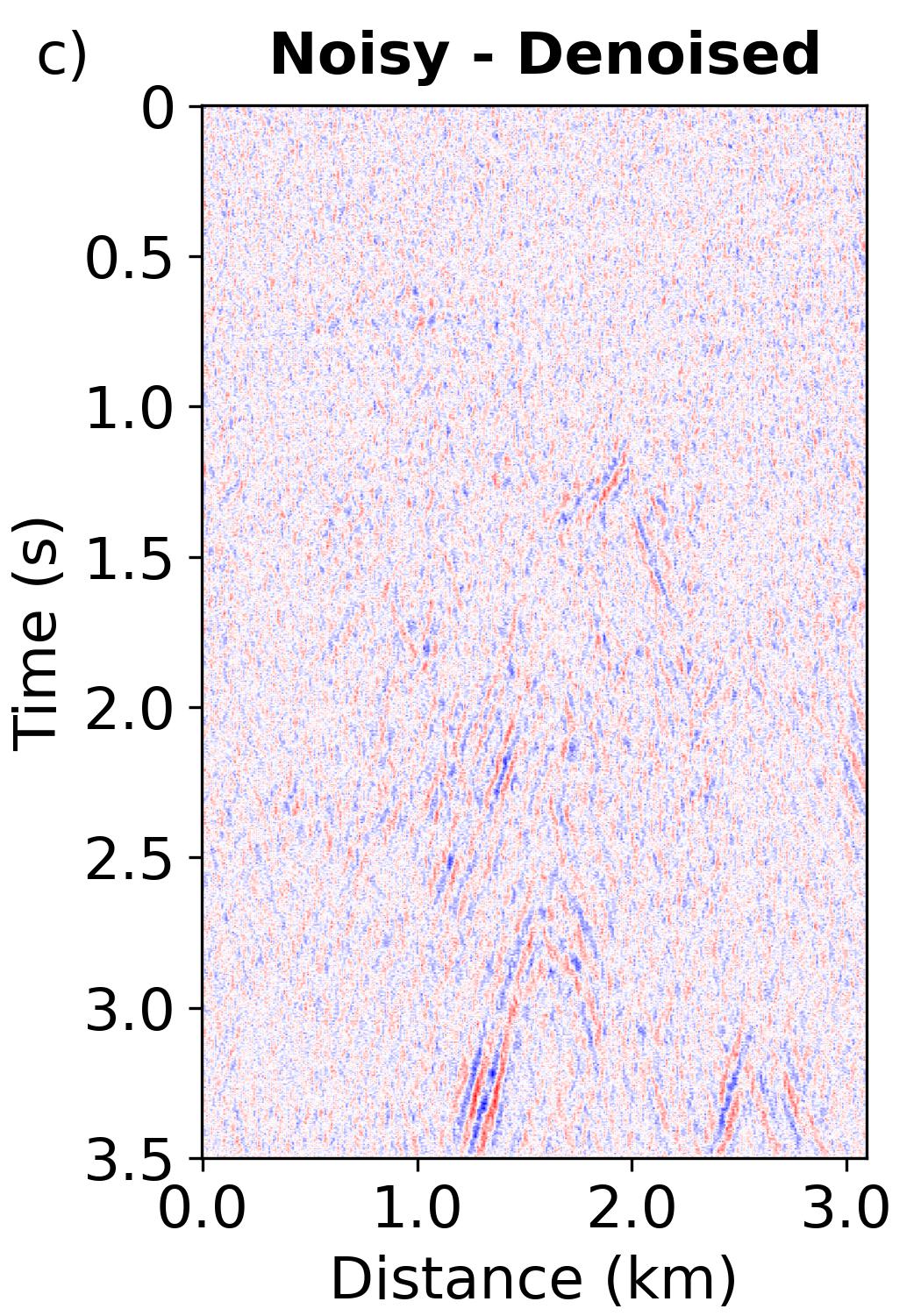} 
\includegraphics[width=0.23\textwidth]{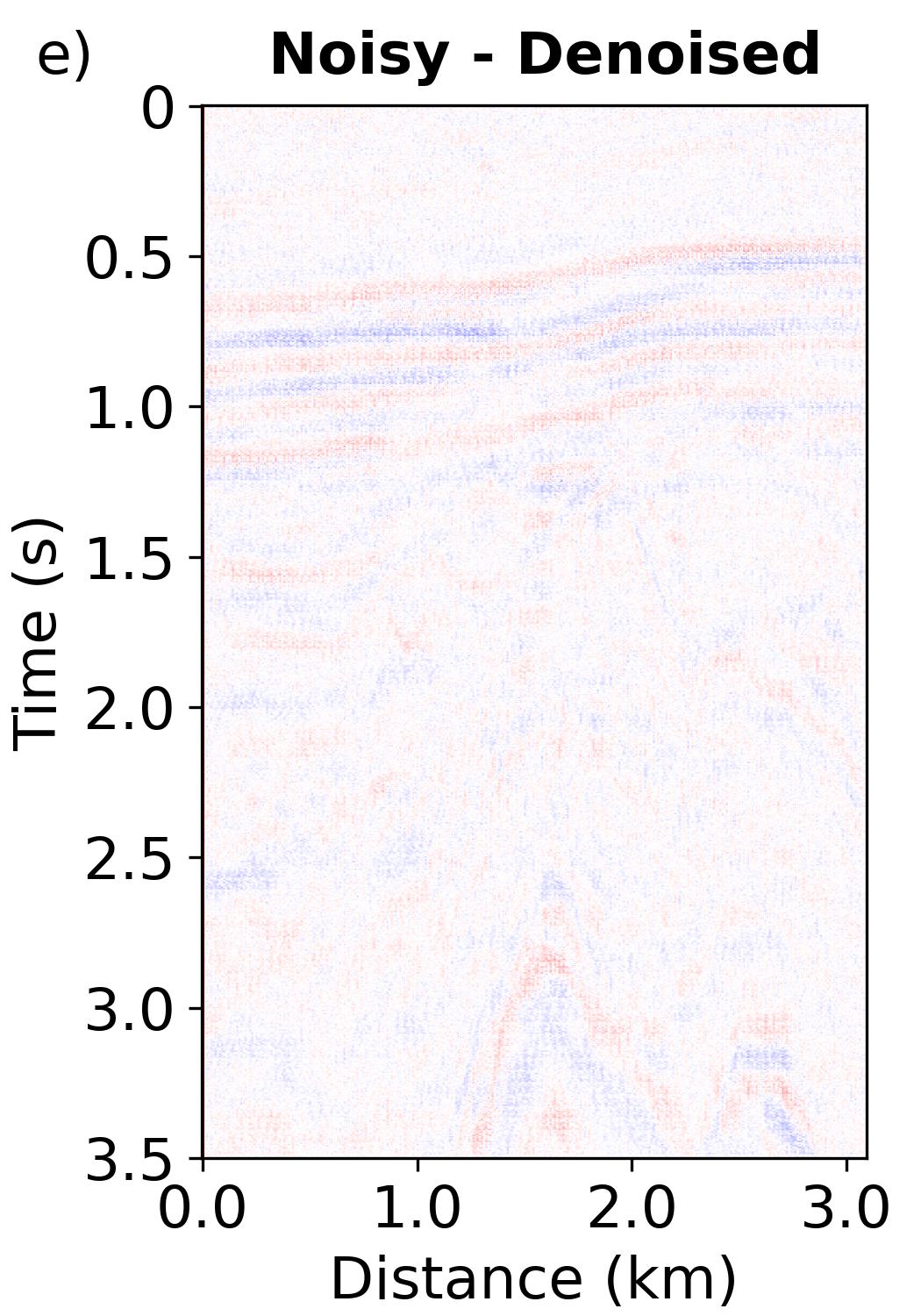} 
\includegraphics[width=0.23\textwidth]{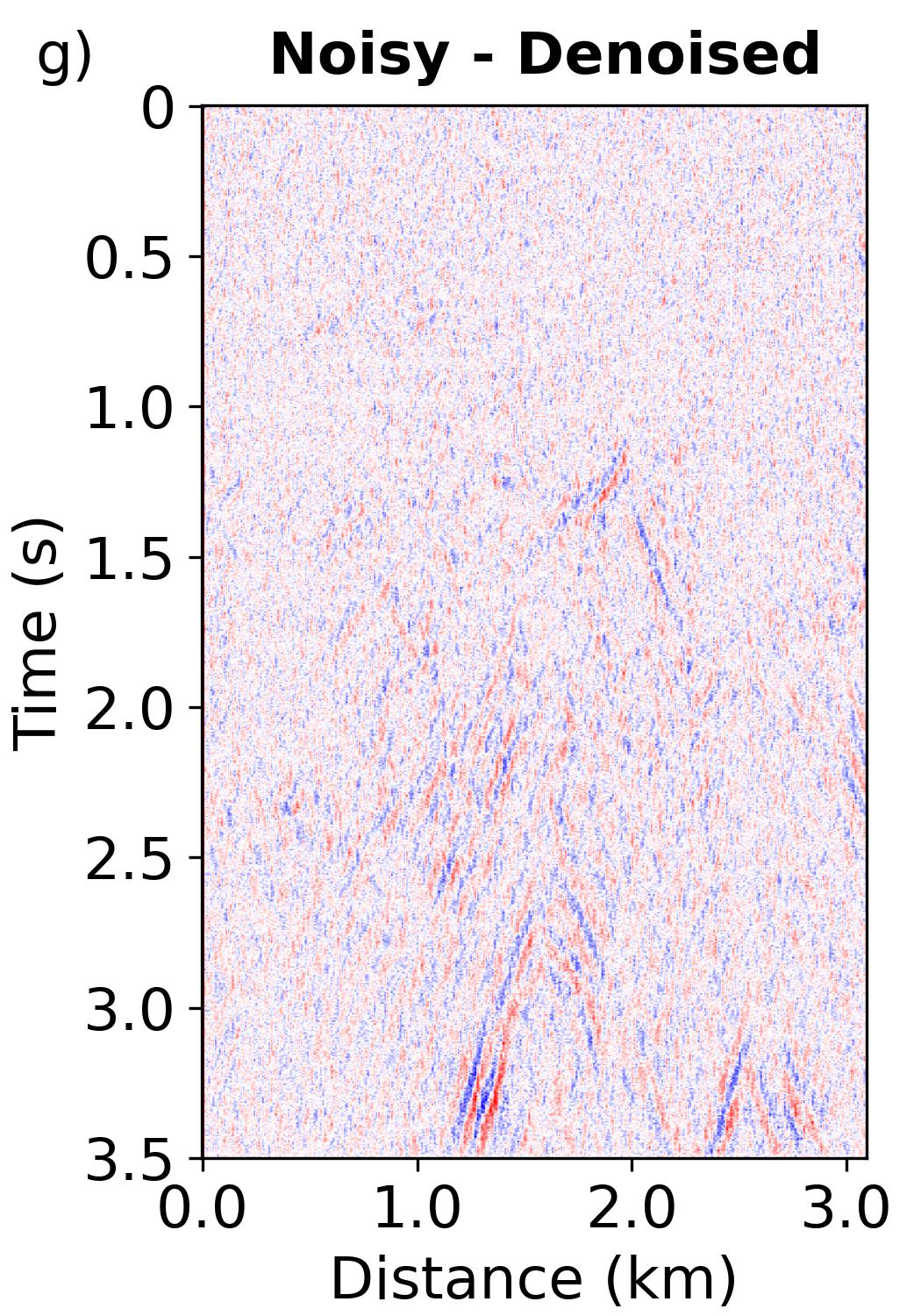}
\includegraphics[width=0.23\textwidth]{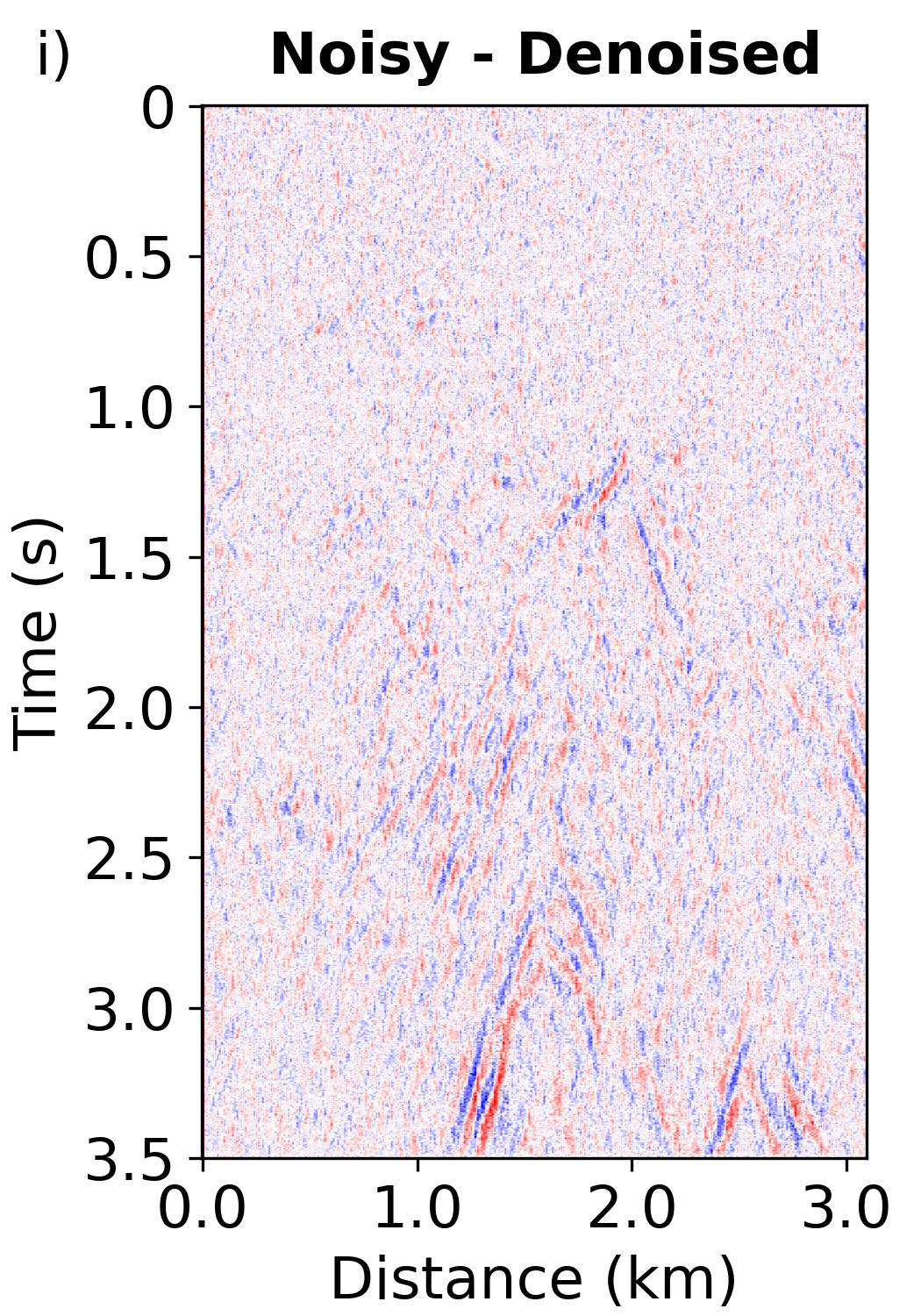}
\caption{Denoising inference on field data comparisons of neural networks trained with meta-learning initialization and random initialization on synthetic data. (a) Input noisy field data. The second rows are the denoising results from the meta initialization-based neural network and random initialization-based neural network, in which, (b) is the prediction results from the meta initialization-based neural network with 10 epoch of training; (d), (f), and (h) are the prediction results from random initialization-base neural network with 10, 100, and 300 epochs of training, respectively. The third row shows the difference between the denoised products and the raw noisy data.}
\label{fig17}
\end{figure}

The MLIN  (MetaL initialization) and RIN (random initialization) are used to denoise a China land dataset, which is known to be polluted by random noise. The resulting denoised results are  displayed in Figure \ref{fig17}. The Meta-Processing algorithm enables the MLIN remove more noise and preserve more effective signals on field data with only 10 epochs of training, as demonstrated by the differences between the original and denoised datasets. On the other hand, RIN requires a large number of epochs of optimization to achieve a slight improvement in denoising performance. Furthermore, some artifacts are introduced in the prediction results of RIN after 100 and 300 epochs of training (see Figure \ref{fig17}f,h), severely contaminating the signal.

\subsubsection{Imaging enhancement}
\begin{figure}[htp]
\centering
\includegraphics[width=0.45\textwidth]{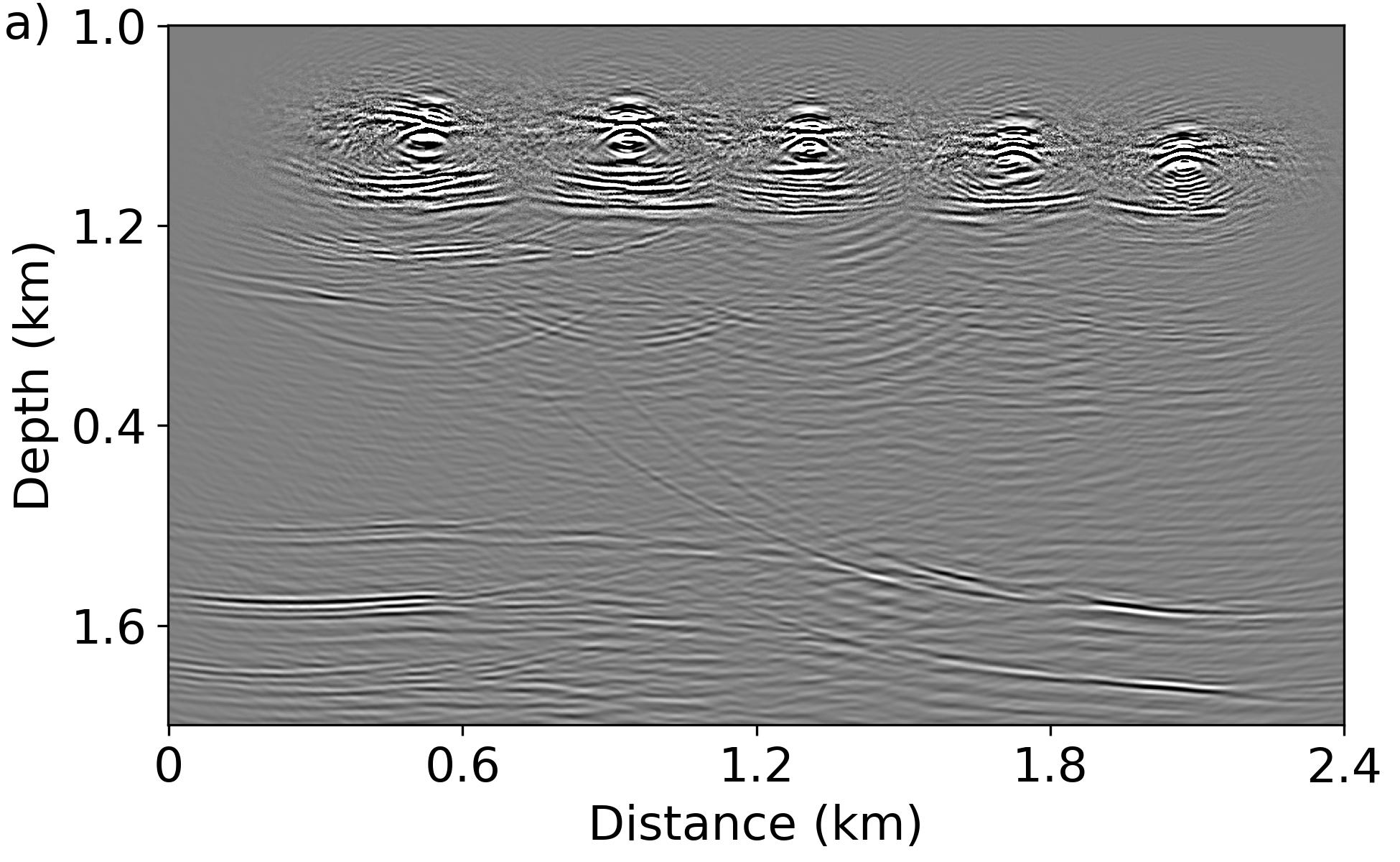} \\
\includegraphics[width=0.45\textwidth]{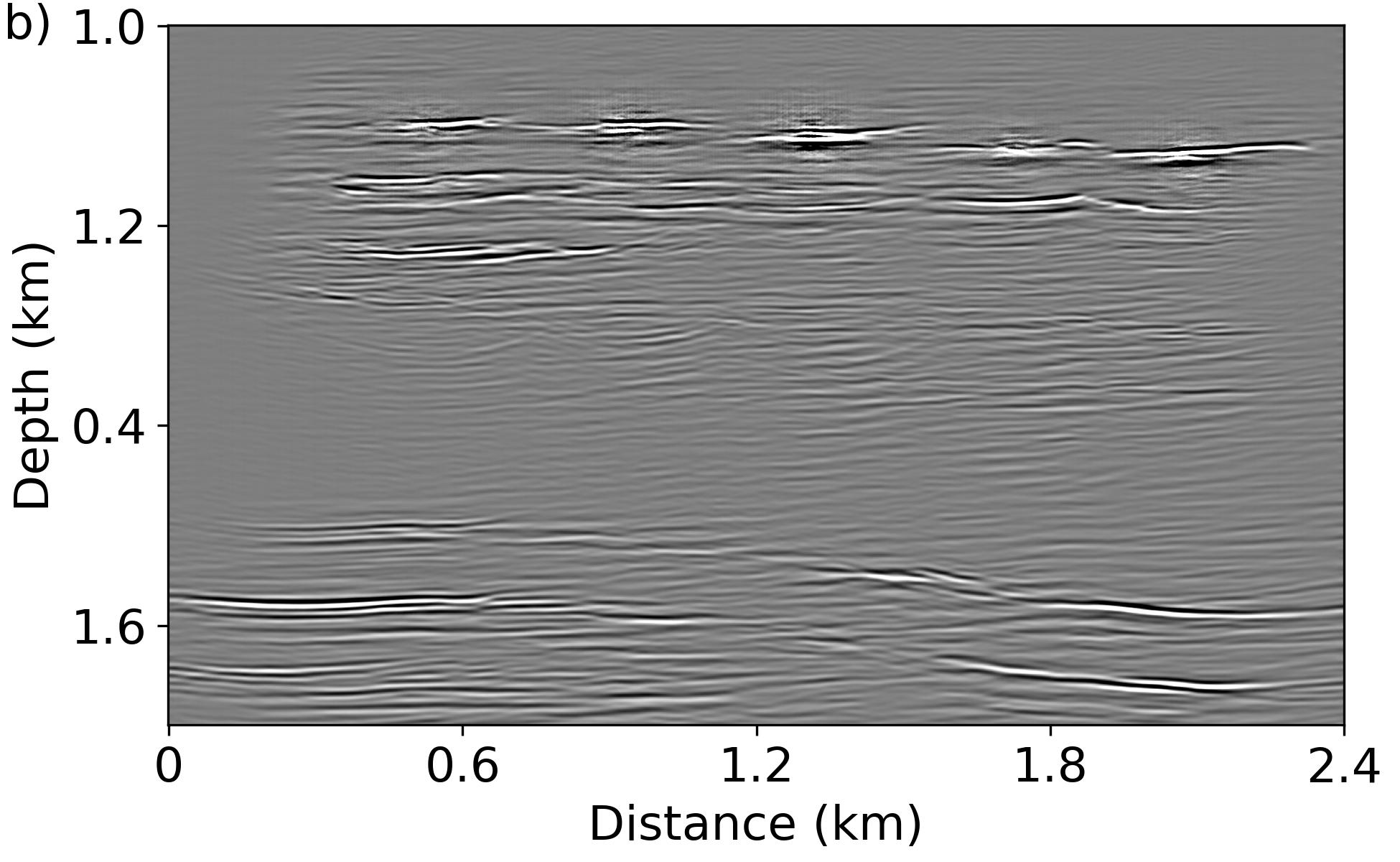} 
\hspace{0.3cm}
\includegraphics[width=0.45\textwidth]{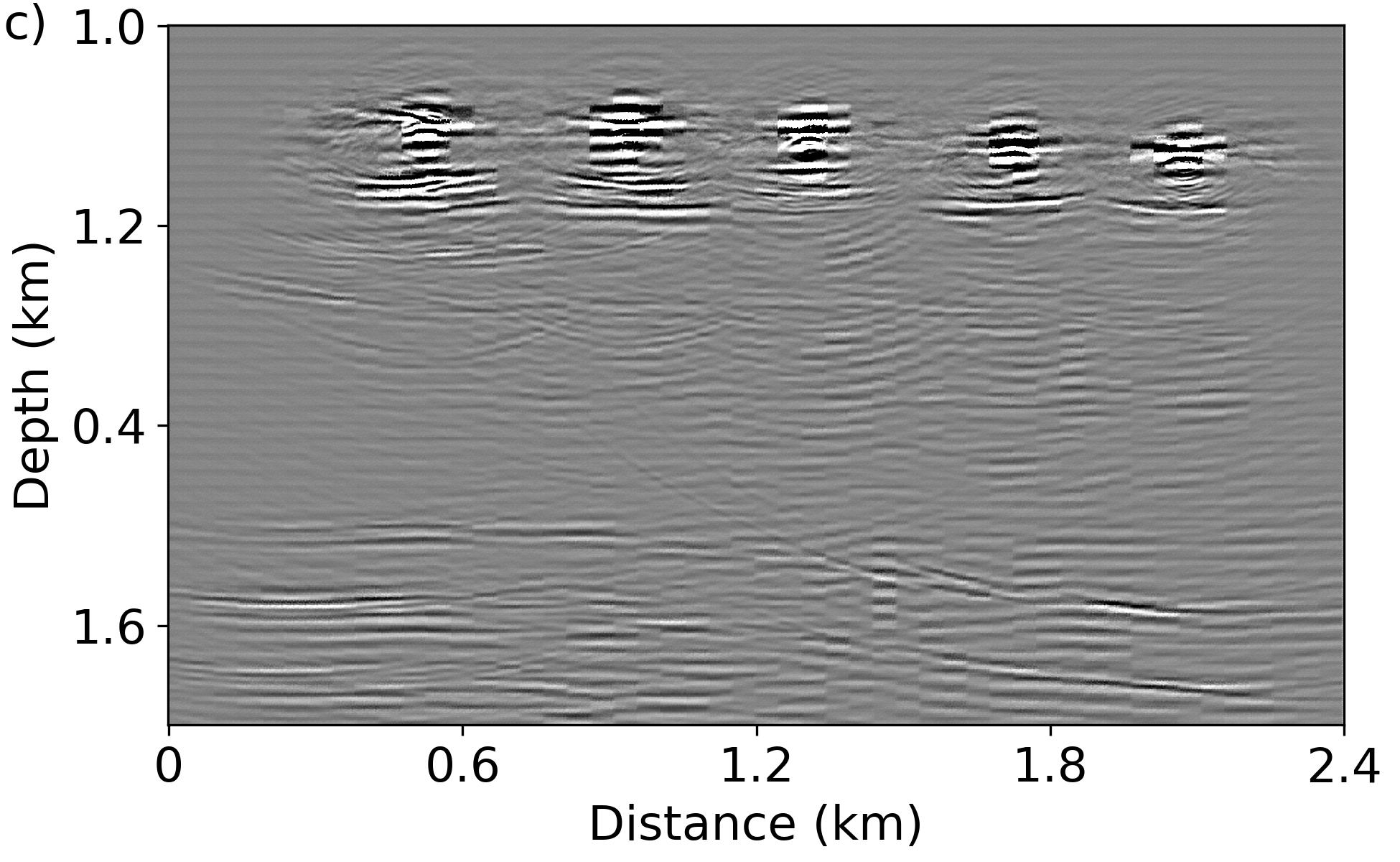} \\
\includegraphics[width=0.45\textwidth]{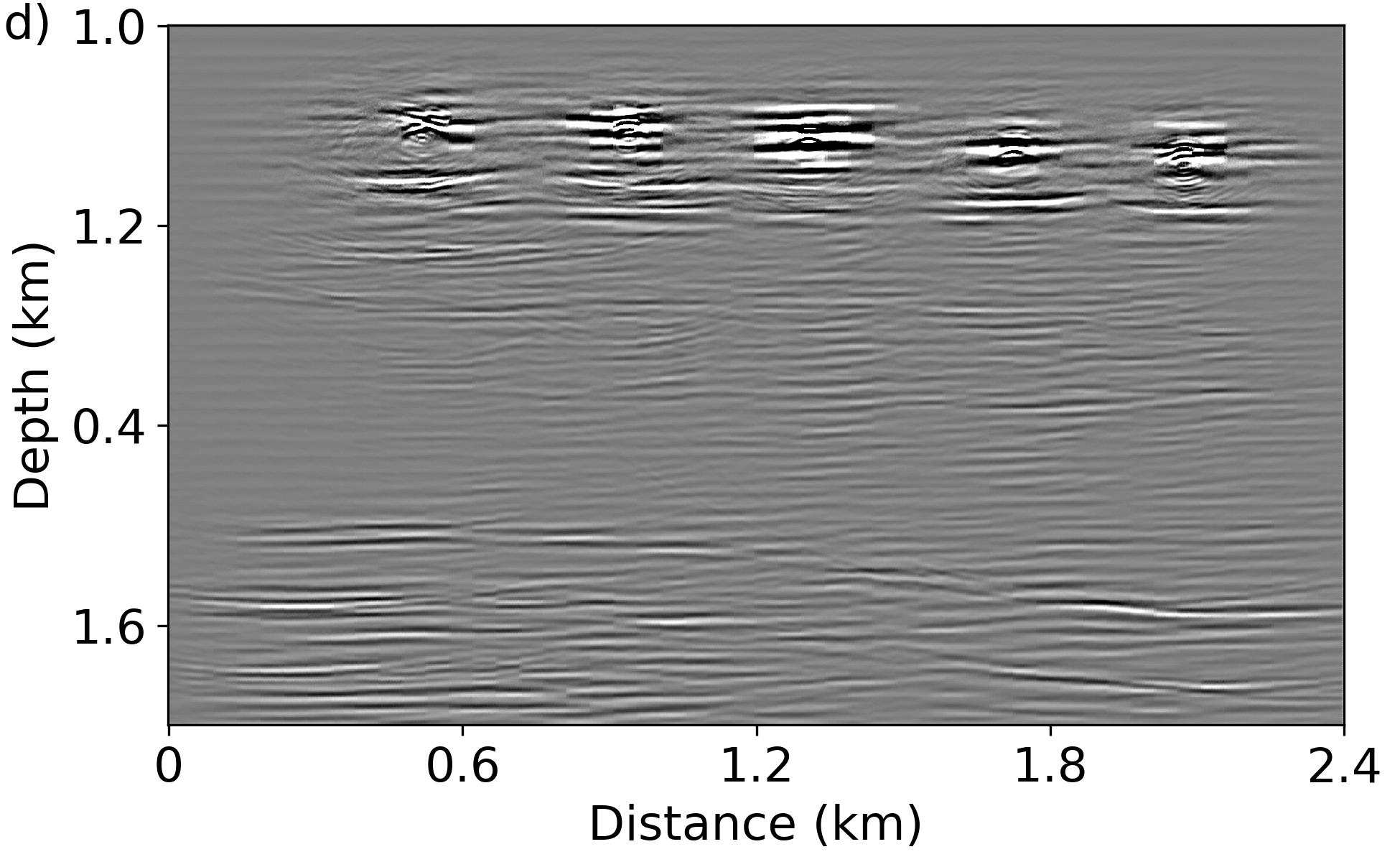} 
\hspace{0.3cm}
\includegraphics[width=0.45\textwidth]{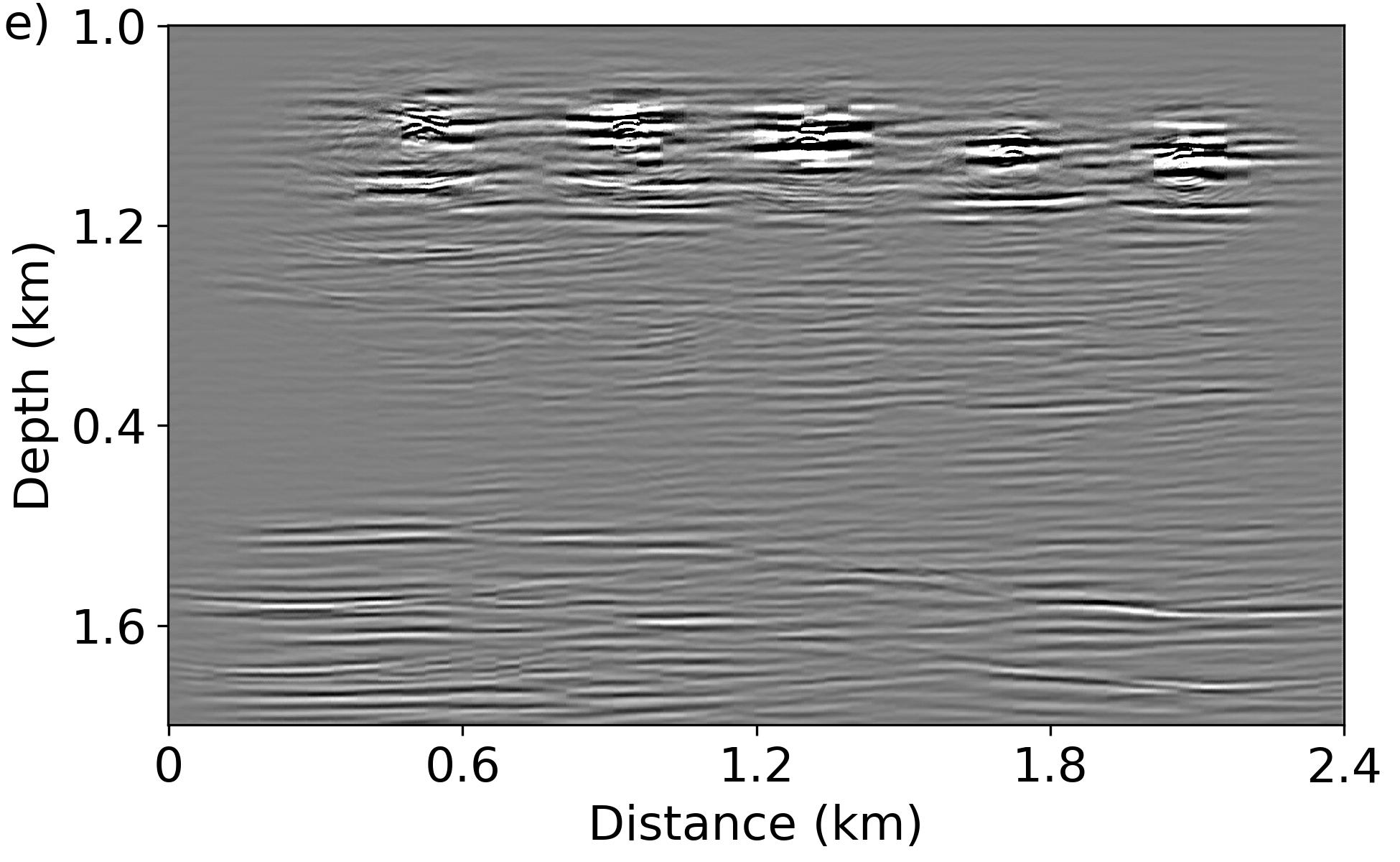}
\caption{Imaging enhancement inference on field data comparisons of neural networks trained with meta-learning initialization and random initialization on synthetic data. (a) Input image. (b) The prediction results from the meta initialization-based neural network with 10 epochs of training. (c), (d), and (e) are the prediction results from random initialization-base neural network with 10, 100, and 300 epochs of training, respectively.}
\label{fig18}
\end{figure}

Then, we assess the performance of the MLIN and RIN on the images acquired from real, sparse OBN surveys. The field data are collected from South China Sea at about 1100 meters water depth, using only five OBNs, with a node spacing of approximately 400 meters. A total of 154 shots are acquired with a spacing of 25 meters. Similar to the synthetic data, we employ common-receiver Gaussian beam migration to generate the input images. The corresponding imaging results, as well as the prediction results of the MLIN and RIN, are displayed in Figure \ref{fig18}. Figure \ref{fig18}a reveals that the sparse acquisition system leads to footprints and arc-shaped artifacts in the images, which considerably reduces image quality. Consequently, the proposed Meta-Processing algorithm helps the NN to overcome these issues effectively with only 10 epochs of optimizations on synthetic data, thereby enhancing the events' continuity and contributing to a significant enhancement in image quality (see Figure \ref{fig18}b). While the RIN removes some of the arc-shaped artifacts, it does not improve the events continuity (see Figure \ref{fig18}c-e). Moreover, it introduces noise and significantly degrades the image resolution, which is unacceptable. 

\subsubsection{Velocity estimation}
\begin{figure}[htp]
\centering
\includegraphics[width=0.45\textwidth]{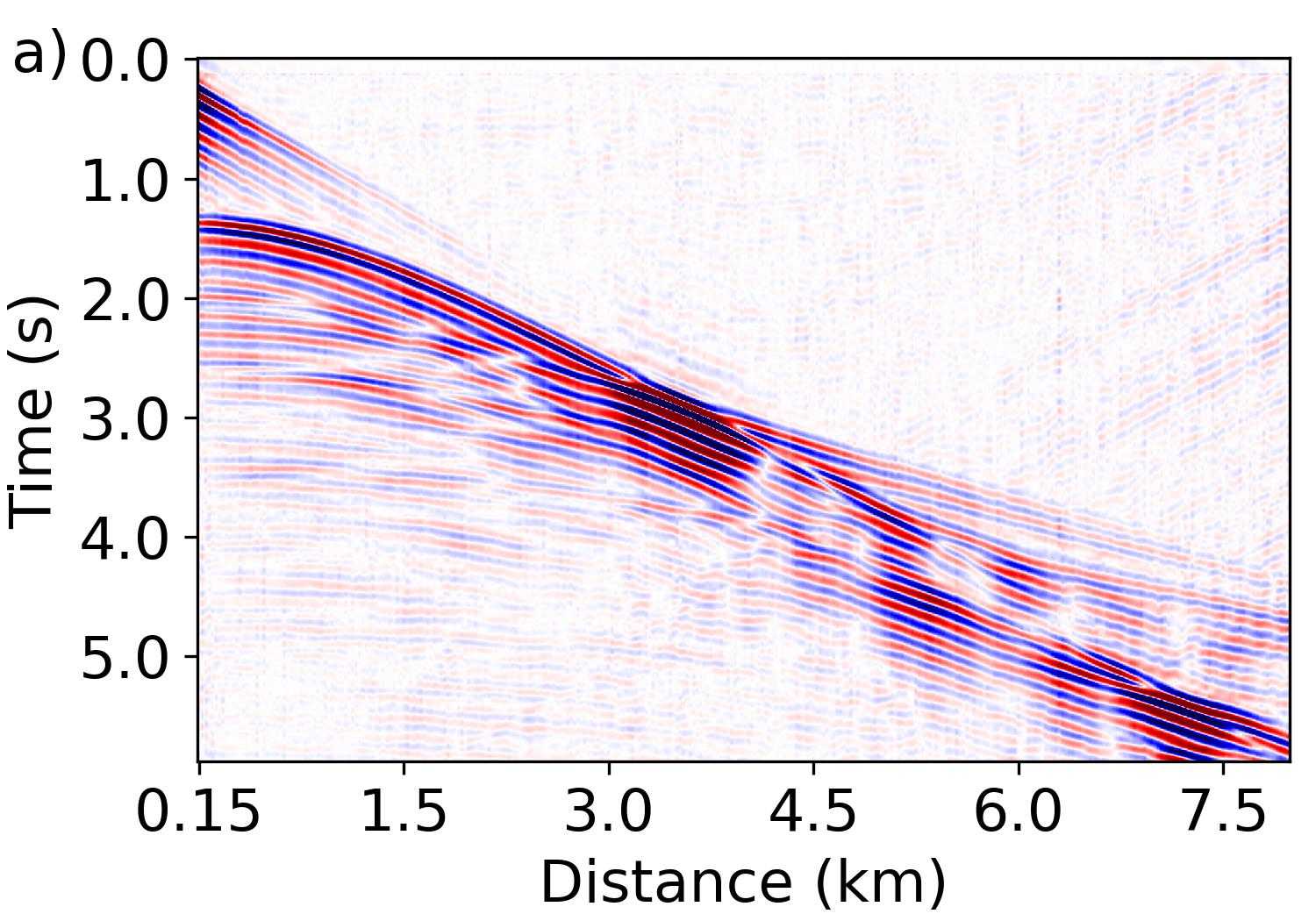} \\
\includegraphics[width=0.45\textwidth]{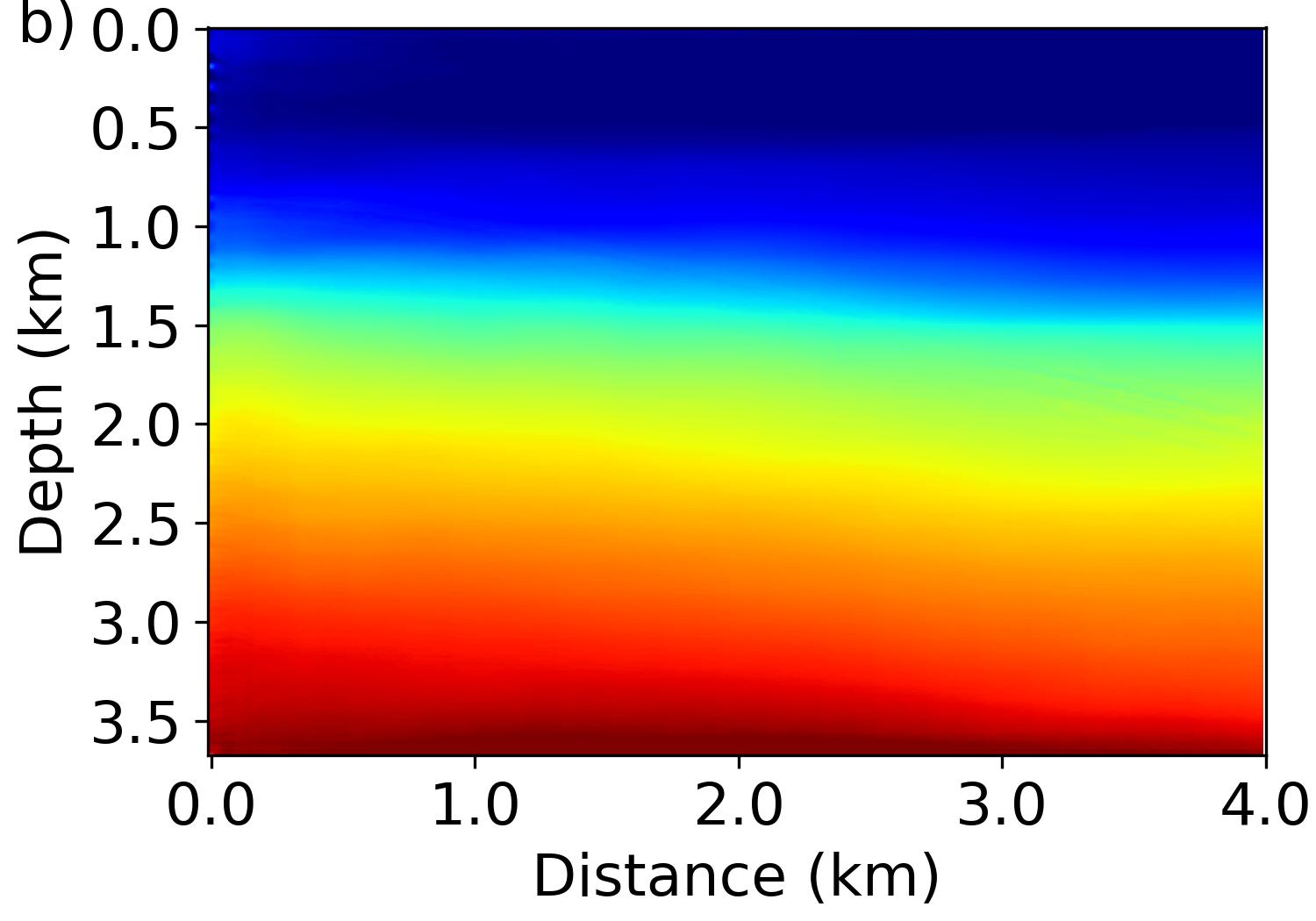} 
\hspace{0.3cm}
\includegraphics[width=0.45\textwidth]{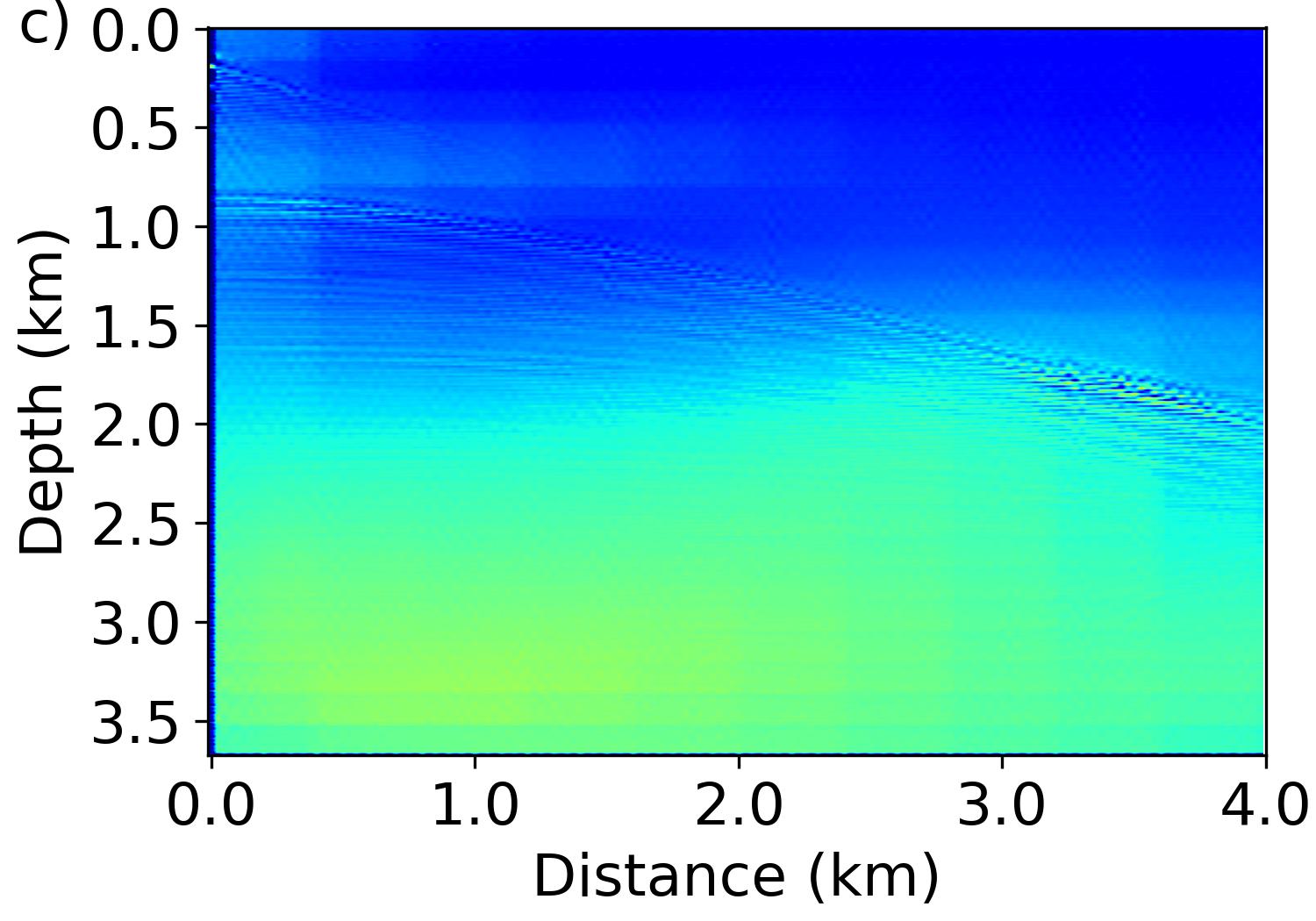} \\
\includegraphics[width=0.45\textwidth]{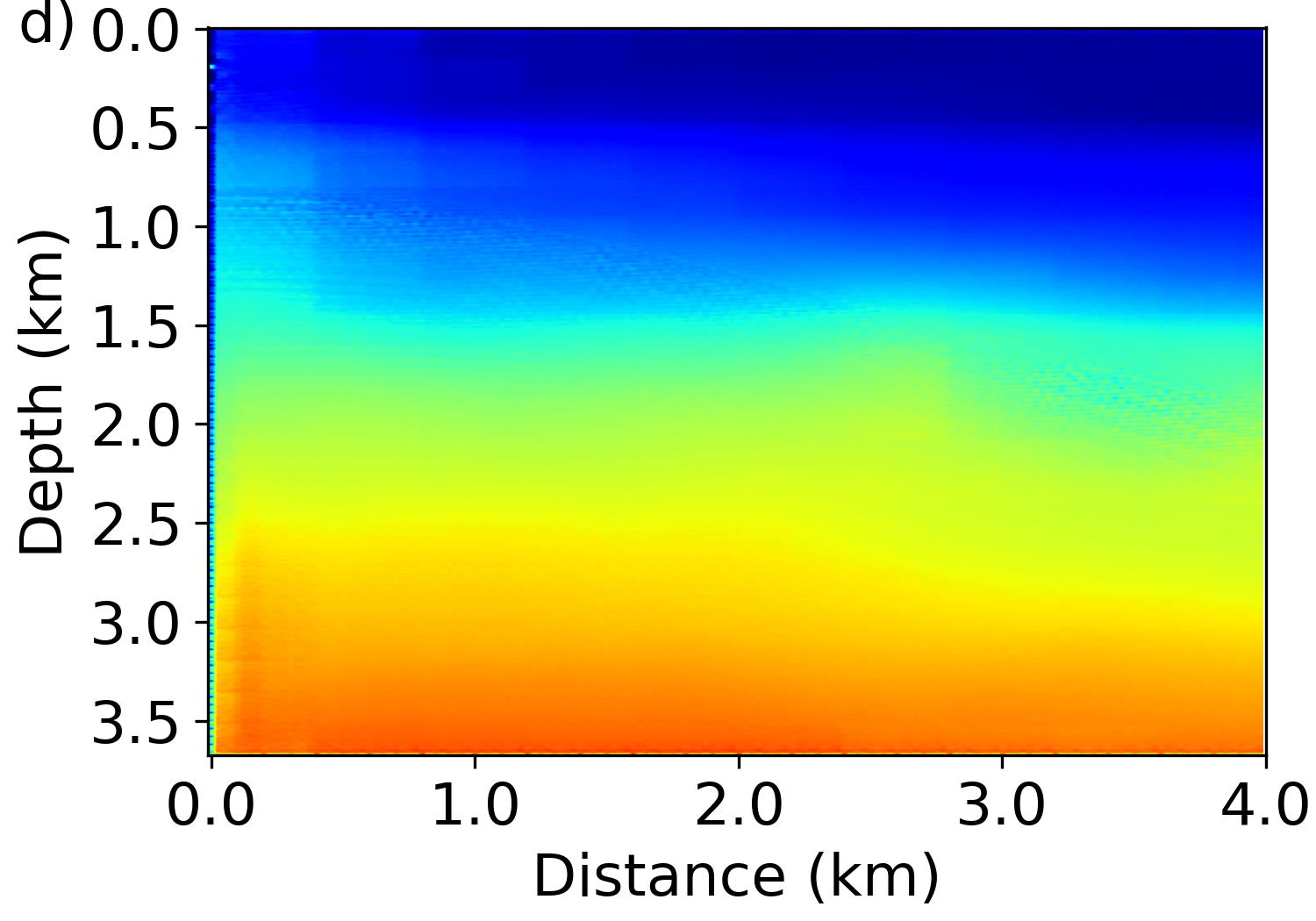} 
\hspace{0.3cm}
\includegraphics[width=0.45\textwidth]{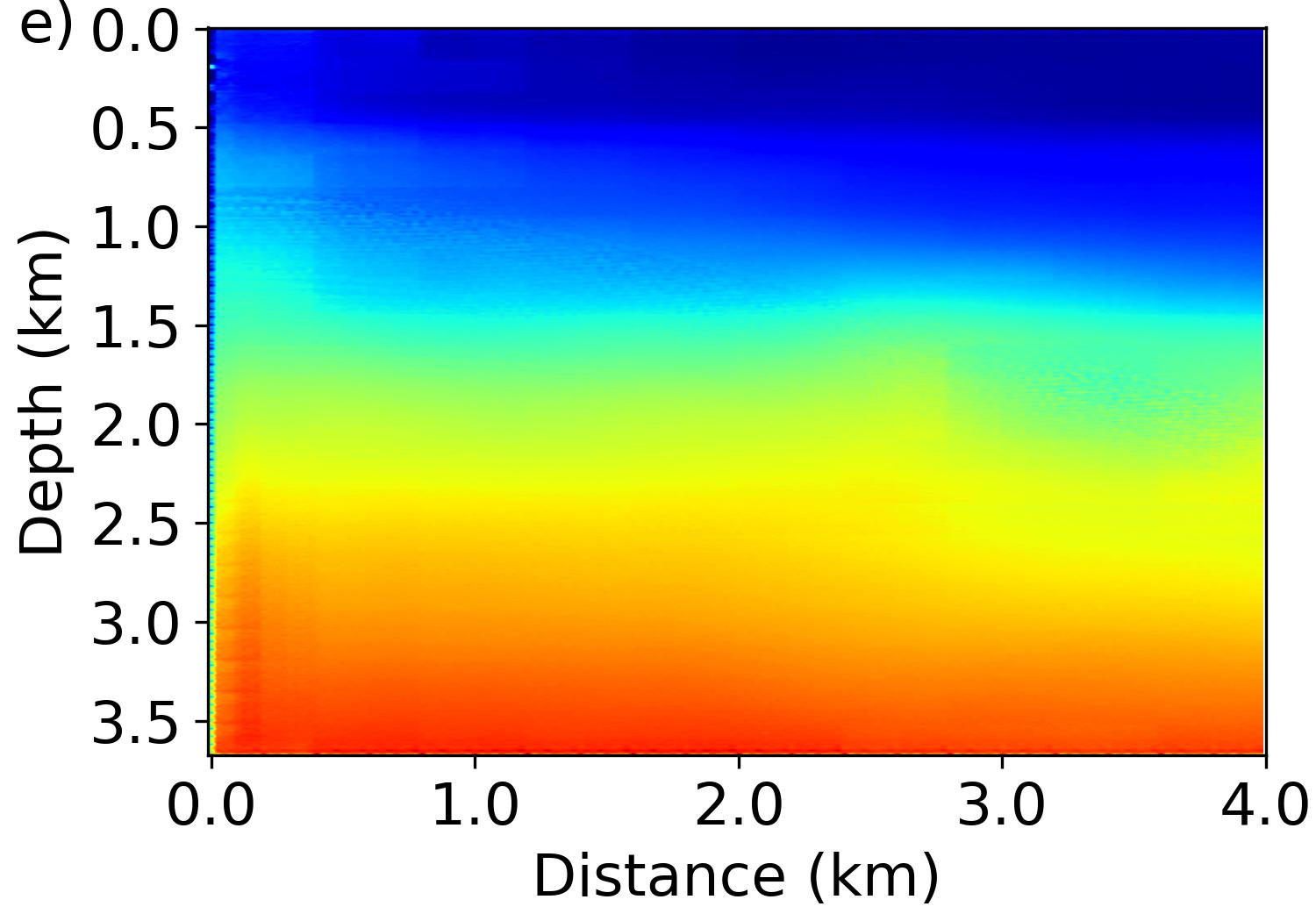}
\includegraphics[width=0.35\textwidth]{Figures/colorbar.jpg}
\caption{Velocity estimation inference on field data comparisons of neural networks trained with meta-learning initialization and random initialization on synthetic data. (a) Input field data. (b) The prediction results from the meta initialization-based neural network with 20 epoch training. (c), (d), and (e) are the prediction results from random initialization-base neural network with 20, 100, and 300 epochs of training, respectively.}
\label{fig19}
\end{figure}

\begin{figure}[htp]
\centering
\includegraphics[width=0.5\textwidth]{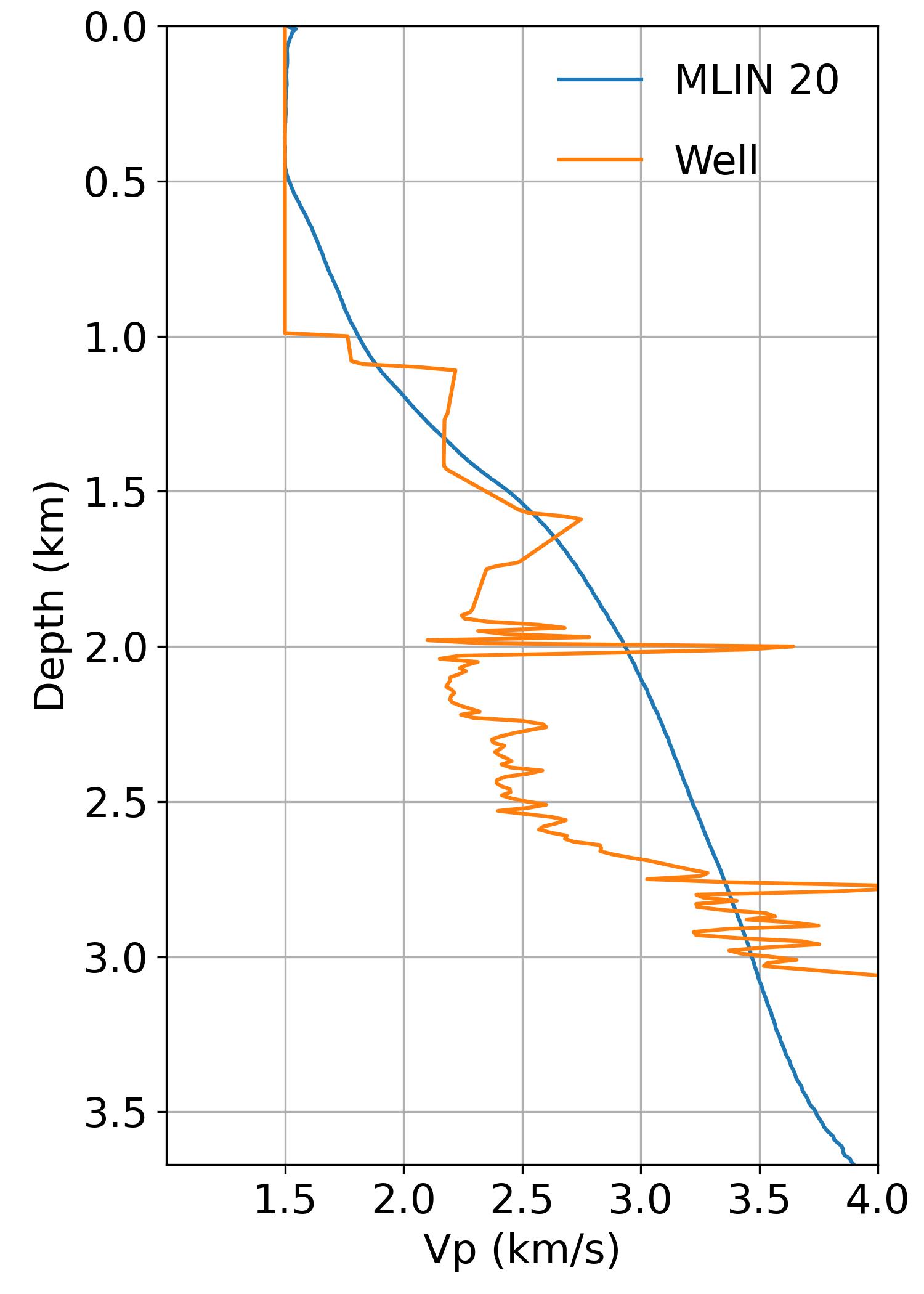}
\caption{Comparison between the predictions from the meta initialization-based neural network with 20 epochs of training and the well data. }
\label{fig20}
\end{figure}

We present the velocity estimation results of the MLIN and RIN on field data. The data was acquired in North West Australia by a streamer containing 324 hydrophones with a 25 m spacing that recorded 1824 shots, with an example of a shot gather shown in Figure \ref{fig19}. We apply the same procedure as for synthetic data to the field data. In Figure \ref{fig19}, panel (a) shows the input single-shot gather, (b) is the prediction result of the MLIN with 20 epochs of training, and (c), (d), and (e) display the prediction results of the RIN after training for 20, 100, and 300 epochs, respectively. As shown in Figure \ref{fig19}b, the MLIN yields a reasonable velocity estimation. However, as seen in the synthetic data, the RIN produces signal footprints and artifacts in the prediction results (see Figure \ref{fig19}c-e), which are obviously unsuitable to be used as an effective velocity model for guiding the next SPTs. To verify the accuracy of the velocity estimated by the MLIN, we compare the predicted velocity, obtained by averaging the predicted velocities of all traces, with the well velocity, as demonstrated in Figure \ref{fig20}. The predicted results are reasonably consistent with the well velocity, although some differences may exist. These differences may be due to the fact that the well velocity is usually lower than the seismic velocity, or field data may be affected by seismic anisotropy, which we do not consider in the synthetic data. \\

\section{Discussion}\label{sec4}
The field of seismic processing has long been a cornerstone in unraveling the mysteries of the Earth's interior. As seismic waves traverse through the Earth's various layers, they carry vital information about its composition, structure, and dynamics. The traditional paradigm of seismic processing often relied on handcrafted algorithms tailored to specific tasks. However, the rapidly evolving landscape of machine learning and NNs has allowed us to reframe this approach. In this study, we present a pioneering framework for multi-task seismic processing using NNs, which not only enhances the efficiency and accuracy of data processing but also unlocks novel insights into the shared features of seismic data across diverse geographical regions and geological settings. 

One of the pivotal aspects of our framework lies in its utilization of MetaL, a technique that empowers models to rapidly adapt to new tasks with limited labeled data. By learning the shared characteristics of seismic data, our approach provides a robust initialization for multi-task seismic processing. This strategic initialization facilitates the seamless adaptation of the model to various SPTs, even when data availability is constrained. Looking back at Figure 18, for example, Meta-Processing removed most of the arc-shaped artifacts and produced a continuous layering of the subsurface on the image, much better than that of the randomly-initialized network. This could lead to a more accurate interpretation of the subsurface and, thus, provides meticulous insight into the deep Earth. Another example was demonstrated in Figure 19, where our proposed framework yielded artifacts-free velocity estimates compared to the results of the randomly-initialized networks, which still contain seismic footprints. These clean $V_{rms}$ estimates are beneficial for the subsequent processing steps, aiding them to produce better imaging of the subsurface. The underlying implication of this achievement is profound – the presence of common features within seismic data suggests that despite the geological diversity and spatial separation, the Earth's interior shares inherent attributes and properties that influence the observed seismic waves. In other words, seismic waves, despite originating from distinct geographical regions and geological contexts, share common characteristics that can be exploited to enhance data processing efficiency and accuracy.

The discovery of shared seismic features through our Meta-Processing framework opens a new avenue for understanding the Earth's solid interior. While each seismic event may seem unique on the surface, our findings hint at an underlying universality in the behavior of seismic waves. The seismic waves recorded across various regions and geological formations appear to encode consistent traits, hinting at fundamental attributes within the Earth's crust and mantle. These shared features may be reflective of universal geological processes or structural arrangements that transcend regional differences, or might be rooted in fundamental principles governing the propagation of seismic waves through the Earth's interior. By delving deeper into these shared characteristics, we might uncover latent relationships between geological properties and seismic wave behaviors, contributing to a deeper comprehension of the Earth's physical properties and processes. \\
  
\section{Conclusion}\label{sec5}
We proposed a unified paradigm for neural network-based seismic processing, Meta-Processing, to provide a powerful technique for various seismic processing tasks (SPTs), which can be efficiently trained on the limited training data. We utilized a modified residual network baseline to replace the conventional convolutional layers of the classic UNet network as our basic network architecture. Within the framework of the Meta-Processing, the network training is performed in two stages: meta-training and meta-testing, each with distinct goals and implementation details. In the meta-training stage, a bilevel gradient descent updating from the support set to the query set is used to optimize the network parameters, aiming to obtain a robust initialization that can quickly adapt to various SPTs. In the meta-testing stage, the optimized network parameters from the meta-training stage are fine-tuned on various SPTs, with the same procedure to conventional supervised learning. The objective in this stage is to achieve rapid convergence and a significant improvement in prediction accuracy. 

We conducted comprehensive numerical tests to demonstrate the performance of our method on various SPTs, including denoising, interpolation, ground-roll attenuation, imaging enhancement, and velocity estimation. Results on both synthetic and field data showed the superiority of our method in terms of convergence speed and prediction accuracy, compared to randomly initialized UNets. Specifically, even when our method was only trained on synthetic data, it did not exhibit significant performance degradation when predicting on field data. In contrast, the randomly initialized network, even after prolonged optimization, resulted in a significant decrease in accuracy when applied to field data. This means that the Meta-Processing algorithm not only extracts shared seismic features but also enhances its adaptability from the synthetic data domain to the field data domain. 

Our framework's success signifies a paradigm shift in seismic processing. Our work demonstrated that a single neural network model can efficiently accommodate diverse tasks, thus enabling a holistic approach to seismic analysis. As a result, this study will help advance neural network-based seismic processing for various practical applications. \\
 
\section*{Code and Data Availability}
The  accompanying codes that support the findings of this study are openly available at \href{https://github.com/DeepWave-KAUST/Meta-Processing}{https://github.com/DeepWave-KAUST/Meta-Processing}. The data in this research are available at \href{https://doi.org/10.5281/zenodo.8187745}{https://doi.org/10.5281/zenodo.8187745}. \\ 
\section*{Acknowledgments}
This publication is based on work supported by the King Abdullah University of Science and Technology (KAUST). The authors thank the DeepWave sponsors for supporting this research. They also thank Guangzhou Marine Geological Survey, Tongji University, and CGG for sharing the field seismic data.

\bibliographystyle{unsrt}  
\bibliography{references}

\end{document}